\documentclass[journal,12pt,draftclsnofoot,onecolumn]{IEEEtran}
\IEEEoverridecommandlockouts

\usepackage{amsmath,amsfonts,mathtools}
\usepackage[noend]{algorithmic}
\usepackage{algorithm}
\usepackage{array}
\usepackage[caption=false,font=footnotesize]{subfig}
\usepackage{textcomp}
\usepackage{stfloats}
\usepackage{url}
\usepackage{color}
\usepackage{verbatim}
\usepackage{graphicx}
\usepackage{cite}
\usepackage[hidelinks]{hyperref}
\usepackage[nolist]{acronym}
\usepackage{bbm}
\usepackage{xifthen}
\def\BibTeX{{\rm B\kern-.05em{\sc i\kern-.025em b}\kern-.08em
    T\kern-.1667em\lower.7ex\hbox{E}\kern-.125emX}}

\graphicspath{{figures/}}

\begin{acronym}
\acro{5G}{fifth generation}
\acro{6G}{sixth generation}
\acro{AMP}{approximate message passing}
\acro{AP}{access point}
\acro{APP}{a posteriori probability}
\acro{AWGN}{additive white Gaussian noise}
\acro{B5G}{beyond fifth generation}
\acro{BPSK}{binary phase-shift-keying}
\acro{CDF}{cumulative distribution function}
\acro{CF-MaMIMO}{cell-free massive multiple-input multiple-output}
\acro{CPU}{central processing unit}
\acro{CSI}{channel state information}
\acro{DER}{detection error rate}
\acro{EP}{expectation propagation}
\acro{GaBP}{Gaussian belief propagation}
\acro{GF-CF-MaMIMO}{grant-free cell-free massive multiple-input multiple-output}
\acro{i.i.d.}{independent and identically distributed}
\acro{IoT}{Internet-of-Things}
\acro{JACD}{joint activity detection, channel estimation, and data detection}
\acro{JACD-EP}{joint activity detection, channel estimation, and data detection via expectation propagation}
\acro{JAC}{joint activity detection and channel estimation}
\acro{JAC-EP}{joint activity detection and channel estimation via expectation propagation}
\acro{JCD}{joint channel estimation and data detection}
\acro{KL}{Kullback-Leibler}
\acro{MaMIMO}{massive multiple-input multiple-output}
\acro{MIMO}{multiple-input multiple-output}
\acro{MAP}{maximum a posteriori}
\acro{MMSE}{minimum mean squared error}
\acro{MMV-AMP}{multiple measurement vector approximate message passing}
\acro{NMSE}{normalized mean squared error}
\acro{QAM}{quadrature amplitude modulation}
\acro{QoS}{quality of service}
\acro{SER}{symbol error rate}
\acro{UE}{user equipment}
\acro{UPP}{uniform point process}
\end{acronym}

\newcommand{\lvec}[1]{\ensuremath{\mathbf{#1}}}  		
\newcommand{\gvec}[1]{\ensuremath{\boldsymbol{#1}}}		
\newcommand{\lmat}[1]{\ensuremath{\mathbf{#1}}}  		
\newcommand{\gmat}[1]{\ensuremath{\boldsymbol{#1}}}		
\newcommand{\trace}[1]{\ensuremath{\mathrm{tr}\!\left\{#1\right\}}}
\newcommand*{\argmin}{\ensuremath{\mathop{\mathrm{arg\,min}}}}
\newcommand*{\argmax}{\ensuremath{\mathop{\mathrm{arg\,max}}}}
\newcommand{\est}[2]{\ensuremath{E_{#1}\!\left\{#2\right\}}}

\newcommand{\Cset}{\mathbb{C}}

\newcommand{\CN}[1]{\ensuremath{\mathcal{CN}\!\left(#1\right)}}

\newcommand{\cat}[1]{\ensuremath{\pi\!\left(#1\right)}}
\newcommand{\diag}[1]{\ensuremath{\mathrm{diag}\!\left\{#1\right\}}}
\newcommand{\proj}[1]{\ensuremath{\mathrm{proj}\!\left\{#1\right\}}}
\newcommand{\ind}[1]{\ensuremath{\mathbbm{1}_{#1}}}
\newcommand{\pilot}[1]{\ensuremath{#1^{p}}}
\newcommand{\data}[1]{\ensuremath{#1^{d}}}
\newcommand{\msg}[2]{\ensuremath{m_{#1;#2}}}

\newcommand{\msgp}[2]{\ensuremath{p_{#1;#2}}}
\newcommand{\mmd}[2]{\ensuremath{q_{#1;#2}}}
\newcommand{\catmsg}[2]{\ensuremath{\pi_{#1;#2}}}
\newcommand{\Mumsg}[2]{\ensuremath{\gvec{\mu}_{#1;#2}}}
\newcommand{\Cmsg}[2]{\ensuremath{\lvec{C}_{#1;#2}}}
\newcommand{\Gammamsg}[2]{\ensuremath{\gvec{\gamma}_{#1;#2}}}
\newcommand{\Lambdamsg}[2]{\ensuremath{\gvec{\Lambda}_{#1;#2}}}
\newcommand{\Mumsga}[2]{\ifthenelse{\isempty{#2}}{\ensuremath{\check{\gvec{\mu}}_{#1}}}{\ensuremath{\check{\gvec{\mu}}_{#1;#2}}}}
\newcommand{\Cmsga}[2]{\ifthenelse{\isempty{#2}}{{\ensuremath{\check{\lvec{C}}_{#1}}}}{\ensuremath{\check{\lvec{C}}_{#1;#2}}}}
\newcommand{\Gammamsga}[2]{\ifthenelse{\isempty{#2}}{\ensuremath{\check{\gvec{\gamma}}_{#1}}}{\ensuremath{\check{\gvec{\gamma}}_{#1;#2}}}}
\newcommand{\Lambdamsga}[2]{\ifthenelse{\isempty{#2}}{\ensuremath{\check{\gvec{\Lambda}}_{#1}}}{\ensuremath{\check{\gvec{\Lambda}}_{#1;#2}}}}
\newcommand{\Mumsgb}[2]{\ifthenelse{\isempty{#2}}{\ensuremath{\hat{\gvec{\mu}}_{#1}}}{\ensuremath{\hat{\gvec{\mu}}_{#1;#2}}}}
\newcommand{\Cmsgb}[2]{\ifthenelse{\isempty{#2}}{{\ensuremath{\hat{\lvec{C}}_{#1}}}}{\ensuremath{\hat{\lvec{C}}_{#1;#2}}}}
\newcommand{\Gammamsgb}[2]{\ifthenelse{\isempty{#2}}{\ensuremath{\hat{\gvec{\gamma}}_{#1}}}{\ensuremath{\hat{\gvec{\gamma}}_{#1;#2}}}}
\newcommand{\Lambdamsgb}[2]{\ifthenelse{\isempty{#2}}{\ensuremath{\hat{\gvec{\Lambda}}_{#1}}}{\ensuremath{\hat{\gvec{\Lambda}}_{#1;#2}}}}
\begin{document}

\title{Distributed Joint User Activity Detection, Channel Estimation, and Data Detection via Expectation Propagation in Cell-Free Massive MIMO
\thanks{This work was funded by the Deutsche Forschungsgemeinschaft (DFG, German Research Foundation) – Project CO 1311/1-1,  Project ID 491320625.}
\thanks{Christian Forsch, Alexander Karataev, and Laura Cottatellucci are with the Institute for Digital Communications, Friedrich-Alexander-Universität Erlangen-Nürnberg, Erlangen, Germany (e-mail: christian.forsch@fau.de; alexander.karataev@fau.de; laura.cottatellucci@fau.de).}
}

\author{Christian~Forsch,~\IEEEmembership{Graduate Student Member,~IEEE,} Alexander~Karataev, and~Laura~Cottatellucci,~\IEEEmembership{Member,~IEEE}}

\maketitle

\begin{abstract}
We consider the uplink of a \acl{GF-CF-MaMIMO} (GF-CF-MaMIMO)\acused{GF-CF-MaMIMO} system.
We propose an algorithm for distributed \ac{JACD} based on \ac{EP} called \acs{JACD-EP}\acused{JACD-EP}.
We develop the algorithm by factorizing the \ac{APP} of activities, channels, and transmitted data, then, mapping functions and variables onto a factor graph, and finally, performing a message passing on the resulting factor graph.
If users with the same pilot sequence are sufficiently distant from each other,  the \ac{JACD-EP} algorithm is able to mitigate the effects of pilot contamination which naturally occurs in grant-free systems due to the large number of potential users and limited signaling resources.
Furthermore, it outperforms state-of-the-art algorithms for \ac{JACD} in \ac{GF-CF-MaMIMO} systems.
\end{abstract}

\begin{IEEEkeywords}
Expectation propagation, activity detection, channel estimation, data detection, grant-free cell-free massive MIMO.
\end{IEEEkeywords}

\acresetall

\section{Introduction}\label{sec:intro}
\Ac{CF-MaMIMO}\acused{MIMO}\acused{MaMIMO} networks are considered a promising enabler for energy efficient \ac{6G} wireless communication systems with ubiquitous coverage and high data rates~\cite{Ngo2017,Ngo2018,Yang2018, Ammar2022}.
Here, a large number of potential \acp{UE} communicate with a \ac{CPU} via distributed  \acp{AP}.
For sporadic burst traffic as in the \ac{IoT}, grant-free random access schemes reduce the communication overhead, decoding latency, and enable an efficient resource utilization~\cite{Liu2018b}.
In this context, \ac{MaMIMO} are especially well-suited~\cite{Liu2018b} and grant-free \ac{CF-MaMIMO} (\acs{GF-CF-MaMIMO})\acused{GF-CF-MaMIMO} systems are considered in, e.g.,~\cite{Iimori2021,Gao2024}. In \ac{GF-CF-MaMIMO}, the task of the \ac{CPU} is to identify the active \acp{UE} and estimate their transmitted data. Furthermore, accurate channel estimation is necessary for reliable data detection.
However, a major problem in grant-free systems is pilot contamination which naturally occurs due to the large number of \acp{UE} and limited signaling resources.
Besides, in contrast to centralized \ac{MaMIMO}, channel hardening and favorable propagation typically do not hold in cell-free systems~\cite{Yin2014,Chen2018,Gholami2020a,Gholami2020b} which further exacerbates  the problem of pilot contamination and renders existing pilot decontamination solutions for centralized \ac{MaMIMO}~\cite{Ngo2012,Yin2013,Cottatellucci2013,Mueller2014,Yin2016} ineffective.
Hence, novel efficient algorithms for \ac{JACD} are necessary to mitigate pilot contamination in grant-free cell-free systems.

In this work, we propose a novel \ac{JACD} message-passing algorithm based on \ac{EP} called \acs{JACD-EP}\acused{JACD-EP}.
\Ac{EP} is an approximate inference technique which iteratively computes a tractable approximation of a factorized probability distribution by projecting each factor onto an exponential family~\cite{Minka2001a,Minka2001b}.
The \ac{EP} algorithm has been already applied, among others, to \ac{JAC} in massive machine-type communications~\cite{Ahn2018} and \ac{JCD} in \ac{CF-MaMIMO}~\cite{Karataev2024}.
For  grant-free communication systems, also other approximate inference techniques were considered such as \ac{AMP} which was applied to \ac{JAC} in~\cite{Liu2018a,Ke2020,Jiang2023}.
Furthermore, bilinear \ac{GaBP} was applied in~\cite{Iimori2021} for \ac{JACD}.

This paper is organized as follows.
In Section~\ref{sec:sys_mod}, we introduce the \ac{GF-CF-MaMIMO} system model and formulate the inference problem.
In Section~\ref{sec:JACD-EP}, we present the novel \ac{JACD-EP} algorithm.
Then, the performance of the proposed algorithm is evaluated in Section~\ref{sec:sims}.
Finally, some conclusions are drawn.

\textit{Notation:}
Lower case, bold lower case, and bold upper case letters, e.g., $x,\lvec{x},\lmat{X}$, represent scalars, vectors and matrices, respectively.
$\lmat{I}_N$ is the $N$-dimensional identity matrix.
$\diag{\cdot}$ is a diagonal matrix with the elements in brackets on the main diagonal.
$\delta(\cdot)$ is the Dirac delta function.
The indicator function $\ind{(\cdot)}$ takes value one if the condition in the subscript is satisfied and zero otherwise.
$(\cdot)^T$ and $(\cdot)^H$ denote the transposition and complex conjugate transposition operation, respectively.
The trace of a matrix $\lmat{X}$ is denoted by $\trace{\lmat{X}}$.
$|\mathcal{S}|$ stands for the cardinality of the set $\mathcal{S}$.
$\est{}{\cdot}$ denotes the expectation operator.
$\CN{\lvec{x}|\gvec{\mu},\lmat{C}}$ represents the circularly-symmetric multivariate complex Gaussian distribution of a complex-valued vector $\lvec{x}$ with mean $\gvec{\mu}$ and covariance matrix $\lmat{C}$.
$\cat{x}$ denotes the categorical distribution of a discrete random variable $x$.
The notation $x\sim p$ indicates that the random variable $x$ follows the distribution $p$.
{The message sent from the factor node $\Psi_\alpha$ to the variable node $\lvec{x}_\beta$ in a factor graph is denoted as $\msg{\Psi_\alpha}{\lvec{x}_\beta}$ and consists of parameters of the distribution $\msgp{\Psi_\alpha}{\lvec{x}_\beta}(\lvec{x}_\beta)$ in the exponential family which are denoted with the same subscript of the message, e.g., mean $\Mumsg{\Psi_\alpha}{\lvec{x}_\beta}$ and covariance matrix $\Cmsg{\Psi_\alpha}{\lvec{x}_\beta}$ for a Gaussian random variable $\lvec{x}_\beta$ with distribution $\msgp{\Psi_\alpha}{\lvec{x}_\beta}(\lvec{x}_\beta)=\CN{\lvec{x}|\Mumsg{\Psi_\alpha}{\lvec{x}_\beta},\Cmsg{\Psi_\alpha}{\lvec{x}_\beta}}$ or probability values $\catmsg{\Psi_\alpha}{\lvec{x}_\beta}(\lvec{x}_\beta)$ for a categorical random variable $\lvec{x}_\beta$ with distribution $\msgp{\Psi_\alpha}{\lvec{x}_\beta}(\lvec{x}_\beta)=\catmsg{\Psi_\alpha}{\lvec{x}_\beta}(\lvec{x}_\beta)$.}
The same holds for variable-to-factor messages $\msg{\lvec{x}_\beta}{\Psi_\alpha}$.

\section{System Model}\label{sec:sys_mod}

\subsection{\acs{GF-CF-MaMIMO}}\label{subsec:GF_CF-MaMIMO}
We consider the uplink of a \ac{GF-CF-MaMIMO} system with $L$ geographically distributed \acp{AP}, serving $K$ synchronized single-antenna \acp{UE}.
Each \ac{AP} is equipped with $N$ antennas.
All \acp{AP} are connected to a \ac{CPU} via fronthaul links.
Only a fraction of the $K$ \acp{UE} are active and transmit data simultaneously.
The received signal at the $l^\text{th}$ \ac{AP} at channel use $t$ $\lvec{y}_{l,t}\in\Cset^{N\times1}$ is given by
\begin{equation}
	\lvec{y}_{l,t} = \lmat{H}_l\lmat{U}\lvec{x}_t + \lvec{n}_{l,t} = \sum_{k=1}^K\lvec{h}_{l,k}u_kx_{kt} + \lvec{n}_{l,t},
	\label{eq:y_lt}
\end{equation}
where  $\lmat{H}_l=[\lvec{h}_{l,1}\cdots\lvec{h}_{l,K}]\!\in\!\Cset^{N\times K}$ is the channel matrix of \ac{AP} $l$ and $\lvec{h}_{l,k}\!\in\!\Cset^{N\times1}$ is the channel between \ac{AP} $l$ and \ac{UE} $k$;
$\lmat{U}=\diag{u_1,\dots,u_K}\in\{0,1\}^{K\times K}$ is the diagonal matrix of user activities whose $k^\text{th}$ diagonal element $u_k$ is equal to one if \ac{UE} $k$ is active and zero otherwise;
$\lvec{x}_t\in\Cset^{K\times1}$ is the transmit vector at channel use $t$ whose $k^\text{th}$ entry $x_{kt}$ denotes the transmit symbol of \ac{UE} $k$ at channel use $t$;
and $\lvec{n}_{l,t}\in\Cset^{N\times1}$ is the vector of \ac{AWGN} at the $l^\text{th}$ \ac{AP} at channel use $t$ with $\lvec{n}_{l,t} \sim\CN{\lvec{n}_{l,t}|\lvec{0}_N,\sigma_n^2\lmat{I}_N}$.
The channel and the user activity are assumed to be constant during $T$ channel uses which correspond to the channel coherence time.
We group the receive, transmit, and \ac{AWGN} vectors over $T$ channel uses in the matrices $\lmat{Y}_l=[\lvec{y}_{l,1}\cdots\lvec{y}_{l,T}]\!\in\!\Cset^{N\times T}$, $\lmat{X}=[\lvec{x}_1\cdots\lvec{x}_T]\!\in\!\Cset^{K\times T}$, and $\lmat{N}_l=[\lvec{n}_{l,1}\cdots\lvec{n}_{l,T}]\!\in\!\Cset^{N\times T}$, respectively.
The channel between \ac{UE} $k$ and \ac{AP} $l$ is assumed to be block Rayleigh fading, i.e., $\lvec{h}_{l,k}\sim p_{h_{l,k}}(\lvec{h}_{l,k})=\CN{\lvec{h}_{l,k}|\lvec{0}_N,\gmat{\Xi}_{l,k}}$ where $\gmat{\Xi}_{l,k}\in\Cset^{N\times N}$ is the  spatial correlation matrix and $\xi_{l,k}=\frac{1}{N}\trace{\gmat{\Xi}_{l,k}}$ is the large-scale fading coefficient.
The activity indicator of user $k$ is drawn from a Bernoulli distribution, $u_k\sim p_u(u_k)=(1-\lambda)\ind{u_k=0}+\lambda\ind{u_k=1}$ where $\lambda$ denotes the probability that a user is active.
The transmit matrix consists of a pilot matrix $\pilot{\lmat{X}}\in\Cset^{K\times T_p}$ {with known pilot symbols $\pilot{x}_{kt}$} and a data matrix $\data{\lmat{X}}\in\mathcal{X}^{K\times T_d}$, i.e., $\lmat{X}=[\pilot{\lmat{X}}\;\data{\lmat{X}}],$ with $T_p+T_d=T$ and $\mathcal{X}$ being the transmit symbol constellation of cardinality $M=|\mathcal{X}|$ which does not contain the zero-symbol, i.e., $0\notin\mathcal{X}$.
The average symbol transmit power is $\sigma_x^2=\est{}{|x_{kt}|^2}$.
A similar decomposition holds for the receive matrix, i.e., $\lmat{Y}_l=[\pilot{\lmat{Y}}_l\;\data{\lmat{Y}}_l]$ with received pilots $\pilot{\lmat{Y}}_l\in\Cset^{N\times T_p}$ and received data $\data{\lmat{Y}}_l\in\Cset^{N\times T_d}$.
Furthermore, the pilot length is much smaller than the number of \acp{UE}, $T_p\ll K$, since the number of \acp{UE} is usually very large and, thus, it is not practical to assign orthogonal pilot sequences to the \acp{UE}.

\subsection{Problem Formulation}\label{subsec:problem}
The received signals of all \acp{AP} $l\in\{1,\dots,L\}$ can be written compactly by the global equation
\begin{equation}
	\lmat{Y} = \lmat{H}\lmat{U}\lmat{X} + \lmat{N},
	\label{eq:Y}
\end{equation}
where, for  $\lmat{A} \equiv \{\lmat{Y},\lmat{H}, \lmat{N} \}$, $\lmat{A}=[\lmat{A}_1^T\cdots\lmat{A}_L^T]^T.$
The task of the receiver is to jointly estimate the user activity, channel, and user data matrices $\lmat{U}$,  $\lmat{H}$, and $\data{\lmat{X}}$, respectively.
The \ac{MAP} estimator is given by
\begin{equation}
	(\hat{\lmat{U}},\hat{\lmat{H}},\data{\hat{\lmat{X}}}) = \argmax_{\lmat{U},\lmat{H},\data{\lmat{X}}}\;p_\text{APP}(\lmat{U},\lmat{H},\data{\lmat{X}}),
	\label{eq:MAP}
\end{equation}
where the \ac{APP} distribution $p_\text{APP}(\lmat{U},\lmat{H},\data{\lmat{X}})$ can be factorized as follows by applying the Bayes theorem,
\begin{align}
	p_\text{APP}(\lmat{U},\lmat{H},\data{\lmat{X}})&=p_{U,H,\data{X}|Y,\pilot{X}}(\lmat{U},\lmat{H},\data{\lmat{X}}|\lmat{Y},\pilot{\lmat{X}})\nonumber\\
	&\propto p_{Y|U,H,X}(\lmat{Y}|\lmat{U},\lmat{H},\lmat{X})\cdot p_{U}(\lmat{U})\cdot p_{H}(\lmat{H})\cdot p_{X}(\lmat{X}).
	\label{eq:APP}
\end{align}
Maximizing~\eqref{eq:APP} with respect to $\lmat{U}$, $\lmat{H}$, and $\data{\lmat{X}}$ is practically not feasible.
Hence, in the following section, we propose an inference technique yielding an approximation of $p_\text{APP}(\lmat{U},\lmat{H},\data{\lmat{X}})$ for low-complexity \ac{JACD}.

\section{\acs{JACD-EP} Algorithm}\label{sec:JACD-EP}
In this section we present the proposed \acs{JACD-EP} algorithm. The algorithm is obtained by introducing a convenient factorization of $p_\text{APP}(\lmat{U},\lmat{H},\data{\lmat{X}})$  which induces a factor graph as shown in Section \ref{subsec:FG}, then selecting parametric representations of distributions from the exponential family to approximate the \ac{APP} distribution, see Section \ref{subsec:EP_approx_fh_load}, and finally,  applying \ac{EP} message-passing rules on a factor graph, e.g., \cite{Ngo2020,Karataev2024}, as detailed in Section \ref{subsec:MP_updates}.

\subsection{Factor Graph Representation}\label{subsec:FG}
\begin{figure}[t]
    \centerline{\includegraphics[width=0.592\textwidth]{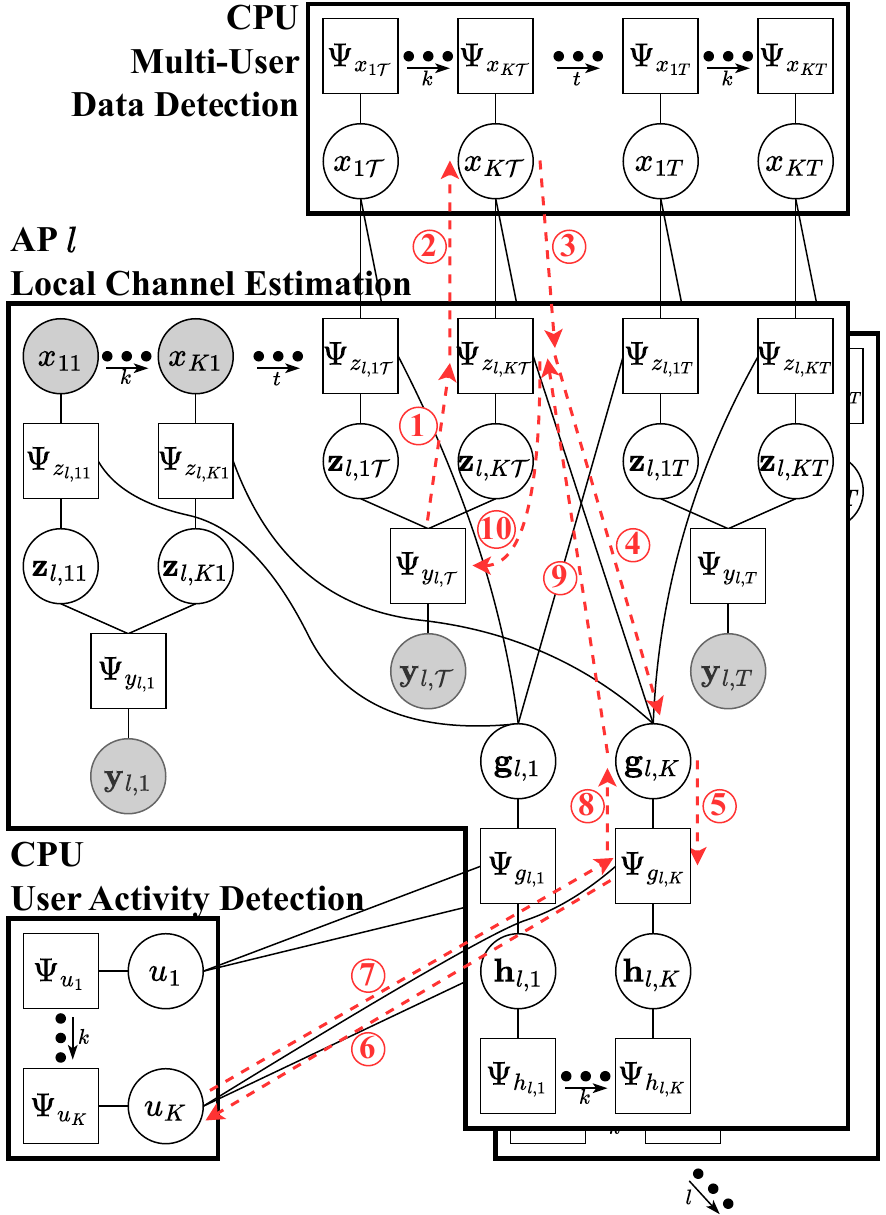}}\vspace*{-6mm}
    \caption{Factor graph for \ac{JACD-EP} with $\mathcal{T}\coloneq T_p+1$. The numbered red dashed arrows show the flow of information according to the scheduling presented in Algorithm~\ref{alg:JACD-EP}. Each number corresponds to one message update in Algorithm~\ref{alg:JACD-EP}.}\vspace*{-8mm}
    \label{fig:FG}
\end{figure}

Similar to \cite{Karataev2024}, to decouple activities, channels, and data of \acp{UE}, we introduce the auxiliary variables $\lvec{g}_{l,k}\coloneq \lvec{h}_{l,k}u_k,$ and $\lvec{z}_{l,kt} \coloneq \lvec{g}_{l,k}x_{kt}.$
We summarize all auxiliary variables in the matrices $\lmat{G}$ and $\lmat{Z}$, respectively.
The joint \ac{MAP} estimator of the user activity, channel, and data as well as the auxiliary variables maximizes the \ac{APP} distribution given by
\begin{equation}
\begin{split}
	p_\text{APP}(\lmat{U},\lmat{H},\data{\lmat{X}},\lmat{G},\lmat{Z})\propto\prod_{l=1}^L\prod_{k=1}^K\prod_{t=1}^T\Big[&p(\lvec{y}_{l,t}|\lvec{z}_{l,1t},\!...,\lvec{z}_{l,Kt})\cdot p(\lvec{z}_{l,kt}|\lvec{g}_{l,k},x_{kt})\\
	&\cdot p(\lvec{g}_{l,k}|\lvec{h}_{l,k},u_k)\cdot\tilde{p}_{u_k}(u_k)\cdot\tilde{p}_{h_{l,k}}(\lvec{h}_{l,k})\cdot p_x(x_{kt})\Big],
	\label{eq:APP_aux}
\end{split}
\end{equation}
which is factorized by taking into account the independence of channel vectors  for different \acp{AP} and \acp{UE}, user activities for different \acp{UE}, and data symbols for different \acp{UE} and time indices.
Furthermore, $\tilde{p}_{u_k}(u_k)$ and $\tilde{p}_{h_{l,k}}(\lvec{h}_{l,k})$ denote improved prior information of the user activity $u_k$ and the channel $\lvec{h}_{l,k}$, respectively, which can be acquired by a pilot-based initialization algorithm.
A possible initialization algorithm is described in Section~\ref{subsec:init_sched_alg}.
We denote the corresponding new prior information for $u_k$ and $\lvec{h}_{l,k}$ by the two probabilities $\tilde{p}_{u_k}(0)$ and $\tilde{p}_{u_k}(1)$, and the mean vector $\tilde{\gvec{\mu}}_{h_{l,k}}$ and covariance matrix $\tilde{\lmat{C}}_{h_{l,k}}$, respectively.
The probability distributions in~\eqref{eq:APP_aux} are represented by factor nodes (rectangles) in the factor graph illustrated in Fig.~\ref{fig:FG} and are given by\vspace*{-3mm}
\begin{alignat}{2}
	&\Psi_{y_{l,t}} &&\coloneq p(\lvec{y}_{l,t}|\lvec{z}_{l,1t},\!...,\lvec{z}_{l,Kt}) = \CN{\lvec{y}_{l,t}\Bigg|\sum_{k=1}^K\lvec{z}_{l,kt},\sigma_n^2\lmat{I}_N},
	\label{eq:p_y_lt}\\
	&\Psi_{z_{l,kt}} &&\coloneq p(\lvec{z}_{l,kt}|\lvec{g}_{l,k},x_{kt}) = \delta(\lvec{z}_{l,kt}-\lvec{g}_{l,k}x_{kt}),
	\label{eq:p_z_lkt}\\
	&\Psi_{g_{l,k}} &&\coloneq p(\lvec{g}_{l,k}|\lvec{h}_{l,k},u_k) = \delta(\lvec{g}_{l,k}-\lvec{h}_{l,k}u_k),
	\label{eq:p_g_lk}\\
	&\Psi_{u_k} &&\coloneq \tilde{p}_{u_k}(u_k) = \tilde{p}_{u_k}(0)\ind{u_k=0}+\tilde{p}_{u_k}(1)\ind{u_k=1},
	\label{eq:p_u_k}\\
	&\Psi_{h_{l,k}} &&\coloneq \tilde{p}_{h_{l,k}}(\lvec{h}_{l,k}) = \CN{\lvec{h}_{l,k}\big|\tilde{\gvec{\mu}}_{h_{l,k}},\tilde{\lmat{C}}_{h_{l,k}}},
	\label{eq:p_h_lk}\\
	&\Psi_{x_{kt}} &&\coloneq p_x(x_{kt}) = \begin{cases}
 {\ind{x_{kt}=\pilot{x}_{kt}}}&\text{for }t\leq T_p,\\
 \frac{1}{M}\ind{x_{kt}\in\mathcal{X}}&\text{for }t>T_p.
        \end{cases}
	\label{eq:p_x_kt}
\end{alignat}
The variables in the above equations correspond to variable nodes (circles) in the factor graph.
It can be observed that the \ac{CPU} combines the information from the \acp{AP} to estimate user activities and data.
However, the channels are estimated locally in each \ac{AP} and there is no need to forward them to the \ac{CPU}.

\subsection{\acs{EP} Approximations and Fronthaul Load}\label{subsec:EP_approx_fh_load}
To apply the \ac{EP} message-passing rules to the factor graph in Fig.~\ref{fig:FG}, we assign a parametric representation of a distribution in the exponential family to each variable node to approximate the corresponding \ac{APP} distribution.
Categorical distributions are chosen for the variables $x_{kt}$ and $u_k$ whereas the variables $\lvec{z}_{l,kt}$, $\lvec{g}_{l,k}$, and $\lvec{h}_{l,k}$ are modeled by multivariate complex Gaussian distributions.
Thus, messages from and towards $x_{kt}$ for $t>T_p$ consist of $M-1$ probabilities while no messages are exchanged for $t\leq T_p$ since the pilot symbols are already known.
For $u_k$, one real value suffices to describe its distribution.
Hence, the total fronthaul load per iteration amounts to $2LK(T_d(M-1)+1)$ real-valued numbers where the factor two stems from the fact that the messages are sent from the \ac{CPU} towards the \acp{AP} and vice versa.
Messages involving $\lvec{z}_{l,kt}$, $\lvec{g}_{l,k}$, and $\lvec{h}_{l,k}$ consist of complex-valued vectors and matrices of dimension $N$ and $N\times N$, respectively, which do not contribute to the fronthaul load since they are only processed within an \ac{AP}.

\subsection{Initialization, Scheduling, and Estimation}\label{subsec:init_sched_alg}
To initialize the \ac{JACD-EP} algorithm, we use an algorithm called \acs{JAC-EP}\acused{JAC-EP} which is obtained by applying the \ac{JACD-EP} algorithm to the pilot part only, i.e., for $t\leq T_p$, with priors $p_{u_k}(u_k)$ and $p_{h_{l,k}}(\lvec{h}_{l,k})$ and taking into account that $p_x(x_{kt})$ is deterministic for $t\leq T_p$ as specified in~\eqref{eq:p_x_kt}.
This initialization algorithm provides the pilot-based prior information $\tilde{p}_{u_k}(u_k)$ and $\tilde{p}_{h_{l,k}}(\lvec{h}_{l,k})\equiv(\tilde{\gvec{\mu}}_{h_{l,k}}, \tilde{\lmat{C}}_{h_{l,k}})$  as output.
Further details on the \ac{JAC-EP} algorithm can be found in Appendix~\ref{app:JAC-EP}.
The soft estimates of the user activities $\tilde{p}_{u_k}(u_k)$ and channels $\tilde{p}_{h_{l,k}}(\lvec{h}_{l,k})$ are then used as new priors for the \ac{JACD-EP} algorithm with joint data detection.
The messages in \ac{JACD-EP} are initialized as follows.
The initial mean vector and covariance matrix of $\msg{\Psi_{g_{l,k}}}{\lvec{g}_{l,k}}$, $\msg{\lvec{g}_{l,k}}{\Psi_{z_{l,kt}}}$, and $\msg{\Psi_{z_{l,kt}}}{\lvec{z}_{l,kt}}$ $\forall k,l,t$ are set according to the prior information on $\lvec{g}_{l,k}$ and $\lvec{z}_{l,kt}$, induced by the prior information on $\tilde{p}_{u_k}(u_k)$, $\tilde{p}_{h_{l,k}}(\lvec{h}_{l,k})$, and $p_x(x_{kt})$ and are given by
\begin{align}
	\Mumsg{\Psi_{g_{l,k}}}{\lvec{g}_{l,k}} &= \Mumsg{\lvec{g}_{l,k}}{\Psi_{z_{l,kt}}} = \tilde{p}_{u_k}(1)\cdot\tilde{\gvec{\mu}}_{h_{l,k}},
	\label{eq:mu_g_Psi_z_init}\\
	\Cmsg{\Psi_{g_{l,k}}}{\lvec{g}_{l,k}} &= \Cmsg{\lvec{g}_{l,k}}{\Psi_{z_{l,kt}}} = \tilde{p}_{u_k}(1)\left(\tilde{\lmat{C}}_{h_{l,k}}+\tilde{\gvec{\mu}}_{h_{l,k}}\tilde{\gvec{\mu}}_{h_{l,k}}^H\cdot\tilde{p}_{u_k}(0)\right),
	\label{eq:C_g_Psi_z_init}
\end{align}
and
\begin{align}
	\Mumsg{\Psi_{z_{l,kt}}}{\lvec{z}_{l,kt}} &= \begin{cases}
 \tilde{p}_{u_k}(1)\cdot\tilde{\gvec{\mu}}_{h_{l,k}}\cdot\pilot{x}_{kt}&\text{for }t\leq T_p,\\
 \lvec{0}&\text{for }t>T_p,
        \end{cases}
	\label{eq:mu_Psi_z_z_init}\\
	\Cmsg{\Psi_{z_{l,kt}}}{\lvec{z}_{l,kt}} &= \begin{cases}
 \tilde{p}_{u_k}(1)\left(\tilde{\lmat{C}}_{h_{l,k}}+\tilde{\gvec{\mu}}_{h_{l,k}}\tilde{\gvec{\mu}}_{h_{l,k}}^H\cdot\tilde{p}_{u_k}(0)\right)|\pilot{x}_{kt}|^2&\text{for }t\leq T_p,\\
 \tilde{p}_{u_k}(1)\left(\tilde{\lmat{C}}_{h_{l,k}}+\tilde{\gvec{\mu}}_{h_{l,k}}\tilde{\gvec{\mu}}_{h_{l,k}}^H\right)\sigma_x^2&\text{for }t>T_p.
        \end{cases}
	\label{eq:C_Psi_z_z_init}
\end{align}
For all other messages, an uninformative initialization is chosen, i.e., a uniform distribution for messages involving categorically distributed variables, and a zero-mean and zero-precision initialization for messages involving Gaussian variables where the precision matrix is the inverse covariance matrix.
Then, we update all the messages according to the scheduling given in Algorithm~\ref{alg:JACD-EP} and is illustrated in Fig.~\ref{fig:FG} by red dashed arrows labeled with sequential numbers.
Finally, the estimates of the user activities, channels, and data are computed as follows,
\begin{align}
	\hat{u}_k &= \argmax_{u_k\in\{0,1\}}\;\hat{p}_{u_k}(u_k),
	\label{eq:estimate_u}\\
        \hat{\lvec{h}}_{l,k} &= \argmax_{\lvec{h}_{l,k}\in\Cset^N}\;\hat{p}_{\lvec{h}_{l,k}}(\lvec{h}_{l,k}) = \frac{1}{{Z}_{l,k}}\left(\catmsg{u_k}{\Psi_{g_{l,k}}}\!(0)\,\vartheta(0)\,\tilde{\gvec{\mu}}_{h_{l,k}} \!+ \catmsg{u_k}{\Psi_{g_{l,k}}}\!(1)\,\vartheta(1)\,\Mumsga{{l,k}}{}\right),
	\label{eq:estimate_h}\\
        \hat{x}_{kt} &= \argmax_{x_{kt}\in\mathcal{X}}\;\hat{p}_{x_{kt}}(x_{kt})\qquad{\text{for }t>T_p},
	\label{eq:estimate_x}
\end{align}
with the approximations of the posterior distributions
\begin{align}
	\hat{p}_{u_k}(u_k) &\propto \tilde{p}_{u_k}(u_k)\cdot\prod_{l=1}^L\catmsg{\Psi_{g_{l,k}}}{u_{k}}(u_k),
	\label{eq:posterior_u}\\
        \hat{p}_{\lvec{h}_{l,k}}(\lvec{h}_{l,k}) &\propto \tilde{p}_{h_{l,k}}(\lvec{h}_{l,k})\cdot\CN{\lvec{h}_{l,k}\big|\Mumsg{\Psi_{g_{l,k}}}{\lvec{h}_{l,k}},\Cmsg{\Psi_{g_{l,k}}}{\lvec{h}_{l,k}}},
        \label{eq:posterior_h}\\
        \hat{p}_{x_{kt}}(x_{kt}) &\propto \prod_{l=1}^L\catmsg{\Psi_{z_{l,kt}}}{x_{kt}}(x_{kt}),
	\label{eq:posterior_x}
\end{align}
and $\catmsg{u_k}{\Psi_{g_{l,k}}}\!(u_k)$, $\vartheta(u_k)$, $\Mumsga{{l,k}}{}$, ${Z}_{l,k}$, $\catmsg{\Psi_{g_{l,k}}}{u_{k}}(u_k)$, $\Mumsg{\Psi_{g_{l,k}}}{\lvec{h}_{l,k}}$, $\Cmsg{\Psi_{g_{l,k}}}{\lvec{h}_{l,k}}$, and $\catmsg{\Psi_{z_{l,kt}}}{x_{kt}}(x_{kt})$ defined in Section~\ref{subsec:MP_updates}.
{The derivation of~\eqref{eq:estimate_h} can be found in~\eqref{eq:app_mu_mmd_Psi_g_h}.}
\begin{algorithm}[t]
\caption{\ac{JACD-EP} Algorithm}
\begin{algorithmic}[1]
\renewcommand{\algorithmicrequire}{\textbf{Input:}}
\renewcommand{\algorithmicensure}{\textbf{Output:}}
\REQUIRE Pilot matrix $\pilot{\lmat{X}}$, transmit power $\sigma_x^2$, received signal $\lmat{Y}$, noise variance $\sigma_n^2$, parameters of prior distributions on user activities $\tilde{p}_{u_k}(u_k)$ and channels $\tilde{p}_{h_{l,k}}(\lvec{h}_{l,k})$, i.e., $\tilde{p}_{u_k}(0)$, $\tilde{p}_{u_k}(1)$, $\tilde{\gvec{\mu}}_{h_{l,k}}$, $\tilde{\lmat{C}}_{h_{l,k}}$.
\ENSURE Estimated activities $\hat{u}_k$, channels $\hat{\lmat{h}}_{l,k}$, and data $\hat{x}_{kt}$.
\STATE $\forall k,l,t$: Initialize $\msg{\Psi_{g_{l,k}}}{\lvec{g}_{l,k}}$ and $\msg{\lvec{g}_{l,k}}{\Psi_{z_{l,kt}}}$ via \eqref{eq:mu_g_Psi_z_init}, \eqref{eq:C_g_Psi_z_init}, and $\msg{\Psi_{z_{l,kt}}}{\lvec{z}_{l,kt}}$ via \eqref{eq:mu_Psi_z_z_init}, \eqref{eq:C_Psi_z_z_init}.
\FOR {$i = 1$ to $i_\text{max}$}
\STATE $\forall k,l,t$: Update $\msg{\Psi_{y_{l,t}}}{\lvec{z}_{l,kt}}$ via \eqref{eq:mu_Psi_y_z}, \eqref{eq:C_Psi_y_z}.
\STATE $\forall k,l,t\!>\!T_p$: Update $\msg{\Psi_{z_{l,kt}}}{x_{kt}}$ via \eqref{eq:m_Psi_z_x}.
\STATE $\forall k,l,t\!>\!T_p$: Update $\msg{x_{kt}}{\Psi_{z_{l,kt}}}$ via \eqref{eq:m_x_Psi_z}.
\STATE $\forall k,l,t$: Update $\msg{\Psi_{z_{l,kt}}}{\lvec{g}_{l,k}}$ via \eqref{eq:C_Psi_z_g}, \eqref{eq:mu_Psi_z_g}.\label{alg_line:m_Psi_z_g}
\STATE $\forall k,l$: Update $\msg{\lvec{g}_{l,k}}{\Psi_{g_{l,k}}}$ via \eqref{eq:C_g_Psi_g}, \eqref{eq:mu_g_Psi_g}.
\STATE $\forall k,l$: Update $\msg{\Psi_{g_{l,k}}}{u_k}$ via \eqref{eq:m_Psi_g_u}.\label{alg_line:m_Psi_g_u}
\STATE $\forall k,l$: Update $\msg{u_k}{\Psi_{g_{l,k}}}$ via \eqref{eq:m_u_Psi_g}.
\STATE $\forall k,l$: Update $\msg{\Psi_{g_{l,k}}}{\lvec{g}_{l,k}}$ via \eqref{eq:C_Psi_g_g}, \eqref{eq:mu_Psi_g_g}.\label{alg_line:m_Psi_g_g}
\STATE $\forall k,l,t$: Update $\msg{\lvec{g}_{l,k}}{\Psi_{z_{l,kt}}}$ via \eqref{eq:C_g_Psi_z}, \eqref{eq:mu_g_Psi_z}.
\STATE $\forall k,l,t$: Update $\msg{\Psi_{z_{l,kt}}}{\lvec{z}_{l,kt}}$ via \eqref{eq:C_Psi_z_z}, \eqref{eq:mu_Psi_z_z}.\label{alg_line:m_Psi_z_z}
\ENDFOR
\RETURN $\hat{u}_k$ calculated via \eqref{eq:estimate_u} $\forall k$.
\RETURN $\hat{\lvec{h}}_{l,k}$ calculated via \eqref{eq:estimate_h} $\forall k,l$.
\RETURN $\hat{x}_{kt}$ calculated via \eqref{eq:estimate_x} $\forall k,{t>T_p}$.
\end{algorithmic} 
\label{alg:JACD-EP}
\end{algorithm}

As in \cite{Karataev2024,Ngo2020}, we apply damping to the factor-to-variable messages with damping parameter $\eta\in[0,1]$, i.e., the updated parameter of a factor-to-variable message is a convex combination of the old parameter and the new parameter.
For example, the updated message $\catmsg{\Psi_{g_{l,k}}}{u_k}(u_k)$ in line~\ref{alg_line:m_Psi_g_u} of Algorithm~\ref{alg:JACD-EP} is given by $\catmsg{\Psi_{g_{l,k}}}{u_k}(u_k)=\eta\cdot \pi_\text{new}+(1-\eta)\cdot \pi_\text{old}$, where $\pi_\text{new}$ is computed according to~\eqref{eq:m_Psi_g_u} and $\pi_\text{old}$ is the message $\catmsg{\Psi_{g_{l,k}}}{u_k}(u_k)$ from the previous iteration.
For the Gaussian factor-to-variable messages, we apply damping to the natural parameters of the Gaussian distribution $\CN{\lvec{x}|\gvec{\mu},\lmat{C}}$, i.e., the precision matrix $\gmat{\Lambda}=\lmat{C}^{-1}$ and the transformed mean vector $\gvec{\gamma}=\lmat{C}^{-1}\gvec{\mu}$.
Furthermore, we update the parameters of $\msg{\Psi_{z_{l,kt}}}{\lvec{g}_{l,k}}$, $\msg{\Psi_{g_{l,k}}}{\lvec{g}_{l,k}}$, and $\msg{\Psi_{z_{l,kt}}}{\lvec{z}_{l,kt}}$ in line~\ref{alg_line:m_Psi_z_g}, \ref{alg_line:m_Psi_g_g}, and \ref{alg_line:m_Psi_z_z} of Algorithm~\ref{alg:JACD-EP} only if the new covariance matrix obtained by~\eqref{eq:C_Psi_z_g}, \eqref{eq:C_Psi_g_g}, and \eqref{eq:C_Psi_z_z}, respectively, is symmetric positive definite.
Otherwise, we keep the parameters from the previous iteration.

\subsection{Message-Passing Update Rules}\label{subsec:MP_updates}
In this section, we present the \ac{EP} message-passing update rules for the factor graph in Fig.~\ref{fig:FG}. Detailed derivations can be found in Appendix~\ref{app:MP_updates}.
The mean vector $\gvec{\mu}$ and the covariance matrix $\lmat{C}$ of a Gaussian random variable can be readily expressed by the natural parameters $\gvec{\gamma}\!=\!\lmat{C}^{-1}\gvec{\mu}$ and $\gmat{\Lambda}\!=\!\lmat{C}^{-1}$.
In the following, we switch between these two representations without explicitly mentioning the transformation, e.g., if $\Mumsg{\Psi_\alpha}{\lvec{x}_\beta}$ and $\Cmsg{\Psi_\alpha}{\lvec{x}_\beta}$ are computed, then $\Gammamsg{\Psi_\alpha}{\lvec{x}_\beta}$ and $\Lambdamsg{\Psi_\alpha}{\lvec{x}_\beta}$ are automatically given and vice versa.

\noindent\textit{Update of $\msg{\Psi_{y_{l,t}}}{\lvec{z}_{l,kt}}$ {(cf.~\eqref{eq:app_mu_Psi_y_z},~\eqref{eq:app_C_Psi_y_z})}:}
\begin{align}
	\Mumsg{\Psi_{y_{l,t}}}{\lvec{z}_{l,kt}} &= \lvec{y}_{l,t}-\sum_{k'\neq k}\Mumsg{\Psi_{z_{l,k't}}}{\lvec{z}_{l,k't}},
	\label{eq:mu_Psi_y_z}\\
	\Cmsg{\Psi_{y_{l,t}}}{\lvec{z}_{l,kt}} &= \sigma_n^2\lmat{I}_N+\sum_{k'\neq k}\Cmsg{\Psi_{z_{l,k't}}}{\lvec{z}_{l,k't}}.
	\label{eq:C_Psi_y_z}
\end{align}

\noindent\textit{Update of $\msg{x_{kt}}{\Psi_{z_{l,kt}}}$ {(cf.~\eqref{eq:app_m_x_Psi_z})}:}
\begin{equation}
	\catmsg{x_{kt}}{\Psi_{z_{l,kt}}}(x_{kt}) \propto \prod_{l'\neq l}\catmsg{\Psi_{z_{l',kt}}}{x_{kt}}(x_{kt}).
	\label{eq:m_x_Psi_z}
\end{equation}

\noindent\textit{Update of $\msg{\lvec{g}_{l,k}}{\Psi_{z_{l,kt}}}$ {(cf.~\eqref{eq:app_C_g_Psi_z},~\eqref{eq:app_mu_g_Psi_z})}:}
\begin{align}
	\Lambdamsg{\lvec{g}_{l,k}}{\Psi_{z_{l,kt}}} &= \Lambdamsg{\Psi_{g_{l,k}}}{\lvec{g}_{l,k}}+\sum_{t'\neq t}\Lambdamsg{\Psi_{z_{l,kt'}}}{\lvec{g}_{l,k}},
	\label{eq:C_g_Psi_z}\\
	\Gammamsg{\lvec{g}_{l,k}}{\Psi_{z_{l,kt}}} &= \Gammamsg{\Psi_{g_{l,k}}}{\lvec{g}_{l,k}}+\sum_{t'\neq t}\Gammamsg{\Psi_{z_{l,kt'}}}{\lvec{g}_{l,k}}.
	\label{eq:mu_g_Psi_z}
\end{align}

\noindent\textit{Update of $\msg{\Psi_{z_{l,kt}}}{x_{kt}}$ {(cf.~\eqref{eq:app_m_Psi_z_x})}:}
\begin{equation}
	\catmsg{\Psi_{z_{l,kt}}}{x_{kt}}(x_{kt}) \propto \theta(x_{kt}),
	\label{eq:m_Psi_z_x}
\end{equation}
with
\begin{equation*}
	\theta(x_{kt}) = \mathcal{CN}(\lvec{0}|\Mumsg{\Psi_{y_{l,t}}}{\lvec{z}_{l,kt}}-\Mumsg{\lvec{g}_{l,k}}{\Psi_{z_{l,kt}}}x_{kt},
 \Cmsg{\Psi_{y_{l,t}}}{\lvec{z}_{l,kt}}+\Cmsg{\lvec{g}_{l,k}}{\Psi_{z_{l,kt}}}|x_{kt}|^2).
	\label{eq:theta_mmd_Psi_z_z}
\end{equation*}

\noindent\textit{Update of $\msg{\Psi_{z_{l,kt}}}{\lvec{z}_{l,kt}}$ {(cf.~\eqref{eq:app_C_Psi_z_z},~\eqref{eq:app_mu_Psi_z_z})}:}
\begin{align}
	\Lambdamsg{\Psi_{z_{l,kt}}}{\lvec{z}_{l,kt}} &= \Lambdamsgb{1_{l,kt}}{}-\Lambdamsg{\Psi_{y_{l,t}}}{\lvec{z}_{l,kt}},
	\label{eq:C_Psi_z_z}\\
	\Gammamsg{\Psi_{z_{l,kt}}}{\lvec{z}_{l,kt}} &= \Gammamsgb{1_{l,kt}}{}-\Gammamsg{\Psi_{y_{l,t}}}{\lvec{z}_{l,kt}},
	\label{eq:mu_Psi_z_z}
\end{align}
with $\Mumsgb{1_{l,kt}}{}\!\!=\!\Mumsga{{l,kt}}{}(\pilot{x}_{kt})$, $\Cmsgb{1_{l,kt}}{}\!\!=\!\Cmsga{{l,kt}}{}(\pilot{x}_{kt})$ for $t\leq T_p$ and
\begin{align}
 \Mumsgb{1_{l,kt}}{} &= \frac{1}{{Z}_{{l,kt}}}\sum_{x_{kt}\in\mathcal{X}}\catmsg{x_{kt}}{\Psi_{z_{l,kt}}}(x_{kt})\cdot\theta(x_{kt})\cdot\Mumsga{{l,kt}}{}(x_{kt}),
	\nonumber\\
	\Cmsgb{1_{l,kt}}{} &= \frac{1}{{Z}_{{l,kt}}}\sum_{x_{kt}\in\mathcal{X}}\catmsg{x_{kt}}{\Psi_{z_{l,kt}}}(x_{kt})\cdot\theta(x_{kt})\cdot\big(\Cmsga{{l,kt}}{}(x_{kt})+\Mumsga{{l,kt}}{}(x_{kt})\cdot\Mumsga{{l,kt}}{}^H(x_{kt})\big) - \Mumsgb{1_{l,kt}}{}\Mumsgb{1_{l,kt}}{}^H,
        \nonumber
\end{align}
for $t>T_p$ with ${Z}_{{l,kt}}$, $\Mumsga{{l,kt}}{}(x_{kt})$, and $\Cmsga{{l,kt}}{}(x_{kt})$ obtained by
\begin{align}
    {Z}_{{l,kt}} &= \sum_{x_{kt}\in\mathcal{X}}\catmsg{x_{kt}}{\Psi_{z_{l,kt}}}(x_{kt})\cdot\theta(x_{kt}),
	\nonumber\\
 \Lambdamsga{{l,kt}}{}(x_{kt}) &= \Lambdamsg{\Psi_{y_{l,t}}}{\lvec{z}_{l,kt}}+\Lambdamsg{\lvec{g}_{l,k}}{\Psi_{z_{l,kt}}}|x_{kt}|^{-2},
	\nonumber\\
	\Gammamsga{{l,kt}}{}(x_{kt}) &= \Gammamsg{\Psi_{y_{l,t}}}{\lvec{z}_{l,kt}}+\Gammamsg{\lvec{g}_{l,k}}{\Psi_{z_{l,kt}}}\frac{x_{kt}}{|x_{kt}|^2}.
	\nonumber
\end{align}

\noindent\textit{Update of $\msg{\Psi_{z_{l,kt}}}{\lvec{g}_{l,k}}$ {(cf.~\eqref{eq:app_C_Psi_z_g},~\eqref{eq:app_mu_Psi_z_g})}:}
\begin{align}
	\Lambdamsg{\Psi_{z_{l,kt}}}{\lvec{g}_{l,k}} &= \Lambdamsgb{2_{l,kt}}{}-\Lambdamsg{\lvec{g}_{l,k}}{\Psi_{z_{l,kt}}},
	\label{eq:C_Psi_z_g}\\
	\Gammamsg{\Psi_{z_{l,kt}}}{\lvec{g}_{l,k}} &= \Gammamsgb{2_{l,kt}}{}-\Gammamsg{\lvec{g}_{l,k}}{\Psi_{z_{l,kt}}},
	\label{eq:mu_Psi_z_g}
\end{align}
with $\Mumsgb{2_{l,kt}}{}=\frac{\Mumsga{{l,kt}}{}(\pilot{x}_{kt})}{\pilot{x}_{kt}}$, $\Cmsgb{2_{l,kt}}{}=\frac{\Cmsga{{l,kt}}{}(\pilot{x}_{kt})}{|\pilot{x}_{kt}|^2}$ for $t\leq T_p$ and
\begin{align}
 \Mumsgb{2_{l,kt}}{} &= \frac{1}{{Z}_{l,kt}}\sum_{x_{kt}\in\mathcal{X}}\catmsg{x_{kt}}{\Psi_{z_{l,kt}}}(x_{kt})\cdot\frac{\theta(x_{kt})}{x_{kt}}\cdot\Mumsga{{l,kt}}{}(x_{kt}),
	\nonumber\\
	\Cmsgb{2_{l,kt}}{} &= \frac{1}{{Z}_{{l,kt}}}\sum_{x_{kt}\in\mathcal{X}}\catmsg{x_{kt}}{\Psi_{z_{l,kt}}}(x_{kt})\cdot\frac{\theta(x_{kt})}{|x_{kt}|^2}\cdot\big(\Cmsga{{l,kt}}{}(x_{kt})+\Mumsga{{l,kt}}{}(x_{kt})\cdot\Mumsga{{l,kt}}{}^H(x_{kt})\big) - \Mumsgb{{l,kt}}{}\Mumsgb{{l,kt}}{}^H,
    \nonumber
\end{align}
for $t>T_p$ with ${Z}_{{l,kt}},\Lambdamsga{{l,kt}}{}(x_{kt}),\Gammamsga{{l,kt}}{}(x_{kt})$ given before.

\noindent\textit{Update of $\msg{u_k}{\Psi_{g_{l,k}}}$ {(cf.~\eqref{eq:app_m_u_Psi_g})}:}
\begin{equation}
	\catmsg{u_k}{\Psi_{g_{l,k}}}(u_k) \propto \tilde{p}_{u_k}(u_k)\cdot\prod_{l'\neq l}\catmsg{\Psi_{g_{l',k}}}{u_{k}}(u_k),
	\label{eq:m_u_Psi_g}
\end{equation}

\noindent\textit{Update of $\msg{\lvec{g}_{l,k}}{\Psi_{g_{l,k}}}$ {(cf.~\eqref{eq:app_C_g_Psi_g},~\eqref{eq:app_mu_g_Psi_g})}:}
\begin{align}
	 \Lambdamsg{\lvec{g}_{l,k}}{\Psi_{g_{l,k}}} &= \sum_{t=1}^T\Lambdamsg{\Psi_{z_{l,kt}}}{\lvec{g}_{l,k}},
	\label{eq:C_g_Psi_g}\\
	\Gammamsg{\lvec{g}_{l,k}}{\Psi_{g_{l,k}}} &= \sum_{t=1}^T\Gammamsg{\Psi_{z_{l,kt}}}{\lvec{g}_{l,k}}.
	\label{eq:mu_g_Psi_g}
\end{align}

\noindent\textit{Update of $\msg{\Psi_{g_{l,k}}}{u_k}$ {(cf.~\eqref{eq:app_m_Psi_g_u})}:}
\begin{equation}
	\catmsg{\Psi_{g_{l,k}}}{u_k}(u_k) \propto \vartheta(u_k),
	\label{eq:m_Psi_g_u}
\end{equation}
with
\begin{equation*}
	\vartheta(u_k) = \mathcal{CN}(\lvec{0}|\Mumsg{\lvec{g}_{l,k}}{\Psi_{g_{l,k}}}\!\!\!-\tilde{\gvec{\mu}}_{h_{l,k}}u_k,\Cmsg{\lvec{g}_{l,k}}{\Psi_{g_{l,k}}}\!\!\!+\tilde{\lmat{C}}_{h_{l,k}}u_k).
	\label{eq:theta_mmd_Psi_g_u}
\end{equation*}

\noindent\textit{Update of $\msg{\Psi_{g_{l,k}}}{\lvec{g}_{l,k}}$ {(cf.~\eqref{eq:app_C_Psi_g_g},~\eqref{eq:app_mu_Psi_g_g})}:}
\begin{align}
	\Lambdamsg{\Psi_{g_{l,k}}}{\lvec{g}_{l,k}} &= \Lambdamsgb{{l,k}}{}-\Lambdamsg{\lvec{g}_{l,k}}{\Psi_{g_{l,k}}},
	\label{eq:C_Psi_g_g}\\
	\Gammamsg{\Psi_{g_{l,k}}}{\lvec{g}_{l,k}} &= \Gammamsgb{{l,k}}{}-\Gammamsg{\lvec{g}_{l,k}}{\Psi_{g_{l,k}}},
	\label{eq:mu_Psi_g_g}
\end{align}
with
\begin{align}
	\Mumsgb{{l,k}}{} &= \frac{1}{{Z}_{{l,k}}}\!\cdot\!\catmsg{u_k}{\Psi_{g_{l,k}}}\!(1)\cdot\vartheta(1)\cdot\Mumsga{{l,k}}{},
	\nonumber\\
	\Cmsgb{{l,k}}{} &= \frac{1}{{Z}_{{l,k}}}\!\cdot\!\catmsg{u_k}{\Psi_{g_{l,k}}}\!(1)\cdot\vartheta(1)\cdot(\Cmsga{{l,k}}{}\!+\!\Mumsga{{l,k}}{}\Mumsga{{l,k}}{}^H) \!-\! \Mumsgb{{l,k}}{}\Mumsgb{{l,k}}{}^H,\nonumber
\end{align}
and ${Z}_{{l,k}}$, $\Lambdamsga{{l,k}}{}$, and $\Gammamsga{{l,k}}{}$ given by
\begin{align}
    {Z}_{{l,k}} &= \catmsg{u_k}{\Psi_{g_{l,k}}}(0)\cdot\vartheta(0)+\catmsg{u_k}{\Psi_{g_{l,k}}}(1)\cdot\vartheta(1)
	\nonumber\\
        \Lambdamsga{{l,k}}{} &= \Lambdamsg{\lvec{g}_{l,k}}{\Psi_{g_{l,k}}}+\tilde{\gmat{\Lambda}}_{h_{l,k}},
	\nonumber\\
	\Gammamsga{{l,k}}{} &= \Gammamsg{\lvec{g}_{l,k}}{\Psi_{g_{l,k}}}+\tilde{\gvec{\gamma}}_{h_{l,k}}\nonumber.
\end{align}

\subsection{Modification for Pilot Contamination}\label{subsec:mod_pc}
Since the number of \acp{UE} is larger than the pilot sequence length, pilot contamination will naturally occur and degrade the performance.
In the following, we propose a simple heuristic approach to account for pilot contamination.
The efficacy of the proposed approach was verified via simulations.

The adverse effect of pilot contamination is particularly pronounced for \acp{UE} that share the same pilot sequence and are close to each other.
For these cases, we propose to add a correction term to the covariance matrix in~\eqref{eq:C_Psi_y_z} for $t\leq T_p$ when generating the prior information via the pilot-based \ac{JAC-EP} initialization algorithm.
We denote the set of \acp{UE} which use the same pilot sequence as \ac{UE} $k$ by  $\mathcal{P}_k$.
By adding the covariance matrix of the received signals from \acp{UE} in $\mathcal{P}_k$ which cause pilot contamination to \ac{UE} $k$ at \ac{AP} $l$ to the covariance matrix in~\eqref{eq:C_Psi_y_z}, we decrease the reliability of information coming from the factor node $\Psi_{y_{l,t}}$ especially when pilot-sharing users are close to the considered \ac{AP}.
Then, the covariance matrix with the correction term is given by
\begin{equation}
	\Cmsg{\Psi_{y_{l,t}}}{\lvec{z}_{l,kt}} = \sigma_n^2\lmat{I}_N+\sum_{k'\in\mathcal{P}_k}\gmat{\Xi}_{l,k}|\pilot{x}_{k't}|^2+\sum_{k'\neq k}\Cmsg{\Psi_{z_{l,k't}}}{\lvec{z}_{l,k't}}.
	\label{eq:C_Psi_y_g_pc}
\end{equation}

\section{Numerical Results}\label{sec:sims}
In this section, we analyze the performance of the \ac{JACD-EP} algorithm via Monte Carlo simulations.

\subsection{Simulation Setup}\label{subsec:sim_setup}
We consider a network of size $400\times400\,$m with $L=16$ \acp{AP} placed  on a regular grid at the following positions $\{(50+i\!\cdot\!100,50+j\!\cdot\!100)\,\text{m}\,|\,i,j\in\{0,1,2,3\}\}$ and  a height of 10 m.
 On the ground, $K=16$ \acp{UE} are deployed according to a \ac{UPP}.
We consider 100 different outcomes of the \ac{UPP}.
For each \ac{UPP} outcome, the large-scale fading coefficients are deterministic and computed according to the 3GPP urban microcell model~\cite{Bjoernson2020} whereas for the random variables, i.e., small-scale fading coefficients,  activities, and pilot sequences, we  generate $10^3$ realizations. 
This allows us to analyze the \ac{QoS} distribution in the network.
The activity probability of all users is set to $\lambda=0.5$, and the transmit power is $\sigma_x^2=16\,$dBm.
Furthermore, random \ac{BPSK} pilot sequences of length $T_p=8$ are utilized and $T_d=\{10,30\}$ 4-\ac{QAM} symbols are transmitted during the data transmission phase.
The noise power at each \ac{AP} is set to $\sigma_n^2=-96\,$dBm.
As benchmark schemes, we consider the centralized linear \ac{MMSE} data detector and the \ac{GaBP} algorithm in~\cite{Iimori2021}.
Both of these schemes are initialized by the multiple measurement vector \acs{AMP} (\acs{MMV-AMP})\acused{MMV-AMP} algorithm in~\cite{Ke2020}.
Besides, we consider the genie-aided centralized linear \ac{MMSE} detector with perfect \ac{UE} activity and channel knowledge.
For all iterative algorithms, the damping parameter is $\eta=0.5$.
Furthermore, the maximum number of iterations for the \ac{MMV-AMP} algorithm is set to 200, whereas the \ac{GaBP}- and \ac{EP}-based algorithms perform 20 iterations.

\subsection{Performance Metrics and Results}\label{subsec:metrics_results}
\begin{figure}[t]
    \centering
    \subfloat[\Ac{CDF} of \ac{DER}.]{\includegraphics[width=0.32\textwidth]{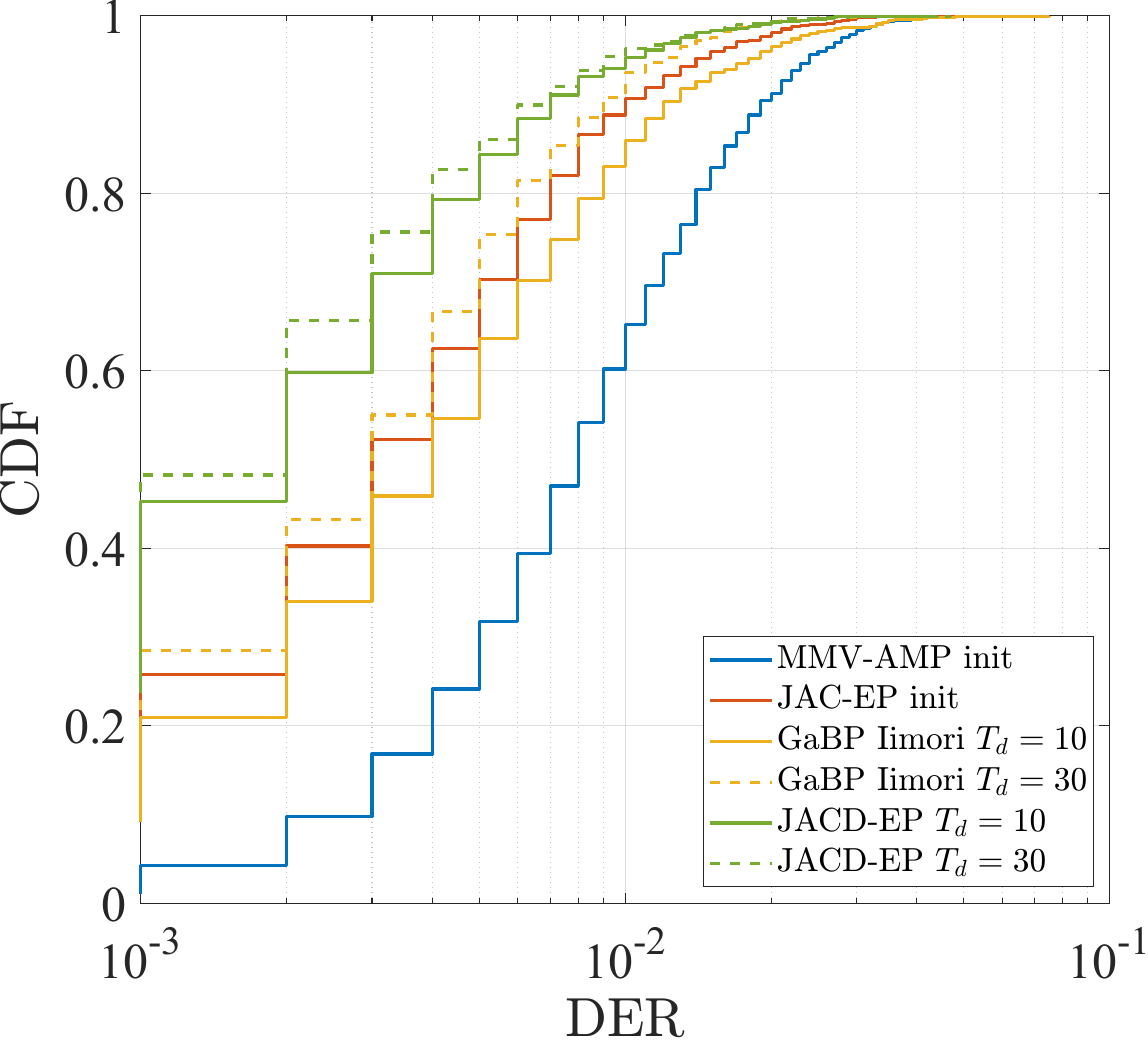}
        \label{fig:DER}}\hfill
    \subfloat[\Ac{CDF} of \ac{NMSE}.]{\includegraphics[width=0.32\textwidth]{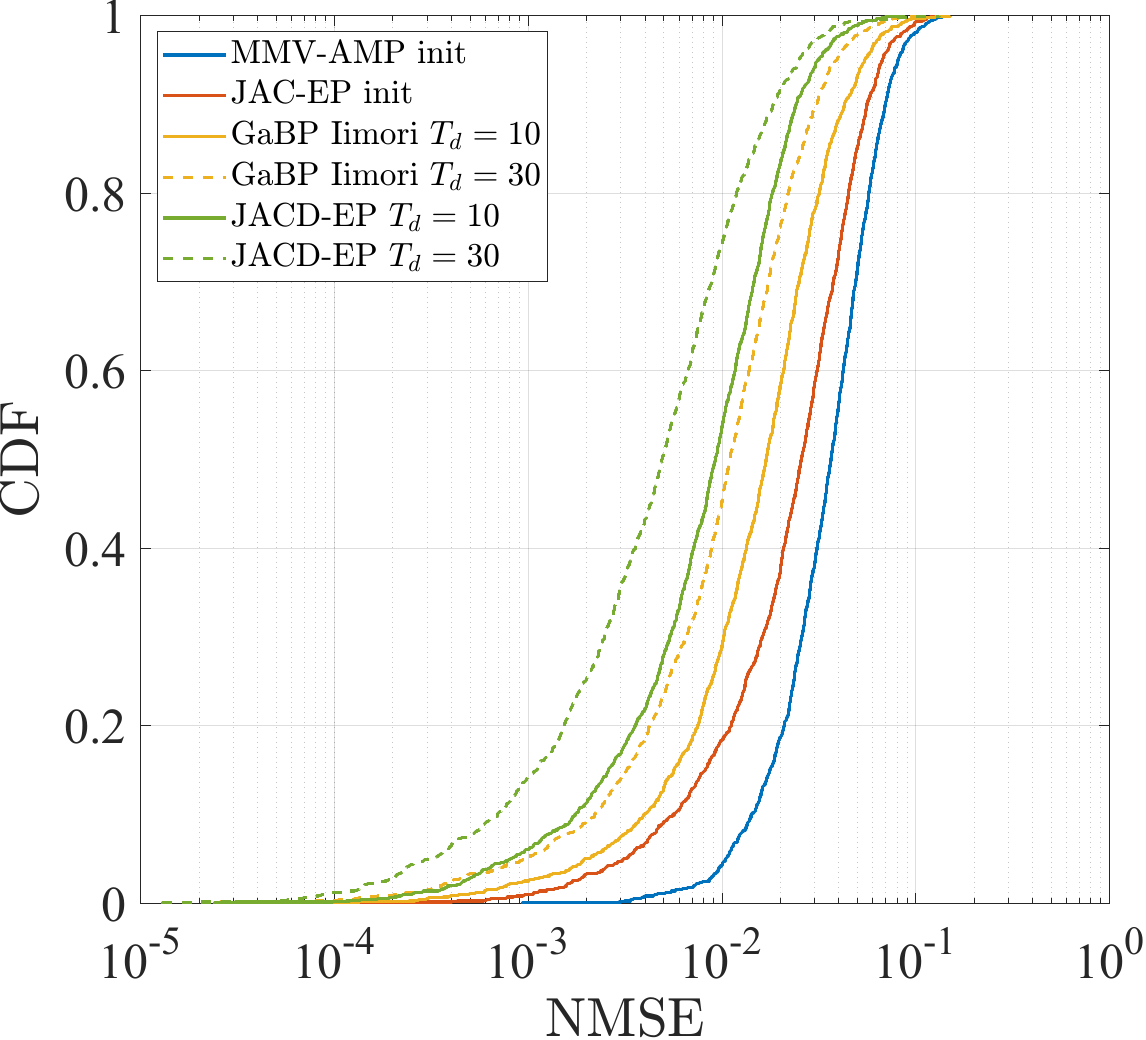}
        \label{fig:NMSE}}\hfill
    \subfloat[\Ac{CDF} of \ac{SER}.]{\includegraphics[width=0.32\textwidth]{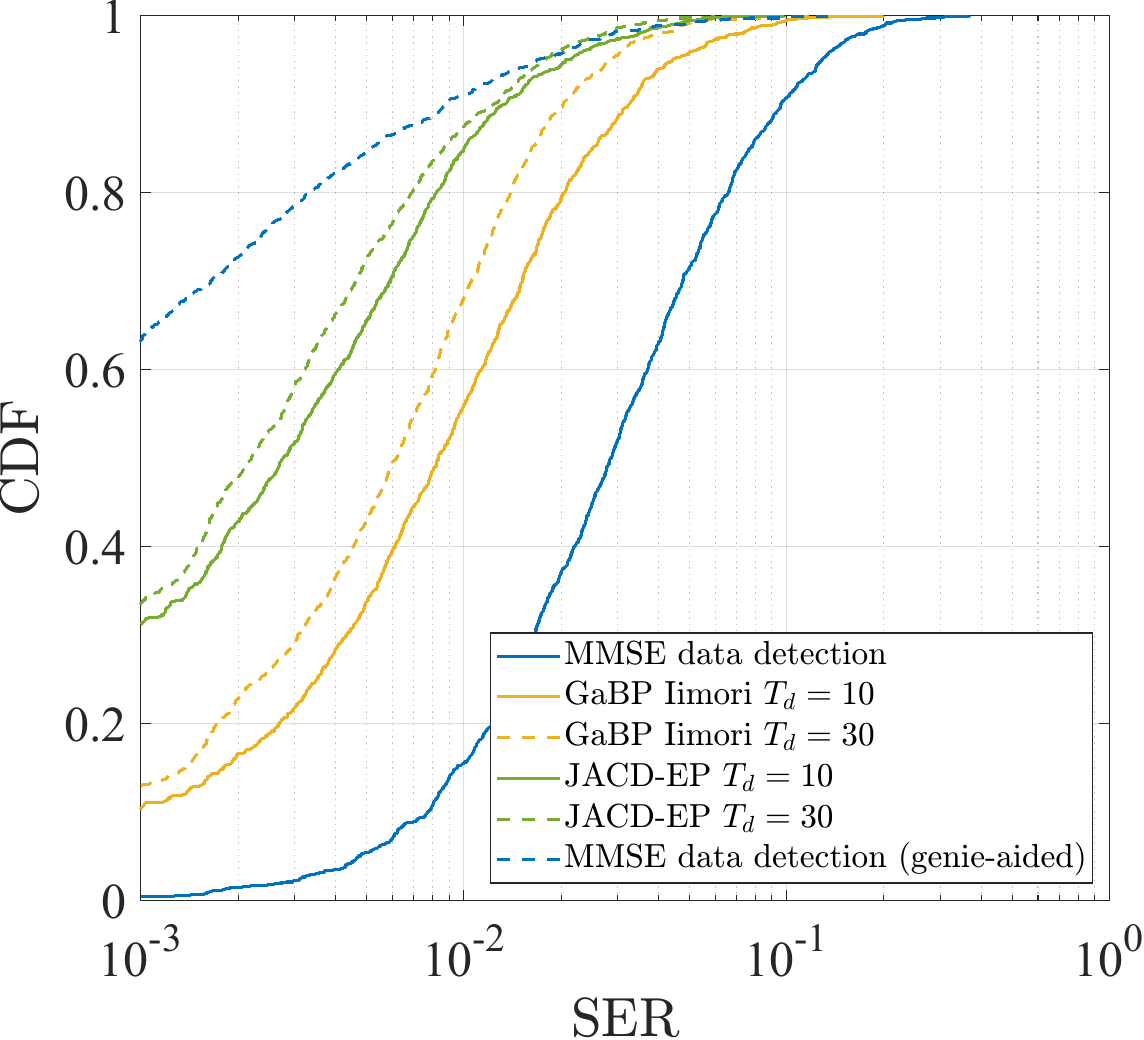}
        \label{fig:SER}}
    \caption{Numerical results for $L=16$, $N=1$, $K=16$, $\lambda=0.5$, $T_p=8$, $T_d=\{10,30\}$, and random \ac{BPSK} pilots.}
    \label{fig:results}
\end{figure}

For assessing the user activity detection, channel estimation, and data detection performance, we consider the empirical \acp{CDF} of the \ac{DER}, the \ac{NMSE}, and the \ac{SER}, respectively.
The empirical \acp{CDF} are obtained considering all \acp{UE} in each \ac{UPP} outcome whereas \ac{DER}, \ac{NMSE}, and \ac{SER}  are computed by averaging over the $10^3$ small-scale fading and \ac{UE} activity realizations generated for each \ac{UPP} outcome.
The \ac{DER} is defined as the ratio of the number of erroneously classified \ac{UE} activities and the total number of \ac{UE} activity realizations, i.e., $\text{DER}\coloneq\est{}{\ind{\hat{u}_k\neq u_k}}$, and includes both misdetections and false alarms.
The corresponding simulation results are illustrated in Fig.~\ref{fig:DER}.
The \ac{NMSE} and \ac{SER} are considered naturally for active users only, i.e., $\text{NMSE}\coloneq\est{}{\frac{||\hat{\lvec{h}}_k-\lvec{h}_k||}{||\lvec{h}_k||}\;\big|\;u_k=1}$, {$\text{SER}\coloneq\est{}{\ind{\hat{x}_{kt}\neq x_{kt}}\;|\;u_k=1,t>T_p}$}.
Furthermore, in the computation of  the \ac{NMSE}, we neglect weak channels, i.e., channels whose large-scale fading coefficient multiplied by the transmit power is smaller than the noise power at the \acp{AP}, to avoid taking into account errors on weak and insignificant channels.
The channel estimation and data detection performance is depicted in Figs.~\ref{fig:NMSE} and~\ref{fig:SER}, respectively.
The results illustrate the advantages of using jointly pilot and data sequences for both data and activity detection and the superior performance of the \ac{JACD-EP} algorithm compared to state-of-the-art benchmark schemes.
Furthermore, it can be observed that an increase of the data length improves the performance with respect to all considered metrics.

\section{Conclusion}\label{sec:concl}
In this paper, we considered the uplink of a \ac{GF-CF-MaMIMO} system and tackled the problem of \ac{JACD}.
We developed the \ac{JACD-EP} algorithm by applying \ac{EP} message passing on a factor graph where we considered accurate categorical probability distributions for user activity and data, and Gaussian probability distributions for the channels.
The proposed algorithm is robust against pilot contamination and outperforms state-of-the-art algorithms in terms of \ac{DER}, \ac{NMSE}, and \ac{SER}.

\appendices
\section{\acs{JAC-EP} Initialization Algorithm}\label{app:JAC-EP}
The \acs{JAC-EP} algorithm can be obtained by running the \ac{JACD-EP} algorithm for the pilot part only, i.e., for $t\leq T_p$, with priors $p_{u_k}(u_k)$ and $p_{h_{l,k}}(\lvec{h}_{l,k})$.
However, since we do not have to estimate the symbols $x_{kt}$ during initialization, the factorization of the \ac{APP} and the corresponding factor graph and algorithm simplify.
In the following, we describe the simplified \ac{JAC-EP} algorithm in more detail.
We use the same system model and notation as in the main body of this document except that we only consider signals in the pilot phase, i.e., $1\leq t\leq T_p$.

\subsection{Factor Graph Representation}\label{app:JAC_FG}
\begin{figure}[t]
    \centerline{\includegraphics[width=0.8\textwidth]{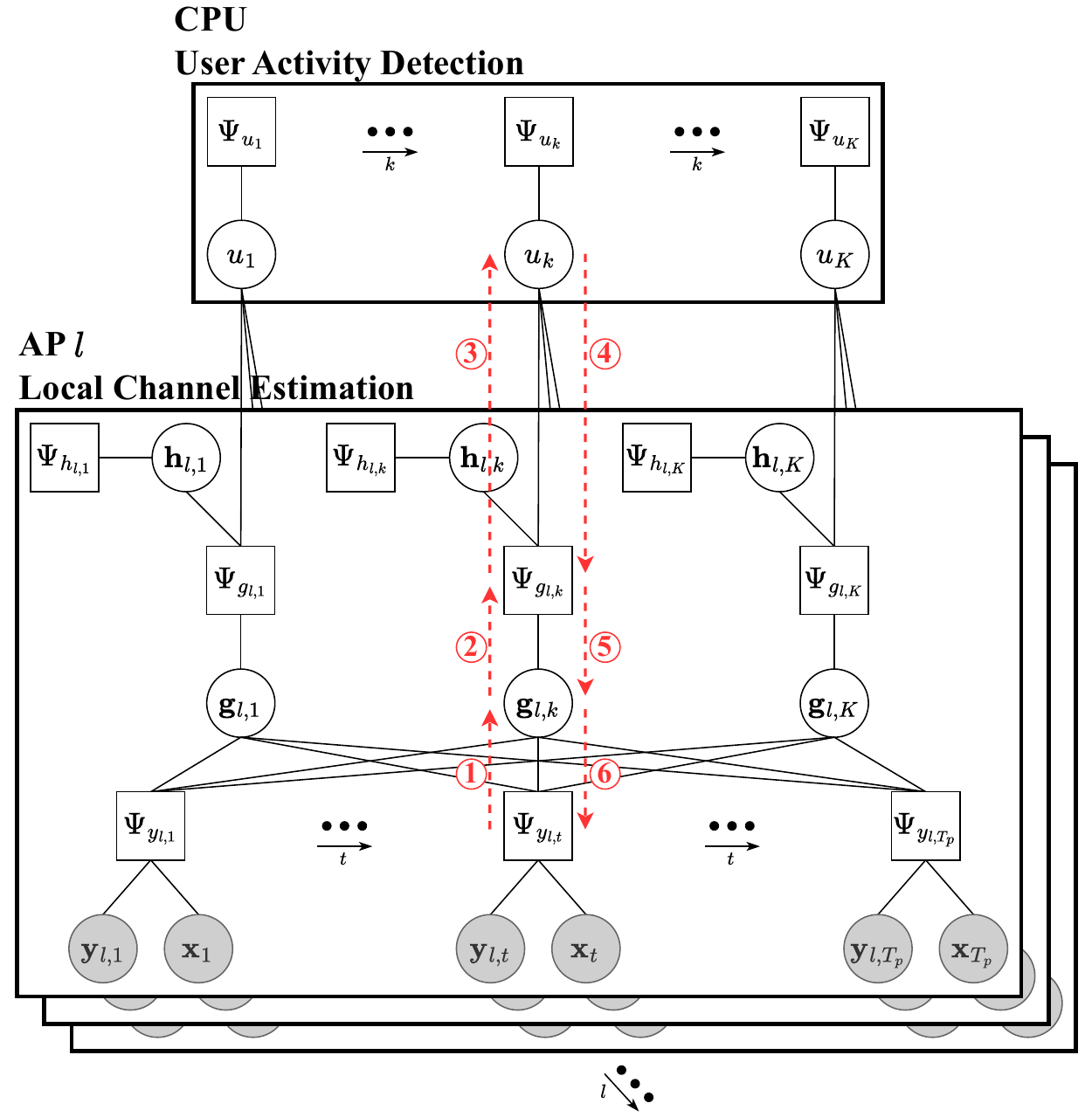}}
    \caption{Factor graph for \ac{JAC-EP}. The numbered red dashed arrows show the flow of information according to the scheduling presented in Algorithm~\ref{alg:app_JAC-EP}. Each number corresponds to one message update in Algorithm~\ref{alg:app_JAC-EP}.}
    \label{fig:app_JAC_FG}
\end{figure}

The task of the \ac{JAC-EP} algorithm is to find initial estimates on the user activities and channels.
Hence, the factorized \ac{APP} distribution with auxiliary variables $\lvec{g}_{l,k}\coloneq \lvec{h}_{l,k}u_k$ is given by
\begin{equation}
\begin{split}
	p_\text{APP}(\lmat{U},\lmat{H},\lmat{G})\propto\prod_{l=1}^L\prod_{k=1}^K\prod_{t=1}^{T_p}\Big[&p(\lvec{y}_{l,t}|\lvec{g}_{l,1},\!...,\lvec{g}_{l,K})\cdot p(\lvec{g}_{l,k}|\lvec{h}_{l,k},u_k)\cdot p_{u_k}(u_k)\cdot p_{h_{l,k}}(\lvec{h}_{l,k})\Big].
	\label{eq:app_JAC_APP_aux}
\end{split}
\end{equation}
In contrast to~\eqref{eq:APP_aux}, the auxiliary variables $\lvec{z}_{l,kt}$ are irrelevant since the pilot symbols $x_{kt}$ are assumed to be known.
The probability distributions in~\eqref{eq:app_JAC_APP_aux} are given by
\begin{align}
	\Psi_{y_{l,t}} &\coloneq p(\lvec{y}_{l,t}|\lvec{g}_{l,1},\!...,\lvec{g}_{l,K}) = \CN{\lvec{y}_{l,t}\Bigg|\sum_{k=1}^K\lvec{g}_{l,k}x_{kt},\sigma_n^2\lmat{I}_N},
	\label{eq:app_JAC_p_y_lt}\\
	\Psi_{u_k} &\coloneq p_{u_k}(u_k) = (1-\lambda)\ind{u_k=0}+\lambda\ind{u_k=1},
	\label{eq:app_JAC_p_u_k}\\
	\Psi_{h_{l,k}} &\coloneq p_{h_{l,k}}(\lvec{h}_{l,k}) = \CN{\lvec{h}_{l,k}\big|\lvec{0},\gmat{\Xi}_{l,k}},
	\label{eq:app_JAC_p_h_lk}
\end{align}
and $\Psi_{g_{l,k}}\coloneq p(\lvec{g}_{l,k}|\lvec{h}_{l,k},u_k)$ is given in~\eqref{eq:p_g_lk}.
The corresponding factor graph is illustrated in Fig.~\ref{fig:app_JAC_FG}.

\subsection{\acs{EP} Approximations and Fronthaul Load}\label{app:JAC_EP_approx_fh_load}
We choose the same approximate exponential family distributions for the variable nodes as in the \ac{JACD-EP} algorithm, i.e., activities $u_k$ are modeled by categorical distributions whereas auxiliary variables $\lvec{g}_{l,k}$ and channels $\lvec{h}_{l,k}$ are modeled by multivariate complex Gaussian distributions.
Therefore, messages from and towards $u_k$ consist of one probability value which yields a fronthaul load of $2LK$ real-valued numbers.
Messages involving $\lvec{g}_{l,k}$ and $\lvec{h}_{l,k}$ consist of complex-valued vectors and matrices of dimension $N$ and $N\times N$, respectively, which do not contribute to the fronthaul load since they are only processed within an \ac{AP}.

\subsection{Initialization, Scheduling, and Estimation}\label{app:JAC_init_sched_alg}
As indicated in Appendix~\ref{app:JAC_FG}, the prior distributions $p_{u_k}(u_k)$ and $p_{h_{l,k}}(\lvec{h}_{l,k})$ are used for initializing the \ac{JAC-EP} algorithm.
The initial mean vector and covariance matrix of $\msg{\Psi_{g_{l,k}}}{\lvec{g}_{l,k}}$ and $\msg{\lvec{g}_{l,k}}{\Psi_{y_{l,t}}}$ $\forall k,l,t\leq T_p$ are set according to the prior information on $\lvec{g}_{l,k}$,
\begin{align}
	\Mumsg{\Psi_{g_{l,k}}}{\lvec{g}_{l,k}} &= \Mumsg{\lvec{g}_{l,k}}{\Psi_{y_{l,t}}} = \lvec{0},
	\label{eq:app_JAC_mu_g_Psi_y_init}\\
	\Cmsg{\Psi_{g_{l,k}}}{\lvec{g}_{l,k}} &= \Cmsg{\lvec{g}_{l,k}}{\Psi_{y_{l,t}}} = \lambda\cdot\gmat{\Xi}_{l,k}.
	\label{eq:app_JAC_C_g_Psi_y_init}
\end{align}
For all other messages, an uninformative initialization is chosen.
Then, we update all the messages according to the scheduling given in Algorithm~\ref{alg:app_JAC-EP} which is illustrated in Fig.~\ref{fig:app_JAC_FG} as red dashed arrows.
Finally, the estimates of the user activities and channels are computed by
\begin{align}
	\hat{u}_k &= \argmax_{u_k\in\{0,1\}}\;\hat{p}_{u_k}(u_k),
	\label{eq:app_JAC_estimate_u}\\
        \hat{\lvec{h}}_{l,k} &= \argmax_{\lvec{h}_{l,k}\in\Cset^N}\;\hat{p}_{\lvec{h}_{l,k}}(\lvec{h}_{l,k}) = \frac{1}{{Z}_{l,k}}\catmsg{u_k}{\Psi_{g_{l,k}}}\!(1)\,\vartheta(1)\,\Mumsga{{l,k}}{},
	\label{eq:app_JAC_estimate_h}
\end{align}
with the approximations of the posterior distributions
\begin{align}
	\hat{p}_{u_k}(u_k) &\propto p_{u_k}(u_k)\cdot\prod_{l=1}^L\catmsg{\Psi_{g_{l,k}}}{u_{k}}(u_k),
	\label{eq:posterior_u}\\
        \hat{p}_{\lvec{h}_{l,k}}(\lvec{h}_{l,k}) &\propto p_{h_{l,k}}(\lvec{h}_{l,k})\cdot\CN{\lvec{h}_{l,k}\big|\Mumsg{\Psi_{g_{l,k}}}{\lvec{h}_{l,k}},\Cmsg{\Psi_{g_{l,k}}}{\lvec{h}_{l,k}}}.
        \label{eq:posterior_h}
\end{align}
and $\catmsg{u_k}{\Psi_{g_{l,k}}}\!(u_k)$, $\vartheta(u_k)$, $\Mumsga{{l,k}}{}$, ${Z}_{l,k}$, $\catmsg{\Psi_{g_{l,k}}}{u_{k}}(u_k)$, $\Mumsg{\Psi_{g_{l,k}}}{\lvec{h}_{l,k}}$, and $\Cmsg{\Psi_{g_{l,k}}}{\lvec{h}_{l,k}}$ defined in Appendix~\ref{app:JAC_MP_updates_}.
{The derivation of~\eqref{eq:app_JAC_estimate_h} can be found in~\eqref{eq:app_JAC_mu_mmd_Psi_g_h}.}
\begin{algorithm}[t]
\caption{\ac{JAC-EP} Algorithm}
\begin{algorithmic}[1]
\renewcommand{\algorithmicrequire}{\textbf{Input:}}
\renewcommand{\algorithmicensure}{\textbf{Output:}}
\REQUIRE Pilot matrix $\pilot{\lmat{X}}$, received pilot signal $\pilot{\lmat{Y}}$, noise variance $\sigma_n^2$, user activity probability $\lambda$, channel correlation matrices $\gmat{\Xi}_{l,k}$.
\ENSURE Estimated activities $\hat{u}_k$ and channels $\hat{\lmat{h}}_{l,k}$.
\STATE $\forall k,l,t\leq T_p$: Initialize $\msg{\Psi_{g_{l,k}}}{\lvec{g}_{l,k}}$ and $\msg{\lvec{g}_{l,k}}{\Psi_{y_{l,t}}}$ via \eqref{eq:app_JAC_mu_g_Psi_y_init}, \eqref{eq:app_JAC_C_g_Psi_y_init}.
\FOR {$i = 1$ to $i_\text{max}$}
\STATE $\forall k,l,t\leq T_p$: Update $\msg{\Psi_{y_{l,t}}}{\lvec{g}_{l,k}}$ via \eqref{eq:app_JAC_mu_Psi_y_g_}, \eqref{eq:app_JAC_C_Psi_y_g_}$/$\eqref{eq:app_JAC_C_Psi_y_g_pc}.
\STATE $\forall k,l$: Update $\msg{\lvec{g}_{l,k}}{\Psi_{g_{l,k}}}$ via \eqref{eq:app_JAC_C_g_Psi_g_}, \eqref{eq:app_JAC_mu_g_Psi_g_}.
\STATE $\forall k,l$: Update $\msg{\Psi_{g_{l,k}}}{u_k}$ via \eqref{eq:app_JAC_m_Psi_g_u_}.
\STATE $\forall k,l$: Update $\msg{u_k}{\Psi_{g_{l,k}}}$ via \eqref{eq:app_JAC_m_u_Psi_g_}.
\STATE $\forall k,l$: Update $\msg{\Psi_{g_{l,k}}}{\lvec{g}_{l,k}}$ via \eqref{eq:app_JAC_C_Psi_g_g_}, \eqref{eq:app_JAC_mu_Psi_g_g_}.\label{alg_line:app_JAC_m_Psi_g_g}
\STATE $\forall k,l,t\leq T_p$: Update $\msg{\lvec{g}_{l,k}}{\Psi_{y_{l,t}}}$ via \eqref{eq:app_JAC_C_g_Psi_y_}, \eqref{eq:app_JAC_mu_g_Psi_y_}.
\ENDFOR
\RETURN $\hat{u}_k$ calculated via \eqref{eq:app_JAC_estimate_u} $\forall k$.
\RETURN $\hat{\lvec{h}}_{l,k}$ calculated via \eqref{eq:app_JAC_estimate_h} $\forall k,l$.
\end{algorithmic} 
\label{alg:app_JAC-EP}
\end{algorithm}

As in the \ac{JACD-EP} algorithm, we apply damping to the factor-to-variable messages with damping parameter $\eta\in[0,1]$
Furthermore, we update the parameters of $\msg{\Psi_{g_{l,k}}}{\lvec{g}_{l,k}}$ in line~\ref{alg_line:app_JAC_m_Psi_g_g} of Algorithm~\ref{alg:app_JAC-EP} only if the new covariance matrix obtained by~\eqref{eq:app_JAC_C_Psi_g_g_} is symmetric positive definite.
Otherwise, we keep the parameters from the previous iteration.

\subsection{Message-Passing Update Rules}\label{app:JAC_MP_updates_}
In this section, we present the \ac{EP} message-passing update rules for the factor graph in Fig.~\ref{fig:app_JAC_FG}. Detailed derivations can be found in Appendix~\ref{app:JAC_MP_updates}.
As in the description of the \ac{JACD-EP} algorithm, we switch between the representation of a Gaussian via the mean vector and covariance matrix and the natural parameters without explicitly mentioning the transformation.

\noindent\textit{Update of $\msg{\lvec{g}_{l,k}}{\Psi_{y_{l,t}}}$ {(cf.~\eqref{eq:app_JAC_C_g_Psi_y},~\eqref{eq:app_JAC_mu_g_Psi_y})}:}
\begin{align}
	\Lambdamsg{\lvec{g}_{l,k}}{\Psi_{y_{l,t}}} &= \Lambdamsg{\Psi_{g_{l,k}}}{\lvec{g}_{l,k}}+\sum_{t'\neq t}\Lambdamsg{\Psi_{y_{l,t'}}}{\lvec{g}_{l,k}}.
	\label{eq:app_JAC_C_g_Psi_y_}\\
	\Gammamsg{\lvec{g}_{l,k}}{\Psi_{y_{l,t}}} &= \Gammamsg{\Psi_{g_{l,k}}}{\lvec{g}_{l,k}}+\sum_{t'\neq t}\Gammamsg{\Psi_{y_{l,t'}}}{\lvec{g}_{l,k}},
	\label{eq:app_JAC_mu_g_Psi_y_}
\end{align}

\noindent\textit{Update of $\msg{\Psi_{y_{l,t}}}{\lvec{g}_{l,k}}$ {(cf.~\eqref{eq:app_JAC_mu_Psi_y_g},~\eqref{eq:app_JAC_C_Psi_y_g})}:}
\begin{align}
	\Mumsg{\Psi_{y_{l,t}}}{\lvec{g}_{l,k}} &= \left(\lvec{y}_{l,t}-\sum_{k'\neq k}\Mumsg{\lvec{g}_{l,k'}}{\Psi_{y_{l,t}}}x_{k't}\right)\cdot x_{kt}^{-1},
	\label{eq:app_JAC_mu_Psi_y_g_}\\
	\Cmsg{\Psi_{y_{l,t}}}{\lvec{g}_{l,k}} &= \left(\sigma_n^2\lmat{I}_N+\sum_{k'\neq k}\Cmsg{\lvec{g}_{l,k'}}{\Psi_{y_{l,t}}}|x_{k't}|^2\right)\cdot |x_{kt}|^{-2}.
	\label{eq:app_JAC_C_Psi_y_g_}
\end{align}
Note that the modification for pilot contamination described in Section~\ref{subsec:mod_pc} can be used to enhance the performance of the \ac{JAC-EP} algorithm.
The update in~\eqref{eq:app_JAC_C_Psi_y_g_} is then given by
\begin{equation}
	\Cmsg{\Psi_{y_{l,t}}}{\lvec{g}_{l,k}} = \left(\sigma_n^2\lmat{I}_N+\sum_{k'\in\mathcal{P}_k}\gmat{\Xi}_{l,k}|\pilot{x}_{k't}|^2+\sum_{k'\neq k}\Cmsg{\lvec{g}_{l,k'}}{\Psi_{y_{l,t}}}|x_{k't}|^2\right)\cdot |x_{kt}|^{-2}.
	\label{eq:app_JAC_C_Psi_y_g_pc}
\end{equation}

\noindent\textit{Update of $\msg{u_k}{\Psi_{g_{l,k}}}$ {(cf.~\eqref{eq:app_JAC_m_u_Psi_g})}:}
\begin{equation}
	\catmsg{u_k}{\Psi_{g_{l,k}}}(u_k) \propto p_{u_k}(u_k)\cdot\prod_{l'\neq l}\catmsg{\Psi_{g_{l',k}}}{u_{k}}(u_k),
	\label{eq:app_JAC_m_u_Psi_g_}
\end{equation}

\noindent\textit{Update of $\msg{\lvec{g}_{l,k}}{\Psi_{g_{l,k}}}$ {(cf.~\eqref{eq:app_JAC_C_g_Psi_g},~\eqref{eq:app_JAC_mu_g_Psi_g})}:}
\begin{align}
	 \Lambdamsg{\lvec{g}_{l,k}}{\Psi_{g_{l,k}}} &= \sum_{t=1}^{T_p}\Lambdamsg{\Psi_{y_{l,t}}}{\lvec{g}_{l,k}},
	\label{eq:app_JAC_C_g_Psi_g_}\\
	\Gammamsg{\lvec{g}_{l,k}}{\Psi_{g_{l,k}}} &= \sum_{t=1}^{T_p}\Gammamsg{\Psi_{y_{l,t}}}{\lvec{g}_{l,k}}.
	\label{eq:app_JAC_mu_g_Psi_g_}
\end{align}

\noindent\textit{Update of $\msg{\Psi_{g_{l,k}}}{u_k}$ {(cf.~\eqref{eq:app_JAC_m_Psi_g_u})}:}
\begin{equation}
	\catmsg{\Psi_{g_{l,k}}}{u_k}(u_k) \propto \vartheta(u_k),
	\label{eq:app_JAC_m_Psi_g_u_}
\end{equation}
with
\begin{equation*}
	\vartheta(u_k) = \mathcal{CN}(\lvec{0}|\Mumsg{\lvec{g}_{l,k}}{\Psi_{g_{l,k}}},\Cmsg{\lvec{g}_{l,k}}{\Psi_{g_{l,k}}}\!\!\!+\gmat{\Xi}_{l,k}u_k).
	\label{eq:app_JAC_theta_mmd_Psi_g_u_}
\end{equation*}

\noindent\textit{Update of $\msg{\Psi_{g_{l,k}}}{\lvec{g}_{l,k}}$ {(cf.~\eqref{eq:app_JAC_C_Psi_g_g},~\eqref{eq:app_JAC_mu_Psi_g_g})}:}
\begin{align}
	\Lambdamsg{\Psi_{g_{l,k}}}{\lvec{g}_{l,k}} &= \Lambdamsgb{{l,k}}{}-\Lambdamsg{\lvec{g}_{l,k}}{\Psi_{g_{l,k}}},
	\label{eq:app_JAC_C_Psi_g_g_}\\
	\Gammamsg{\Psi_{g_{l,k}}}{\lvec{g}_{l,k}} &= \Gammamsgb{{l,k}}{}-\Gammamsg{\lvec{g}_{l,k}}{\Psi_{g_{l,k}}},
	\label{eq:app_JAC_mu_Psi_g_g_}
\end{align}
with
\begin{align}
	\Mumsgb{{l,k}}{} &= \frac{1}{{Z}_{{l,k}}}\!\cdot\!\catmsg{u_k}{\Psi_{g_{l,k}}}\!(1)\cdot\vartheta(1)\cdot\Mumsga{{l,k}}{},
	\nonumber\\
	\Cmsgb{{l,k}}{} &= \frac{1}{{Z}_{{l,k}}}\!\cdot\!\catmsg{u_k}{\Psi_{g_{l,k}}}\!(1)\cdot\vartheta(1)\cdot(\Cmsga{{l,k}}{}\!+\!\Mumsga{{l,k}}{}\Mumsga{{l,k}}{}^H) \!-\! \Mumsgb{{l,k}}{}\Mumsgb{{l,k}}{}^H.
	\nonumber\\
    {Z}_{{l,k}} &= \catmsg{u_k}{\Psi_{g_{l,k}}}(0)\cdot\vartheta(0)+\catmsg{u_k}{\Psi_{g_{l,k}}}(1)\cdot\vartheta(1)
	\nonumber\\
        \Lambdamsga{{l,k}}{} &= \Lambdamsg{\lvec{g}_{l,k}}{\Psi_{g_{l,k}}}+\gmat{\Xi}_{l,k}^{-1},
	\nonumber\\
	\Gammamsga{{l,k}}{} &= \Gammamsg{\lvec{g}_{l,k}}{\Psi_{g_{l,k}}}\nonumber.
\end{align}

\section{Properties of Gaussian Distributions}\label{app:Gaussian}
\subsection{Gaussian Product Lemma}
The product of two Gaussian distributions yields an unnormalized Gaussian distribution given by~\cite{Bromiley2003,Ngo2020}
\begin{equation}
	\CN{\lvec{x}|\gvec{\mu}_1,\lmat{C}_1} \cdot \CN{\lvec{x}|\gvec{\mu}_2,\lmat{C}_2} = \CN{\lvec{x}|\gvec{\mu},\lmat{C}} \cdot \CN{\lvec{0}|\gvec{\mu}_1-\gvec{\mu}_2,\lmat{C}_1+\lmat{C}_2}
	\label{eq:app_Gaussian_product}
\end{equation}
with
\begin{align}
	\lmat{C} &= \left(\lmat{C}_1^{-1}+\lmat{C}_2^{-1}\right)^{-1},
	\label{eq:app_Gaussian_product_C}\\
	\gvec{\mu} &= \lmat{C}\left(\lmat{C}_1^{-1}\gvec{\mu}_1+\lmat{C}_2^{-1}\gvec{\mu}_2\right).
	\label{eq:app_Gaussian_product_mu}
\end{align}

\subsection{Gaussian Quotient Lemma}
The quotient of two Gaussian distributions is proportional to a Gaussian distribution given by~\cite{Ngo2020}
\begin{equation}
	\CN{\lvec{x}|\gvec{\mu}_1,\lmat{C}_1} / \CN{\lvec{x}|\gvec{\mu}_2,\lmat{C}_2} \propto \CN{\lvec{x}|\gvec{\mu},\lmat{C}}
	\label{eq:app_Gaussian_quotient}
\end{equation}
with
\begin{align}
	\lmat{C} &= \left(\lmat{C}_1^{-1}-\lmat{C}_2^{-1}\right)^{-1},\\
	\gvec{\mu} &= \lmat{C}\left(\lmat{C}_1^{-1}\gvec{\mu}_1-\lmat{C}_2^{-1}\gvec{\mu}_2\right).
\end{align}
This can be verified by considering $\CN{\lvec{x}|\gvec{\mu}_1,\lmat{C}_1}\propto\CN{\lvec{x}|\gvec{\mu}_2,\lmat{C}_2}\cdot\CN{\lvec{x}|\gvec{\mu},\lmat{C}}$ and applying the Gaussian product lemma.

\subsection{Gaussian Scaling Lemma}
{The probability function of $\lvec{y}=c\,\lvec{x}$ with $c$ being a scalar constant and $\lvec{x}\in\Cset^N$ being an $N$-dimensional Gaussian random vector, $\lvec{x}\sim\CN{\lvec{x}|\gvec{\mu},\lmat{C}}$, is a Gaussian distribution with mean $c\,\gvec{\mu}$ and covariance matrix $|c|^2\,\lmat{C}$ which can be written in terms of the probability distribution of $\lvec{x}$, i.e.,  $\CN{\lvec{x}|\gvec{\mu},\lmat{C}}=\CN{c^{-1}\,\lvec{y}|\gvec{\mu},\lmat{C}}$, as
\begin{equation}
	\CN{\lvec{y}\,|\,c\,\gvec{\mu},|c|^2\,\lmat{C}} = |c|^{-2N}\cdot\CN{c^{-1}\,\lvec{y}|\gvec{\mu},\lmat{C}}.
	\label{eq:app_Gaussian_scaling}
\end{equation}
}
This can be shown by plugging in the definition of the Gaussian distribution and rearranging the result.

\section{Expectation Propagation on Graphs}\label{app:EP_graph}
In this section, we present the general message-passing update rules for \ac{EP} on graphs.
Details can be found in~\cite{Karataev2024} and references therein.

\begin{figure}[t]
    \centering
    \subfloat[Variable-to-factor message update.]{\includegraphics[width=0.8\textwidth]{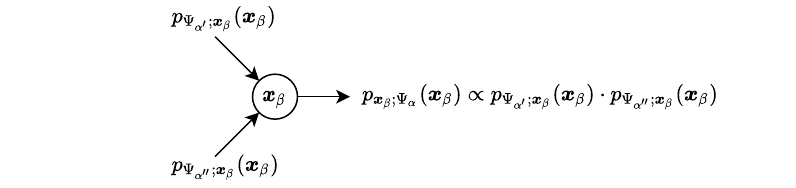}
        \label{fig:EP_MP_var_to_fac_update}}\\
    \subfloat[Factor-to-variable message update.]{\includegraphics[width=0.8\textwidth]{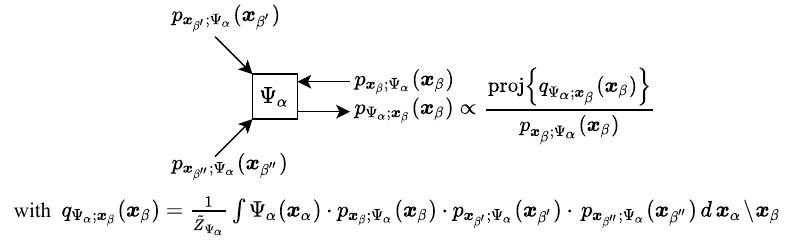}
        \label{fig:EP_MP_fac_to_var_update}}
    \caption{Illustration of the \ac{EP} message-passing update rules.}
    \label{fig:EP_MP_updates}
\end{figure}

{
The variable-to-factor message update is obtained by computing the parameters of the following distribution~\cite{Karataev2024},
}
\begin{equation}
	\msgp{\lvec{x}_\beta}{\Psi_\alpha}(\lvec{x}_\beta) \propto \prod_{\alpha'\in N_\beta\setminus\alpha}\msgp{\Psi_{\alpha'}}{\lvec{x}_\beta}(\lvec{x}_\beta),
	\label{eq:EP_var_to_fac_message}
\end{equation}
where $N_\beta$ denotes the set of indices $\alpha$ of all factor nodes $\Psi_\alpha$ that are connected to the variable node $\lvec{x}_\beta$.
{
The factor-to-variable message is updated by determining the parameters of the following distribution~\cite{Karataev2024},
}
\begin{equation}
	\msgp{\Psi_\alpha}{\lvec{x}_\beta}(\lvec{x}_\beta) \propto \frac{\text{proj}\left\{\mmd{\Psi_{\alpha}}{\lvec{x}_{\beta}}(\lvec{x}_\beta)\right\}}{\msgp{\lvec{x}_\beta}{\Psi_\alpha}(\lvec{x}_\beta)},
	\label{eq:EP_fac_to_var_message}
\end{equation}
where the distribution $\mmd{\Psi_{\alpha}}{\lvec{x}_{\beta}}(\lvec{x}_\beta)$ is given by
\begin{equation}
	\mmd{\Psi_{\alpha}}{\lvec{x}_{\beta}}(\lvec{x}_\beta) = \frac{1}{\tilde{Z}_{\Psi_\alpha}}\int\Psi_\alpha(\lvec{x}_\alpha)\prod_{\beta'\in N_\alpha}\msgp{\lvec{x}_{\beta'}}{\Psi_\alpha}(\lvec{x}_{\beta'})\,\mathrm{d}\lvec{x}_\alpha\!\setminus\!\lvec{x}_\beta.
	\label{eq:message_projection}
\end{equation}
Here, $\tilde{Z}_{\Psi_\alpha}$ is a normalization constant, $N_\alpha$ denotes the set of indices $\beta$ of all variable nodes $\lvec{x}_\beta$ that are connected to the factor node $\Psi_\alpha$, and $\lvec{x}_\alpha$ is a vector containing all variables connected to $\Psi_\alpha$, i.e., $\lvec{x}_\alpha=\bigcup_{\beta\in N_\alpha}\lvec{x}_{\beta}$.
Furthermore, $\text{proj}\{\cdot\}$ denotes the projection operator defined as
\begin{equation}
	\text{proj}\{f(\lvec{x})\} = \argmin_{g(\lvec{x})\in\mathcal{F}}D_{KL}(f(\lvec{x})||g(\lvec{x})),
	\label{eq:projection}
\end{equation}
where $D_{KL}(f(\lvec{x})||g(\lvec{x}))$ is the \ac{KL} divergence between $f$ and $g$ and $\mathcal{F}$ is an exponential family distribution.
This minimization is done by \textit{moment matching} which corresponds to matching the mean vector and covariance matrix of $f$ and $g$ if $\mathcal{F}$ represents Gaussian distributions and to matching the probability values of $f$ and $g$ evaluated at the feasible points of the discrete random variable $\lvec{x}$ if $\mathcal{F}$ represents categorical distributions.
The message-passing update rules~\eqref{eq:EP_var_to_fac_message} and~\eqref{eq:EP_fac_to_var_message} are illustrated in Fig.~\ref{fig:EP_MP_updates}.
The approximate posterior distribution $\hat{p}_{\lvec{x}_\beta}(\lvec{x}_\beta)$ of the variable $\lvec{x}_\beta$ can be computed via~\cite{Karataev2024}
\begin{align}
	\hat{p}_{\lvec{x}_\beta}(\lvec{x}_\beta) &\propto \prod_{\alpha\in N_\beta} \msgp{\Psi_\alpha}{\lvec{x}_\beta}(\lvec{x}_\beta).
	\label{eq:approx_factor_beta_factorization}
\end{align}

\section{Derivation of Message-Passing Update Rules for \acs{JACD-EP}}\label{app:MP_updates}
In the following, we apply the \ac{EP} message-passing rules presented in Appendix~\ref{app:EP_graph} to the factor graph in Fig.~\ref{fig:FG} and show the detailed derivations for the message updates.
{
Here, the message updates are implicitly given by the computation of the corresponding probability distributions.
}
Additionally, we use the properties of Gaussian distributions presented in Appendix~\ref{app:Gaussian}.

\subsection{Message Updates for Leaf Nodes $\Psi_{u_k}$, $\Psi_{h_{l,k}}$, and $\Psi_{x_{kt}}$}\label{subsubsec:MP_leaf}
Since the prior distributions $\tilde{p}_{u_k}(u_k)$, $\tilde{p}_{h_{l,k}}(\lvec{h}_{l,k})$, and $p_x(x_{kt})$ are in the same exponential family as the approximate distributions of $u_k$, $\lvec{h}_{l,k}$, and $x_{kt}$, respectively, the corresponding message updates simplify significantly.
The factor-to-variable messages are constant and consist of the prior information on the variables $u_k$, $\lvec{h}_{l,k}$, and $x_{kt}$ {which correspond to the distributions}
\begin{align}
	\msgp{\Psi_{u_k}}{u_k}(u_k) &= \tilde{p}_{u_k}(u_k),
	\label{eq:app_m_Psi_u_u}\\
	\msgp{\Psi_{h_{l,k}}}{\lvec{h}_{l,k}}(\lvec{h}_{l,k}) &= \tilde{p}_{h_{l,k}}(\lvec{h}_{l,k}),
	\label{eq:app_m_Psi_h_h}\\
	\msgp{\Psi_{x_{kt}}}{x_{kt}}(x_{kt}) &= p_x(x_{kt}).
	\label{eq:app_m_Psi_x_x}
\end{align}
This result is obtained by applying the message-passing rule in~\eqref{eq:EP_fac_to_var_message} while taking into account the fact that exponential family distributions are closed under multiplication.
This makes the projection operation in~\eqref{eq:EP_fac_to_var_message} superfluous.
Hence, {the distributions corresponding to the updated messages} are directly given by the factors $\Psi_{u_k}$, $\Psi_{h_{l,k}}$, and $\Psi_{x_{kt}}$, respectively, which correspond to the priors according to~\eqref{eq:p_u_k}-\eqref{eq:p_x_kt}.
The variable-to-factor messages for the leaf nodes are irrelevant in the unfolding of the algorithm and omitted here.

\subsection{Message Updates for $\Psi_{y_{l,t}}$}\label{subsubsec:MP_Psi_y}

\subsubsection{Incoming messages to factor node $\Psi_{y_{l,t}}$}
\begin{equation}
	\msgp{\lvec{z}_{l,kt}}{\Psi_{y_{l,t}}}(\lvec{z}_{l,kt}) = \msgp{\Psi_{z_{l,kt}}}{\lvec{z}_{l,kt}}(\lvec{z}_{l,kt}).
	\label{eq:app_m_z_Psi_y}
\end{equation}

\subsubsection{Outgoing messages from factor node $\Psi_{y_{l,t}}$}
\hfill\\\underline{Message Update $\msg{\Psi_{y_{l,t}}}{\lvec{z}_{l,kt}}$}\\
\begin{equation}
	\msgp{\Psi_{y_{l,t}}}{\lvec{z}_{l,kt}}(\lvec{z}_{l,kt}) \propto \frac{\proj{\mmd{\Psi_{y_{l,t}}}{\lvec{z}_{l,kt}}(\lvec{z}_{l,kt})}}{\msgp{\lvec{z}_{l,kt}}{\Psi_{y_{l,t}}}(\lvec{z}_{l,kt})},
	\label{eq:app_m_Psi_y_z0}
\end{equation}
with
\begin{align}
	&\mmd{\Psi_{y_{l,t}}}{\lvec{z}_{l,kt}}(\lvec{z}_{l,kt})\nonumber\\
	&\quad\propto \int\!\!\cdots\!\!\int \CN{\lvec{y}_{l,t}\Bigg|\sum_{k'=1}^K\lvec{z}_{l,k't},\sigma_n^2\lmat{I}_N}\cdot\msgp{\lvec{z}_{l,kt}}{\Psi_{y_{l,t}}}(\lvec{z}_{l,kt})\nonumber\cdot\prod_{k'\neq k}\msgp{\lvec{z}_{l,k't}}{\Psi_{y_{l,t}}}(\lvec{z}_{l,k't})\,d\lvec{z}_{l,k't}\nonumber\\
	&\quad\overset{(a)}{=} \msgp{\lvec{z}_{l,kt}}{\Psi_{y_{l,t}}}(\lvec{z}_{l,kt})\cdot\int\!\!\cdots\!\!\int \CN{\lvec{z}_{l,k''t}\Bigg|\lvec{y}_{l,t}-\sum_{k'\neq k''}\lvec{z}_{l,k't},\sigma_n^2\lmat{I}_N}\nonumber\\
	&\qquad\cdot\prod_{k'\neq k}\CN{\lvec{z}_{l,k't}\big|\Mumsg{\Psi_{z_{l,k't}}}{\lvec{z}_{l,k't}},\Cmsg{\Psi_{z_{l,k't}}}{\lvec{z}_{l,k't}}}\,d\lvec{z}_{l,k't}\nonumber\\
	&\quad\overset{(b)}{=} \msgp{\lvec{z}_{l,kt}}{\Psi_{y_{l,t}}}(\lvec{z}_{l,kt})\cdot\int\!\!\cdots\!\!\int \CN{\lvec{z}_{l,k''t}|\gvec{\mu}_\text{tmp},\lmat{C}_\text{tmp}}\,d\lvec{z}_{l,k''t}\nonumber\\
	&\qquad\cdot\CN{\lvec{0}\Bigg|\lvec{y}_{l,t}-\sum_{k'\neq k''}\lvec{z}_{l,k't}-\Mumsg{\Psi_{z_{l,k''t}}}{\lvec{z}_{l,k''t}},\sigma_n^2\lmat{I}_N+\Cmsg{\Psi_{z_{l,k''t}}}{\lvec{z}_{l,k''t}}}\nonumber\\
	&\qquad\cdot\prod_{k'\neq\{k,k''\}}\CN{\lvec{z}_{l,k't}\big|\Mumsg{\Psi_{z_{l,k't}}}{\lvec{z}_{l,k't}},\Cmsg{\Psi_{z_{l,k't}}}{\lvec{z}_{l,k't}}}\,d\lvec{z}_{l,k't}\nonumber\\
	&\quad\overset{(c)}{=} \msgp{\lvec{z}_{l,kt}}{\Psi_{y_{l,t}}}(\lvec{z}_{l,kt})\cdot\int\!\!\cdots\!\!\int \CN{\lvec{y}_{l,t}\Bigg|\sum_{k'\neq\{k''\}}\lvec{z}_{l,k't}+\Mumsg{\Psi_{z_{l,k''t}}}{\lvec{z}_{l,k''t}},\sigma_n^2\lmat{I}_N+\Cmsg{\Psi_{z_{l,k''t}}}{\lvec{z}_{l,k''t}}}\nonumber\\
	&\qquad\cdot\prod_{k'\neq\{k,k''\}}\CN{\lvec{z}_{l,k't}\big|\Mumsg{\Psi_{z_{l,k't}}}{\lvec{z}_{l,k't}},\Cmsg{\Psi_{z_{l,k't}}}{\lvec{z}_{l,k't}}}\,d\lvec{z}_{l,k't}\nonumber\\
	&\quad= \dots\nonumber\\
	&\quad= \msgp{\lvec{z}_{l,kt}}{\Psi_{y_{l,t}}}(\lvec{z}_{l,kt})\cdot\CN{\lvec{z}_{l,kt}\Big|\lvec{y}_{l,t}-\sum_{k'\neq k}\Mumsg{\Psi_{z_{l,k't}}}{\lvec{z}_{l,k't}},\sigma_n^2\lmat{I}_N+\sum_{k'\neq k}\Cmsg{\Psi_{z_{l,k't}}}{\lvec{z}_{l,k't}}},
	\label{eq:app_mmd_Psi_y_z}
\end{align}
where $(a)$ is obtained by a basic transformation of the Gaussian distribution and using~\eqref{eq:app_m_z_Psi_y}, $(b)$ is obtained by the Gaussian product rule~\eqref{eq:app_Gaussian_product} with $\gvec{\mu}_1=\lvec{y}_{l,t}-\sum_{k'\neq k''}\lvec{z}_{l,k't}$, $\gvec{\mu}_2=\Mumsg{\Psi_{z_{l,k''t}}}{\lvec{z}_{l,k''t}}$, $\lmat{C}_1=\sigma_n^2\lmat{I}_N$, and $\lmat{C}_2=\Cmsg{\Psi_{z_{l,k''t}}}{\lvec{z}_{l,k''t}}$ which yields $\lmat{C}_\text{tmp}$ and $\gvec{\mu}_\text{tmp}$ according to~\eqref{eq:app_Gaussian_product_C} and~\eqref{eq:app_Gaussian_product_mu}, respectively, and $(c)$ is obtained by integrating over $\lvec{z}_{l,k''t}$ and applying a basic transformation of the Gaussian distribution.
The final result in~\eqref{eq:app_mmd_Psi_y_z} is obtained by repeatedly applying the above described steps for all $k'\neq k$.
Since $\msgp{\lvec{z}_{l,kt}}{\Psi_{y_{l,t}}}(\lvec{z}_{l,kt})$ is Gaussian distributed, we can conclude by utilizing the Gaussian product lemma that $\mmd{\Psi_{y_{l,t}}}{\lvec{z}_{l,kt}}(\lvec{z}_{l,kt})$ is Gaussian distributed as well.
Hence, the projection operation in~\eqref{eq:app_m_Psi_y_z0} is superfluous since it projects $\mmd{\Psi_{y_{l,t}}}{\lvec{z}_{l,kt}}(\lvec{z}_{l,kt})$ onto a Gaussian distribution which is the \ac{EP} exponential family approximation choice of $\lvec{z}_{l,kt}$.
Therefore, the denominator of~\eqref{eq:app_m_Psi_y_z0} cancels with the first term in~\eqref{eq:app_mmd_Psi_y_z}, and the final message update rule is given by {the distribution}
\begin{equation}
	\msgp{\Psi_{y_{l,t}}}{\lvec{z}_{l,kt}}(\lvec{z}_{l,kt}) = \CN{\lvec{z}_{l,kt}|\Mumsg{\Psi_{y_{l,t}}}{\lvec{z}_{l,kt}},\Cmsg{\Psi_{y_{l,t}}}{\lvec{z}_{l,kt}}},
	\label{eq:app_m_Psi_y_z}
\end{equation}
with
\begin{align}
	\Mumsg{\Psi_{y_{l,t}}}{\lvec{z}_{l,kt}} &= \lvec{y}_{l,t}-\sum_{k'\neq k}\Mumsg{\Psi_{z_{l,k't}}}{\lvec{z}_{l,k't}},
	\label{eq:app_mu_Psi_y_z}\\
	\Cmsg{\Psi_{y_{l,t}}}{\lvec{z}_{l,kt}} &= \sigma_n^2\lmat{I}_N+\sum_{k'\neq k}\Cmsg{\Psi_{z_{l,k't}}}{\lvec{z}_{l,k't}}.
	\label{eq:app_C_Psi_y_z}
\end{align}

\subsection{Message Updates for $\Psi_{z_{l,kt}}$}\label{subsubsec:MP_Psi_z}

\subsubsection{Incoming messages to factor node $\Psi_{z_{l,kt}}$}
\begin{align}
	\msgp{x_{kt}}{\Psi_{z_{l,kt}}}(x_{kt}) &\propto \msgp{\Psi_{x_{kt}}}{x_{kt}}(x_{kt})\cdot\prod_{l'\neq l}\msgp{\Psi_{z_{l',kt}}}{x_{kt}}(x_{kt}),\nonumber\\
    &\propto \prod_{l'\neq l}\msgp{\Psi_{z_{l',kt}}}{x_{kt}}(x_{kt}),
	\label{eq:app_m_x_Psi_z}\\
	\msgp{\lvec{z}_{l,kt}}{\Psi_{z_{l,kt}}}(\lvec{z}_{l,kt}) &= \msgp{\Psi_{y_{l,t}}}{\lvec{z}_{l,kt}}(\lvec{z}_{l,kt}),
	\label{eq:app_m_z_Psi_z}\\
	\msgp{\lvec{g}_{l,k}}{\Psi_{z_{l,kt}}}(\lvec{g}_{l,k}) &\propto \msgp{\Psi_{g_{l,k}}}{\lvec{g}_{l,k}}(\lvec{g}_{l,k})\cdot\prod_{t'\neq t}\msgp{\Psi_{z_{l,kt'}}}{\lvec{g}_{l,k}}(\lvec{g}_{l,k})\nonumber\\
	&\propto \CN{\lvec{g}_{l,k}|\Mumsg{\lvec{g}_{l,k}}{\Psi_{z_{l,kt}}},\Cmsg{\lvec{g}_{l,k}}{\Psi_{z_{l,kt}}}},
	\label{eq:app_m_g_Psi_z}
\end{align}
with
\begin{align}
	\Cmsg{\lvec{g}_{l,k}}{\Psi_{z_{l,kt}}} &= \left(\Cmsg{\Psi_{g_{l,k}}}{\lvec{g}_{l,k}}^{-1}+\sum_{t'\neq t}\Cmsg{\Psi_{z_{l,kt'}}}{\lvec{g}_{l,k}}^{-1}\right)^{-1},
	\label{eq:app_C_g_Psi_z}\\
	\Mumsg{\lvec{g}_{l,k}}{\Psi_{z_{l,kt}}} &= \Cmsg{\lvec{g}_{l,k}}{\Psi_{z_{l,kt}}}\Bigg(\Cmsg{\Psi_{g_{l,k}}}{\lvec{g}_{l,k}}^{-1}\Mumsg{\Psi_{g_{l,k}}}{\lvec{g}_{l,k}}+\sum_{t'\neq t}\Cmsg{\Psi_{z_{l,kt'}}}{\lvec{g}_{l,k}}^{-1}\Mumsg{\Psi_{z_{l,kt'}}}{\lvec{g}_{l,k}}\Bigg),
	\label{eq:app_mu_g_Psi_z}
\end{align}
which is obtained by applying the Gaussian product lemma multiple times.
The update in~\eqref{eq:app_m_x_Psi_z} is obtained by noticing that $\msgp{\Psi_{x_{kt}}}{x_{kt}}(x_{kt})$~\eqref{eq:app_m_Psi_x_x} is uninformative according to~\eqref{eq:p_x_kt} for $t>T_p$.

\subsubsection{Outgoing messages from factor node $\Psi_{z_{l,kt}}$}
\hfill\\\underline{Message Update $\msg{\Psi_{z_{l,kt}}}{x_{kt}}$}\\
\begin{equation}
	\msgp{\Psi_{z_{l,kt}}}{x_{kt}}(x_{kt}) \propto \frac{\proj{\mmd{\Psi_{z_{l,kt}}}{x_{kt}}(x_{kt})}}{\msgp{x_{kt}}{\Psi_{z_{l,kt}}}(x_{kt})},
	\label{eq:app_m_Psi_z_x0}
\end{equation}
with
\begin{align}
	\mmd{\Psi_{z_{l,kt}}}{x_{kt}}(x_{kt})&\propto \int\int\delta(\lvec{z}_{l,kt}-\lvec{g}_{l,k}x_{kt})\cdot\msgp{x_{kt}}{\Psi_{z_{l,kt}}}(x_{kt})\cdot\msgp{\lvec{z}_{l,kt}}{\Psi_{z_{l,kt}}}(\lvec{z}_{l,kt})\nonumber\\
	&\qquad\quad\cdot\msgp{\lvec{g}_{l,k}}{\Psi_{z_{l,kt}}}(\lvec{g}_{l,k})\,d\lvec{z}_{l,kt}\,d\lvec{g}_{l,k}\nonumber\\
	&\overset{(a)}{=} \msgp{x_{kt}}{\Psi_{z_{l,kt}}}(x_{kt})\cdot\int\msgp{\lvec{z}_{l,kt}}{\Psi_{z_{l,kt}}}(\lvec{g}_{l,k}x_{kt})\cdot\msgp{\lvec{g}_{l,k}}{\Psi_{z_{l,kt}}}(\lvec{g}_{l,k})\,d\lvec{g}_{l,k}\nonumber\\
	&\overset{(b)}{=} \msgp{x_{kt}}{\Psi_{z_{l,kt}}}(x_{kt})\cdot\int\CN{\lvec{g}_{l,k}x_{kt}|\Mumsg{\Psi_{y_{l,t}}}{\lvec{z}_{l,kt}},\Cmsg{\Psi_{y_{l,t}}}{\lvec{z}_{l,kt}}}\nonumber\\
	&\qquad\cdot\CN{\lvec{g}_{l,k}|\Mumsg{\lvec{g}_{l,k}}{\Psi_{z_{l,kt}}},\Cmsg{\lvec{g}_{l,k}}{\Psi_{z_{l,kt}}}}\,d\lvec{g}_{l,k}\nonumber\\
	&\overset{(c)}{=} \msgp{x_{kt}}{\Psi_{z_{l,kt}}}(x_{kt})\cdot\int|x_{kt}|^{-2N}\cdot\CN{\lvec{g}_{l,k}|\gvec{\mu}_\text{tmp},\lmat{C}_\text{tmp}}\nonumber\\
	&\qquad\cdot\CN{\lvec{0}|\Mumsg{\Psi_{y_{l,t}}}{\lvec{z}_{l,kt}}x_{kt}^{-1}-\Mumsg{\lvec{g}_{l,k}}{\Psi_{z_{l,kt}}},\Cmsg{\Psi_{y_{l,t}}}{\lvec{z}_{l,kt}}|x_{kt}|^{-2}+\Cmsg{\lvec{g}_{l,k}}{\Psi_{z_{l,kt}}}}\,d\lvec{g}_{l,k}\nonumber\\
	&= \msgp{x_{kt}}{\Psi_{z_{l,kt}}}(x_{kt})\cdot|x_{kt}|^{-2N}\nonumber\\
	&\quad\cdot\CN{\lvec{0}|\Mumsg{\Psi_{y_{l,t}}}{\lvec{z}_{l,kt}}x_{kt}^{-1}-\Mumsg{\lvec{g}_{l,k}}{\Psi_{z_{l,kt}}},\Cmsg{\Psi_{y_{l,t}}}{\lvec{z}_{l,kt}}|x_{kt}|^{-2}+\Cmsg{\lvec{g}_{l,k}}{\Psi_{z_{l,kt}}}}\nonumber\\
	&= \msgp{x_{kt}}{\Psi_{z_{l,kt}}}(x_{kt})\cdot\theta(x_{kt}),
	\label{eq:app_mmd_Psi_z_x}
\end{align}
where $(a)$ is obtained by the sifting property of the Dirac delta function~\cite{Candan2021}, $(b)$ is obtained by using~\eqref{eq:app_m_z_Psi_z}, and $(c)$ is obtained by utilizing the Gaussian scaling lemma on $\mathcal{CN}\Big(\lvec{g}_{l,k}x_{kt}|\Mumsg{\Psi_{y_{l,t}}}{\lvec{z}_{l,kt}},$ $\Cmsg{\Psi_{y_{l,t}}}{\lvec{z}_{l,kt}}\Big)$ and then the Gaussian product lemma with $\gvec{\mu}_1=\Mumsg{\Psi_{y_{l,t}}}{\lvec{z}_{l,kt}}x_{kt}^{-1}$, $\gvec{\mu}_2=\Mumsg{\lvec{g}_{l,k}}{\Psi_{z_{l,kt}}}$, $\lmat{C}_1=\Cmsg{\Psi_{y_{l,t}}}{\lvec{z}_{l,kt}}|x_{kt}|^{-2}$, and $\lmat{C}_2=\Cmsg{\lvec{g}_{l,k}}{\Psi_{z_{l,kt}}}$ which yields $\lmat{C}_\text{tmp}$ and $\gvec{\mu}_\text{tmp}$ according to~\eqref{eq:app_Gaussian_product_C} and~\eqref{eq:app_Gaussian_product_mu}, respectively.
The final result in~\eqref{eq:app_mmd_Psi_z_x} is obtained by applying the Gaussian scaling lemma once again with
\begin{equation}
	\theta(x_{kt}) = \CN{\lvec{0}|\Mumsg{\Psi_{y_{l,t}}}{\lvec{z}_{l,kt}}-\Mumsg{\lvec{g}_{l,k}}{\Psi_{z_{l,kt}}}x_{kt},\Cmsg{\Psi_{y_{l,t}}}{\lvec{z}_{l,kt}}+\Cmsg{\lvec{g}_{l,k}}{\Psi_{z_{l,kt}}}|x_{kt}|^2}.
	\label{eq:app_theta_mmd_Psi_z_z}
\end{equation}
Evaluating~\eqref{eq:app_mmd_Psi_z_x} at $x_{kt}\in\mathcal{X}$ for $t>T_p$ yields a categorical distribution for $\mmd{\Psi_{z_{l,kt}}}{x_{kt}}(x_{kt})$.
Hence, the projection operation in~\eqref{eq:app_m_Psi_z_x0} is superfluous since it projects $\mmd{\Psi_{z_{l,kt}}}{x_{kt}}(x_{kt})$ onto a categorical distribution, and the denominator of~\eqref{eq:app_m_Psi_z_x0} cancels with the first term in~\eqref{eq:app_mmd_Psi_z_x}.
Thus, the final message update rule is given by {the distribution}
\begin{equation}
	\msgp{\Psi_{z_{l,kt}}}{x_{kt}}(x_{kt}) \propto \theta(x_{kt}).
	\label{eq:app_m_Psi_z_x}
\end{equation}

\noindent\underline{Message Update $\msg{\Psi_{z_{l,kt}}}{\lvec{z}_{l,kt}}$}\\
\begin{equation}
	\msgp{\Psi_{z_{l,kt}}}{\lvec{z}_{l,kt}}(\lvec{z}_{l,kt}) \propto \frac{\proj{\mmd{\Psi_{z_{l,kt}}}{\lvec{z}_{l,kt}}(\lvec{z}_{l,kt})}}{\msgp{\lvec{z}_{l,kt}}{\Psi_{z_{l,kt}}}(\lvec{z}_{l,kt})},
	\label{eq:app_m_Psi_z_z0}
\end{equation}
with
\begin{align}
	&\mmd{\Psi_{z_{l,kt}}}{\lvec{z}_{l,kt}}(\lvec{z}_{l,kt})\nonumber\\
	&\quad\propto \sum_{x_{kt}}\int\delta(\lvec{z}_{l,kt}-\lvec{g}_{l,k}x_{kt})\cdot\msgp{x_{kt}}{\Psi_{z_{l,kt}}}(x_{kt})\cdot\msgp{\lvec{z}_{l,kt}}{\Psi_{z_{l,kt}}}(\lvec{z}_{l,kt})\nonumber\cdot\msgp{\lvec{g}_{l,k}}{\Psi_{z_{l,kt}}}(\lvec{g}_{l,k})\,d\lvec{g}_{l,k}\nonumber\\
	&\quad\overset{(a)}{=} \sum_{x_{kt}}|x_{kt}|^{-2N}\cdot\msgp{x_{kt}}{\Psi_{z_{l,kt}}}(x_{kt})\cdot\msgp{\Psi_{y_{l,t}}}{\lvec{z}_{l,kt}}(\lvec{z}_{l,kt})\cdot\msgp{\lvec{g}_{l,k}}{\Psi_{z_{l,kt}}}\left(\frac{\lvec{z}_{l,kt}}{x_{kt}}\right)\nonumber\\
	&\quad= \sum_{x_{kt}}\msgp{x_{kt}}{\Psi_{z_{l,kt}}}(x_{kt})\cdot\CN{\lvec{z}_{l,kt}|\Mumsga{\Psi_{z_{l,kt}}}{\lvec{z}_{l,kt}}(x_{kt}),\Cmsga{\Psi_{z_{l,kt}}}{\lvec{z}_{l,kt}}(x_{kt})}\cdot\theta(x_{kt}),
	\label{eq:app_mmd_Psi_z_z}
\end{align}
where $(a)$ is obtained by applying the scaling and sifting property of the Dirac delta function~\cite{Candan2021} and using~\eqref{eq:app_m_z_Psi_z}.
The final result in~\eqref{eq:app_mmd_Psi_z_z} is obtained by the Gaussian scaling and product lemma with $\theta(x_{kt})$ given in~\eqref{eq:app_theta_mmd_Psi_z_z} and
\begin{align}
	\Cmsga{\Psi_{z_{l,kt}}}{\lvec{z}_{l,kt}}(x_{kt}) &= \left(\Cmsg{\Psi_{y_{l,t}}}{\lvec{z}_{l,kt}}^{-1}+\Cmsg{\lvec{g}_{l,k}}{\Psi_{z_{l,kt}}}^{-1}|x_{kt}|^{-2}\right)^{-1},
	\label{eq:app_C_tmp_mmd_Psi_z_z}\\
	\Mumsga{\Psi_{z_{l,kt}}}{\lvec{z}_{l,kt}}(x_{kt}) &= \Cmsga{\Psi_{z_{l,kt}}}{\lvec{z}_{l,kt}}(x_{kt})\Bigg(\Cmsg{\Psi_{y_{l,t}}}{\lvec{z}_{l,kt}}^{-1}\Mumsg{\Psi_{y_{l,t}}}{\lvec{z}_{l,kt}}+\Cmsg{\lvec{g}_{l,k}}{\Psi_{z_{l,kt}}}^{-1}\Mumsg{\lvec{g}_{l,k}}{\Psi_{z_{l,kt}}}\frac{x_{kt}}{|x_{kt}|^2}\Bigg).
	\label{eq:app_mu_tmp_mmd_Psi_z_z}
\end{align}
The normalization constant for $\mmd{\Psi_{z_{l,kt}}}{\lvec{z}_{l,kt}}(\lvec{z}_{l,kt})$ is given by
\begin{align}
	\tilde{Z}_{\Psi_{z_{l,kt}}} &= \int\sum_{x_{kt}}\msgp{x_{kt}}{\Psi_{z_{l,kt}}}(x_{kt})\cdot\CN{\lvec{z}_{l,kt}|\Mumsga{\Psi_{z_{l,kt}}}{\lvec{z}_{l,kt}}(x_{kt}),\Cmsga{\Psi_{z_{l,kt}}}{\lvec{z}_{l,kt}}(x_{kt})}\cdot\theta(x_{kt})\,d\lvec{z}_{l,kt}\nonumber\\
	&= \sum_{x_{kt}}\msgp{x_{kt}}{\Psi_{z_{l,kt}}}(x_{kt})\cdot\theta(x_{kt}).
	\label{eq:app_Z_Psi_z}
\end{align}
According to~\eqref{eq:app_mmd_Psi_z_z}, $\mmd{\Psi_{z_{l,kt}}}{\lvec{z}_{l,kt}}(\lvec{z}_{l,kt})$ is a Gaussian mixture with mean vector and covariance matrix
\begin{align}
	\Mumsgb{\Psi_{z_{l,kt}}}{\lvec{z}_{l,kt}} &= \frac{1}{\tilde{Z}_{\Psi_{z_{l,kt}}}}\sum_{x_{kt}}\msgp{x_{kt}}{\Psi_{z_{l,kt}}}(x_{kt})\cdot\theta(x_{kt})\cdot\Mumsga{\Psi_{z_{l,kt}}}{\lvec{z}_{l,kt}}(x_{kt}),
	\label{eq:app_mu_mmd_Psi_z_z}\\
\begin{split}
	\Cmsgb{\Psi_{z_{l,kt}}}{\lvec{z}_{l,kt}} &= \frac{1}{\tilde{Z}_{\Psi_{z_{l,kt}}}}\sum_{x_{kt}}\msgp{x_{kt}}{\Psi_{z_{l,kt}}}(x_{kt})\cdot\theta(x_{kt})\cdot\Big(\Cmsga{\Psi_{z_{l,kt}}}{\lvec{z}_{l,kt}}(x_{kt})\\
	&\quad+\Mumsga{\Psi_{z_{l,kt}}}{\lvec{z}_{l,kt}}(x_{kt})\Mumsga{\Psi_{z_{l,kt}}}{\lvec{z}_{l,kt}}^H(x_{kt})\Big) - \Mumsgb{\Psi_{z_{l,kt}}}{\lvec{z}_{l,kt}}\Mumsgb{\Psi_{z_{l,kt}}}{\lvec{z}_{l,kt}}^H.
	\label{eq:app_C_mmd_Psi_z_z}
\end{split}
\end{align}
Note that in the pilot phase for $t\leq T_p$, the transmitted symbol $\pilot{x}_{kt}$ is already known.
Hence, $\mmd{\Psi_{z_{l,kt}}}{\lvec{z}_{l,kt}}(\lvec{z}_{l,kt})$ reduces to a Gaussian distribution for $t\leq T_p$ with mean vector $\Mumsgb{\Psi_{z_{l,kt}}}{\lvec{z}_{l,kt}}=\Mumsga{\Psi_{z_{l,kt}}}{\lvec{z}_{l,kt}}(\pilot{x}_{kt})$ and covariance matrix $\Cmsgb{\Psi_{z_{l,kt}}}{\lvec{z}_{l,kt}}=\Cmsga{\Psi_{z_{l,kt}}}{\lvec{z}_{l,kt}}(\pilot{x}_{kt})$.
The final message update can be computed according to~\eqref{eq:app_m_Psi_z_z0} using moment matching to find the projected distribution and the Gaussian quotient lemma,
\begin{equation}
	\msgp{\Psi_{z_{l,kt}}}{\lvec{z}_{l,kt}}(\lvec{z}_{l,kt}) = \CN{\lvec{z}_{l,kt}|\Mumsg{\Psi_{z_{l,kt}}}{\lvec{z}_{l,kt}},\Cmsg{\Psi_{z_{l,kt}}}{\lvec{z}_{l,kt}}},
	\label{eq:app_m_Psi_z_z}
\end{equation}
with
\begin{align}
	\Cmsg{\Psi_{z_{l,kt}}}{\lvec{z}_{l,kt}} &= \left(\Cmsgb{\Psi_{z_{l,kt}}}{\lvec{z}_{l,kt}}^{-1}-\Cmsg{\Psi_{y_{l,t}}}{\lvec{z}_{l,kt}}^{-1}\right)^{-1},
	\label{eq:app_C_Psi_z_z}\\
	\Mumsg{\Psi_{z_{l,kt}}}{\lvec{z}_{l,kt}} &= \Cmsg{\Psi_{z_{l,kt}}}{\lvec{z}_{l,kt}}\left(\Cmsgb{\Psi_{z_{l,kt}}}{\lvec{z}_{l,kt}}^{-1}\Mumsgb{\Psi_{z_{l,kt}}}{\lvec{z}_{l,kt}}-\Cmsg{\Psi_{y_{l,t}}}{\lvec{z}_{l,kt}}^{-1}\Mumsg{\Psi_{y_{l,t}}}{\lvec{z}_{l,kt}}\right).
	\label{eq:app_mu_Psi_z_z}
\end{align}

\noindent\underline{Message Update $\msg{\Psi_{z_{l,kt}}}{\lvec{g}_{l,k}}$}\\
\begin{equation}
	\msgp{\Psi_{z_{l,kt}}}{\lvec{g}_{l,k}}(\lvec{g}_{l,k}) \propto \frac{\proj{\mmd{\Psi_{z_{l,kt}}}{\lvec{g}_{l,k}}(\lvec{g}_{l,k})}}{\msgp{\lvec{g}_{l,k}}{\Psi_{z_{l,kt}}}(\lvec{g}_{l,k})},
	\label{eq:app_m_Psi_z_g0}
\end{equation}
with
\begin{align}
	&\mmd{\Psi_{z_{l,kt}}}{\lvec{g}_{l,k}}(\lvec{g}_{l,k})\nonumber\\
	&\quad\propto \sum_{x_{kt}}\int\delta(\lvec{z}_{l,kt}-\lvec{g}_{l,k}x_{kt})\cdot\msgp{x_{kt}}{\Psi_{z_{l,kt}}}(x_{kt})\cdot\msgp{\lvec{z}_{l,kt}}{\Psi_{z_{l,kt}}}(\lvec{z}_{l,kt})\cdot\msgp{\lvec{g}_{l,k}}{\Psi_{z_{l,kt}}}(\lvec{g}_{l,k})\,d\lvec{z}_{l,kt}\nonumber\\
	&\quad\overset{(a)}{=} \sum_{x_{kt}}\msgp{x_{kt}}{\Psi_{z_{l,kt}}}(x_{kt})\cdot\msgp{\Psi_{y_{l,t}}}{\lvec{z}_{l,kt}}(\lvec{g}_{l,k}x_{kt})\cdot\msgp{\lvec{g}_{l,k}}{\Psi_{z_{l,kt}}}(\lvec{g}_{l,k})\nonumber\\
	&\quad= \sum_{x_{kt}}\msgp{x_{kt}}{\Psi_{z_{l,kt}}}(x_{kt})\cdot\CN{\lvec{g}_{l,k}|\Mumsga{\Psi_{z_{l,kt}}}{\lvec{g}_{l,k}}(x_{kt}),\Cmsga{\Psi_{z_{l,kt}}}{\lvec{g}_{l,k}}(x_{kt})}\cdot\theta(x_{kt}),
	\label{eq:app_mmd_Psi_z_g}
\end{align}
where $(a)$ is obtained by the sifting property of the Dirac delta function and using~\eqref{eq:app_m_z_Psi_z}.
The final result in~\eqref{eq:app_mmd_Psi_z_g} is obtained by the Gaussian scaling and product lemma with $\theta$ given in~\eqref{eq:app_theta_mmd_Psi_z_z} and
\begin{align}
	\Cmsga{\Psi_{z_{l,kt}}}{\lvec{g}_{l,k}}(x_{kt}) &= \left(\Cmsg{\Psi_{y_{l,t}}}{\lvec{z}_{l,kt}}^{-1}|x_{kt}|^2+\Cmsg{\lvec{g}_{l,k}}{\Psi_{z_{l,kt}}}^{-1}\right)^{-1},
	\label{eq:app_C_tmp_mmd_Psi_z_g}\\
	\Mumsga{\Psi_{z_{l,kt}}}{\lvec{g}_{l,k}}(x_{kt}) &= \Cmsga{\Psi_{z_{l,kt}}}{\lvec{g}_{l,k}}(x_{kt})\Bigg(\Cmsg{\Psi_{y_{l,t}}}{\lvec{z}_{l,kt}}^{-1}\Mumsg{\Psi_{y_{l,t}}}{\lvec{z}_{l,kt}}\frac{|x_{kt}|^2}{x_{kt}}+\Cmsg{\lvec{g}_{l,k}}{\Psi_{z_{l,kt}}}^{-1}\Mumsg{\lvec{g}_{l,k}}{\Psi_{z_{l,kt}}}\Bigg).
	\label{eq:app_mu_tmp_mmd_Psi_z_g}
\end{align}
According to~\eqref{eq:app_mmd_Psi_z_g}, $\mmd{\Psi_{z_{l,kt}}}{\lvec{g}_{l,k}}(\lvec{g}_{l,k})$ is a Gaussian mixture with mean vector and covariance matrix
\begin{align}
	\Mumsgb{\Psi_{z_{l,kt}}}{\lvec{g}_{l,k}} &= \frac{1}{\tilde{Z}_{\Psi_{z_{l,kt}}}}\sum_{x_{kt}}\msgp{x_{kt}}{\Psi_{z_{l,kt}}}(x_{kt})\cdot\theta(x_{kt})\cdot\Mumsga{\Psi_{z_{l,kt}}}{\lvec{g}_{l,k}}(x_{kt}),
	\label{eq:app_mu_mmd_Psi_z_g}\\
\begin{split}
	\Cmsgb{\Psi_{z_{l,kt}}}{\lvec{g}_{l,k}} &= \frac{1}{\tilde{Z}_{\Psi_{z_{l,kt}}}}\sum_{x_{kt}}\msgp{x_{kt}}{\Psi_{z_{l,kt}}}(x_{kt})\cdot\theta(x_{kt})\cdot\Big(\Cmsga{\Psi_{z_{l,kt}}}{\lvec{g}_{l,k}}(x_{kt})\\
	&\quad+\Mumsga{\Psi_{z_{l,kt}}}{\lvec{g}_{l,k}}(x_{kt})\Mumsga{\Psi_{z_{l,kt}}}{\lvec{g}_{l,k}}^H(x_{kt})\Big) - \Mumsgb{\Psi_{z_{l,kt}}}{\lvec{g}_{l,k}}\Mumsgb{\Psi_{z_{l,kt}}}{\lvec{g}_{l,k}}^H.
	\label{eq:app_C_mmd_Psi_z_g}
\end{split}
\end{align}
Note that in the pilot phase for $t\leq T_p$, the transmitted symbol $\pilot{x}_{kt}$ is already known.
Hence, $\mmd{\Psi_{z_{l,kt}}}{\lvec{g}_{l,k}}(\lvec{g}_{l,k})$ reduces to a Gaussian distribution for $t\leq T_p$ with mean vector $\Mumsgb{\Psi_{z_{l,kt}}}{\lvec{g}_{l,k}}=\Mumsga{\Psi_{z_{l,kt}}}{\lvec{g}_{l,k}}(\pilot{x}_{kt})$ and covariance matrix $\Cmsgb{\Psi_{z_{l,kt}}}{\lvec{g}_{l,k}}=\Cmsga{\Psi_{z_{l,kt}}}{\lvec{g}_{l,k}}(\pilot{x}_{kt})$.
The final message update can be computed according to~\eqref{eq:app_m_Psi_z_g0} using moment matching and the Gaussian quotient lemma,
\begin{equation}
	\msgp{\Psi_{z_{l,kt}}}{\lvec{g}_{l,k}}(\lvec{g}_{l,k}) = \CN{\lvec{g}_{l,k}|\Mumsg{\Psi_{z_{l,kt}}}{\lvec{g}_{l,k}},\Cmsg{\Psi_{z_{l,kt}}}{\lvec{g}_{l,k}}},
	\label{eq:app_m_Psi_z_g}
\end{equation}
with
\begin{align}
	\Cmsg{\Psi_{z_{l,kt}}}{\lvec{g}_{l,k}} &= \left(\Cmsgb{\Psi_{z_{l,kt}}}{\lvec{g}_{l,k}}^{-1}-\Cmsg{\lvec{g}_{l,k}}{\Psi_{z_{l,kt}}}^{-1}\right)^{-1},
	\label{eq:app_C_Psi_z_g}\\
	\Mumsg{\Psi_{z_{l,kt}}}{\lvec{g}_{l,k}} &= \Cmsg{\Psi_{z_{l,kt}}}{\lvec{g}_{l,k}}\left(\Cmsgb{\Psi_{z_{l,kt}}}{\lvec{g}_{l,k}}^{-1}\Mumsgb{\Psi_{z_{l,kt}}}{\lvec{g}_{l,k}}-\Cmsg{\lvec{g}_{l,k}}{\Psi_{z_{l,kt}}}^{-1}\Mumsg{\lvec{g}_{l,k}}{\Psi_{z_{l,kt}}}\right).
	\label{eq:app_mu_Psi_z_g}
\end{align}

\subsection{Message Updates for $\Psi_{g_{l,k}}$}\label{subsubsec:MP_Psi_g}

\subsubsection{Incoming messages to factor node $\Psi_{g_{l,k}}$}
\begin{align}
	\msgp{u_k}{\Psi_{g_{l,k}}}(u_k) &\propto \msgp{\Psi_{u_k}}{u_k}(u_k)\cdot\prod_{l'\neq l}\msgp{\Psi_{g_{l',k}}}{u_{k}}(u_k),
	\label{eq:app_m_u_Psi_g}\\
	\msgp{\lvec{h}_{l,k}}{\Psi_{g_{l,k}}}(\lvec{h}_{l,k}) &= \msgp{\Psi_{h_{l,k}}}{\lvec{h}_{l,k}}(\lvec{h}_{l,k}),
	\label{eq:app_m_h_Psi_g}\\
	 \msgp{\lvec{g}_{l,k}}{\Psi_{g_{l,k}}}(\lvec{g}_{l,k})&\propto \prod_{t=1}^T\msgp{\Psi_{z_{l,kt}}}{\lvec{g}_{l,k}}(\lvec{g}_{l,k})\nonumber\\
	&\propto \CN{\lvec{g}_{l,k}|\Mumsg{\lvec{g}_{l,k}}{\Psi_{g_{l,k}}},\Cmsg{\lvec{g}_{l,k}}{\Psi_{g_{l,k}}}},
	\label{eq:app_m_g_Psi_g}
\end{align}
with
\begin{align}
	\Cmsg{\lvec{g}_{l,k}}{\Psi_{g_{l,k}}} &= \left(\sum_{t=1}^T\Cmsg{\Psi_{z_{l,kt}}}{\lvec{g}_{l,k}}^{-1}\right)^{-1},
	\label{eq:app_C_g_Psi_g}\\
	\Mumsg{\lvec{g}_{l,k}}{\Psi_{g_{l,k}}} &= \Cmsg{\lvec{g}_{l,k}}{\Psi_{g_{l,k}}}\left(\sum_{t=1}^T\Cmsg{\Psi_{z_{l,kt}}}{\lvec{g}_{l,k}}^{-1}\Mumsg{\Psi_{z_{l,kt}}}{\lvec{g}_{l,k}}\right),
	\label{eq:app_mu_g_Psi_g}
\end{align}
which is obtained by applying the Gaussian product lemma multiple times.

\subsubsection{Outgoing messages from factor node $\Psi_{g_{l,k}}$}
\hfill\\\underline{Message Update $\msg{\Psi_{g_{l,k}}}{u_k}$}\\
\begin{equation}
	\msgp{\Psi_{g_{l,k}}}{u_k}(u_k) \propto \frac{\proj{\mmd{\Psi_{g_{l,k}}}{u_k}(u_k)}}{\msgp{u_k}{\Psi_{g_{l,k}}}(u_k)},
	\label{eq:app_m_Psi_g_u0}
\end{equation}
with
\begin{align}
	\mmd{\Psi_{g_{l,k}}}{u_k}(u_k)&\propto \int\int\delta(\lvec{g}_{l,k}-\lvec{h}_{l,k}u_k)\cdot\msgp{u_k}{\Psi_{g_{l,k}}}(u_k)\cdot\msgp{\lvec{g}_{l,k}}{\Psi_{g_{l,k}}}(\lvec{g}_{l,k})\cdot\msgp{\lvec{h}_{l,k}}{\Psi_{g_{l,k}}}(\lvec{h}_{l,k})\,d\lvec{g}_{l,k}\,d\lvec{h}_{l,k}\nonumber\\
	&\overset{(a)}{=} \msgp{u_k}{\Psi_{g_{l,k}}}(u_k)\cdot\int\msgp{\lvec{g}_{l,k}}{\Psi_{g_{l,k}}}(\lvec{h}_{l,k}u_k)\cdot\msgp{\Psi_{h_{l,k}}}{\lvec{h}_{l,k}}(\lvec{h}_{l,k})\,d\lvec{h}_{l,k}\nonumber\\
	&= \msgp{u_k}{\Psi_{g_{l,k}}}(u_k)\cdot\vartheta(u_k),
	\label{eq:app_mmd_Psi_g_u}
\end{align}
where $(a)$ is obtained by the sifting property of the Dirac delta function and using~\eqref{eq:app_m_h_Psi_g}.
The final result in~\eqref{eq:app_mmd_Psi_g_u} is obtained by considering the fact that $u_k$ is a binary random variable with $u_k\in\{0,1\}$, utilizing~\eqref{eq:app_m_Psi_h_h} and the Gaussian product rule, and, then, integrating over $\lvec{h}_{l,k}$ with
\begin{equation}
	\vartheta(u_k) = \CN{\lvec{0}|\Mumsg{\lvec{g}_{l,k}}{\Psi_{g_{l,k}}}-\tilde{\gvec{\mu}}_{h_{l,k}}u_k,\Cmsg{\lvec{g}_{l,k}}{\Psi_{g_{l,k}}}+\tilde{\lmat{C}}_{h_{l,k}}u_k}.
	\label{eq:app_theta_mmd_Psi_g_u}
\end{equation}
The projection operation in~\eqref{eq:app_m_Psi_g_u0} is superfluous since $\mmd{\Psi_{g_{l,k}}}{u_k}(u_k)$~\eqref{eq:app_mmd_Psi_g_u} is already categorically distributed.
Hence, the denominator of~\eqref{eq:app_m_Psi_g_u0} cancels with the first term in~\eqref{eq:app_mmd_Psi_g_u}.
The final message update rule is given by {the distribution}
\begin{equation}
	\msgp{\Psi_{g_{l,k}}}{u_k}(u_k) \propto \vartheta(u_k).
	\label{eq:app_m_Psi_g_u}
\end{equation}

\noindent\underline{Message Update $\msg{\Psi_{g_{l,k}}}{\lvec{h}_{l,k}}$}\\
\begin{equation}
	\msgp{\Psi_{g_{l,k}}}{\lvec{h}_{l,k}}(\lvec{h}_{l,k}) \propto \frac{\proj{\mmd{\Psi_{g_{l,k}}}{\lvec{h}_{l,k}}(\lvec{h}_{l,k})}}{\msgp{\lvec{h}_{l,k}}{\Psi_{g_{l,k}}}(\lvec{h}_{l,kt})},
	\label{eq:app_m_Psi_g_h0}
\end{equation}
with
\begin{align}
	\mmd{\Psi_{g_{l,k}}}{\lvec{h}_{l,k}}(\lvec{h}_{l,k})&\propto \sum_{u_k}\int\delta(\lvec{g}_{l,k}-\lvec{h}_{l,k}u_k)\cdot\msgp{u_k}{\Psi_{g_{l,k}}}(u_k)\cdot\msgp{\lvec{g}_{l,k}}{\Psi_{g_{l,k}}}(\lvec{g}_{l,k})\cdot\msgp{\lvec{h}_{l,k}}{\Psi_{g_{l,k}}}(\lvec{h}_{l,k})\,d\lvec{g}_{l,k}\nonumber\\
	&\overset{(a)}{=} \msgp{u_k}{\Psi_{g_{l,k}}}(0)\cdot\msgp{\lvec{g}_{l,k}}{\Psi_{g_{l,k}}}(\lvec{0})\cdot\msgp{\Psi_{h_{l,k}}}{\lvec{h}_{l,k}}(\lvec{h}_{l,k})+\msgp{u_k}{\Psi_{g_{l,k}}}(1)\cdot\msgp{\lvec{g}_{l,k}}{\Psi_{g_{l,k}}}(\lvec{h}_{l,k})\nonumber\\
	&\quad\cdot\msgp{\Psi_{h_{l,k}}}{\lvec{h}_{l,k}}(\lvec{h}_{l,k})\nonumber\\
\begin{split}
	&= \msgp{u_k}{\Psi_{g_{l,k}}}(0)\cdot\vartheta(0)\cdot\msgp{\Psi_{h_{l,k}}}{\lvec{h}_{l,k}}(\lvec{h}_{l,k})+\msgp{u_k}{\Psi_{g_{l,k}}}(1)\cdot\vartheta(1)\\
	&\quad\cdot\CN{\lvec{h}_{l,k}|\Mumsga{\Psi_{g_{l,k}}}{\lvec{h}_{l,k}},\Cmsga{\Psi_{g_{l,k}}}{\lvec{h}_{l,k}}},
\end{split}
	\label{eq:app_mmd_Psi_g_h}
\end{align}
where $(a)$ is obtained by making the sum explicit for $u_k$ and  utilizing the sifting property of the Dirac delta function and~\eqref{eq:app_m_h_Psi_g}.
The final result in~\eqref{eq:app_mmd_Psi_g_h} is obtained by using~\eqref{eq:app_m_Psi_h_h}, \eqref{eq:app_theta_mmd_Psi_g_u}, and the Gaussian multiplication lemma with
\begin{align}
	\Cmsga{\Psi_{g_{l,k}}}{\lvec{h}_{l,k}} &= \left(\Cmsg{\lvec{g}_{l,k}}{\Psi_{g_{l,k}}}^{-1}+\tilde{\lmat{C}}_{h_{l,k}}^{-1}\right)^{-1},
	\label{eq:app_C_tmp_mmd_Psi_g_h}\\
	\Mumsga{\Psi_{g_{l,k}}}{\lvec{h}_{l,k}} &= \Cmsga{\Psi_{g_{l,k}}}{\lvec{h}_{l,k}}\Bigg(\Cmsg{\lvec{g}_{l,k}}{\Psi_{g_{l,k}}}^{-1}\Mumsg{\lvec{g}_{l,k}}{\Psi_{g_{l,k}}}+\tilde{\lmat{C}}_{h_{l,k}}^{-1}\tilde{\gvec{\mu}}_{h_{l,k}}\Bigg).
	\label{eq:app_mu_tmp_mmd_Psi_g_h}
\end{align}
The normalization constant for $\mmd{\Psi_{g_{l,k}}}{\lvec{h}_{l,k}}(\lvec{h}_{l,k})$ is given by
\begin{align}
	\tilde{Z}_{\Psi_{g_{l,k}}} &= \int\msgp{u_k}{\Psi_{g_{l,k}}}(0)\cdot\vartheta(0)\cdot\msgp{\Psi_{h_{l,k}}}{\lvec{h}_{l,k}}(\lvec{h}_{l,k})\nonumber\\
	&\qquad+ \msgp{u_k}{\Psi_{g_{l,k}}}(1)\cdot\vartheta(1)\cdot\CN{\lvec{h}_{l,k}|\Mumsga{\Psi_{g_{l,k}}}{\lvec{h}_{l,k}},\Cmsga{\Psi_{g_{l,k}}}{\lvec{h}_{l,k}}}\,d\lvec{h}_{l,k}\nonumber\\
	&= \msgp{u_k}{\Psi_{g_{l,k}}}(0)\cdot\vartheta(0)+\msgp{u_k}{\Psi_{g_{l,k}}}(1)\cdot\vartheta(1).
	\label{eq:app_Z_Psi_g}
\end{align}
According to~\eqref{eq:app_mmd_Psi_g_h}, $\mmd{\Psi_{g_{l,k}}}{\lvec{h}_{l,k}}(\lvec{h}_{l,k})$ is a Gaussian mixture in two components with mean vector and covariance matrix
\begin{align}
	\Mumsgb{\Psi_{g_{l,k}}}{\lvec{h}_{l,k}} &= \frac{1}{\tilde{Z}_{\Psi_{g_{l,k}}}}\Big(\msgp{u_k}{\Psi_{g_{l,k}}}(0)\cdot\vartheta(0)\cdot\tilde{\gvec{\mu}}_{h_{l,k}}+\msgp{u_k}{\Psi_{g_{l,k}}}(1)\cdot\vartheta(1)\cdot\Mumsga{\Psi_{g_{l,k}}}{\lvec{h}_{l,k}}\Big),
	\label{eq:app_mu_mmd_Psi_g_h}\\
\begin{split}
	\Cmsgb{\Psi_{g_{l,k}}}{\lvec{h}_{l,k}} &= \frac{1}{\tilde{Z}_{\Psi_{z_{l,kt}}}}\Big(\msgp{u_k}{\Psi_{g_{l,k}}}(0)\cdot\vartheta(0)\cdot\big(\tilde{\lmat{C}}_{h_{l,k}}+\tilde{\gvec{\mu}}_{h_{l,k}}\tilde{\gvec{\mu}}_{h_{l,k}}^H\big)+\msgp{u_k}{\Psi_{g_{l,k}}}(1)\cdot\vartheta(1)\\
	&\qquad\qquad\quad\cdot\big(\Cmsga{\Psi_{g_{l,k}}}{\lvec{h}_{l,k}}+\Mumsga{\Psi_{g_{l,k}}}{\lvec{h}_{l,k}}\Mumsga{\Psi_{g_{l,k}}}{\lvec{h}_{l,k}}^H\big)\Big)-\Mumsgb{\Psi_{g_{l,k}}}{\lvec{g}_{l,k}}\Mumsgb{\Psi_{g_{l,k}}}{\lvec{g}_{l,k}}^H.
	\label{eq:app_C_mmd_Psi_g_h}
\end{split}
\end{align}
Note that the estimated posterior distribution~\eqref{eq:approx_factor_beta_factorization} of $\lvec{h}_{l,k}$ is proportional to the product of $\msgp{\Psi_{g_{l,k}}}{\lvec{h}_{l,k}}(\lvec{h}_{l,k})$ and $\msgp{\Psi_{h_{l,k}}}{\lvec{h}_{l,k}}(\lvec{h}_{l,kt})$ which is proportional to $\proj{\mmd{\Psi_{g_{l,k}}}{\lvec{h}_{l,k}}(\lvec{h}_{l,k})}$ according to~\eqref{eq:app_m_h_Psi_g} and~\eqref{eq:app_m_Psi_g_h0}.
Hence, the final estimate of $\lvec{h}_{l,k}$ can be computed via~\eqref{eq:app_mu_mmd_Psi_g_h} after the last iteration of the \ac{EP} algorithm.

By applying moment matching and the Gaussian quotient lemma to~\eqref{eq:app_m_Psi_g_h0}, we obtain for the final message update the following distribution,
\begin{equation}
	\msgp{\Psi_{g_{l,k}}}{\lvec{h}_{l,k}}(\lvec{h}_{l,k}) = \CN{\lvec{h}_{l,k}|\Mumsg{\Psi_{g_{l,k}}}{\lvec{h}_{l,k}},\Cmsg{\Psi_{g_{l,k}}}{\lvec{h}_{l,k}}},
	\label{eq:app_m_Psi_g_h}
\end{equation}
with
\begin{align}
	\Cmsg{\Psi_{g_{l,k}}}{\lvec{h}_{l,k}} &= \left(\Cmsgb{\Psi_{g_{l,k}}}{\lvec{h}_{l,k}}^{-1}-\tilde{\lmat{C}}_{h_{l,k}}^{-1}\right)^{-1},
	\label{eq:app_C_Psi_g_h}\\
	\Mumsg{\Psi_{g_{l,k}}}{\lvec{h}_{l,k}} &= \Cmsg{\Psi_{g_{l,k}}}{\lvec{h}_{l,k}}\left(\Cmsgb{\Psi_{g_{l,k}}}{\lvec{h}_{l,k}}^{-1}\Mumsgb{\Psi_{g_{l,k}}}{\lvec{h}_{l,k}}-\tilde{\lmat{C}}_{h_{l,k}}^{-1}\tilde{\gvec{\mu}}_{h_{l,k}}\right).
	\label{eq:app_mu_Psi_g_h}
\end{align}

\noindent\underline{Message Update $\msg{\Psi_{g_{l,k}}}{\lvec{g}_{l,k}}$}\\
\begin{equation}
	\msgp{\Psi_{g_{l,k}}}{\lvec{g}_{l,k}}(\lvec{g}_{l,k}) \propto \frac{\proj{\mmd{\Psi_{g_{l,k}}}{\lvec{g}_{l,k}}(\lvec{g}_{l,k})}}{\msgp{\lvec{g}_{l,k}}{\Psi_{g_{l,k}}}(\lvec{g}_{l,kt})},
	\label{eq:app_m_Psi_g_g0}
\end{equation}
with
\begin{align}
	\mmd{\Psi_{g_{l,k}}}{\lvec{g}_{l,k}}(\lvec{g}_{l,k})&\propto \sum_{u_k}\int\delta(\lvec{g}_{l,k}-\lvec{h}_{l,k}u_k)\cdot\msgp{u_k}{\Psi_{g_{l,k}}}(u_k)\cdot\msgp{\lvec{g}_{l,k}}{\Psi_{g_{l,k}}}(\lvec{g}_{l,k})\cdot\msgp{\lvec{h}_{l,k}}{\Psi_{g_{l,k}}}(\lvec{h}_{l,k})\,d\lvec{h}_{l,k}\nonumber\\
	&\overset{(a)}{=} \msgp{u_k}{\Psi_{g_{l,k}}}(0)\cdot\msgp{\lvec{g}_{l,k}}{\Psi_{g_{l,k}}}(\lvec{g}_{l,k})\cdot\delta(\lvec{g}_{l,k})+\msgp{u_k}{\Psi_{g_{l,k}}}(1)\cdot\msgp{\lvec{g}_{l,k}}{\Psi_{g_{l,k}}}(\lvec{g}_{l,k})\nonumber\\
	&\quad\cdot\msgp{\Psi_{h_{l,k}}}{\lvec{h}_{l,k}}(\lvec{g}_{l,k})\nonumber\\
\begin{split}
	&= \msgp{u_k}{\Psi_{g_{l,k}}}(0)\cdot\msgp{\lvec{g}_{l,k}}{\Psi_{g_{l,k}}}(\lvec{g}_{l,k})\cdot\delta(\lvec{g}_{l,k})+ \msgp{u_k}{\Psi_{g_{l,k}}}(1)\cdot\vartheta(1)\\
	&\quad\cdot\CN{\lvec{g}_{l,k}|\Mumsga{\Psi_{g_{l,k}}}{\lvec{g}_{l,k}},\Cmsga{\Psi_{g_{l,k}}}{\lvec{g}_{l,k}}},
	\label{eq:app_mmd_Psi_g_g}
\end{split}
\end{align}
where $(a)$ is obtained by making the sum explicit for $u_k$ and  utilizing the sifting property of the Dirac delta function and~\eqref{eq:app_m_h_Psi_g}.
The final result in~\eqref{eq:app_mmd_Psi_g_g} is obtained by using~\eqref{eq:app_m_Psi_h_h}, \eqref{eq:app_theta_mmd_Psi_g_u}, and the Gaussian multiplication lemma with
\begin{align}
	\Cmsga{\Psi_{g_{l,k}}}{\lvec{g}_{l,k}} &= \Cmsga{\Psi_{g_{l,k}}}{\lvec{h}_{l,k}},
	\label{eq:app_C_tmp_mmd_Psi_g_g}\\
	\Mumsga{\Psi_{g_{l,k}}}{\lvec{g}_{l,k}} &= \Mumsga{\Psi_{g_{l,k}}}{\lvec{h}_{l,k}}.
	\label{eq:app_mu_tmp_mmd_Psi_g_g}
\end{align}
The mean vector and covariance matrix of the Bernoulli-Gaussian distribution $\mmd{\Psi_{g_{l,k}}}{\lvec{g}_{l,k}}(\lvec{g}_{l,k})$ in~\eqref{eq:app_mmd_Psi_g_g} is given by
\begin{align}
	\Mumsgb{\Psi_{g_{l,k}}}{\lvec{g}_{l,k}} &= \frac{1}{\tilde{Z}_{\Psi_{g_{l,k}}}}\cdot\msgp{u_k}{\Psi_{g_{l,k}}}(1)\cdot\vartheta(1)\cdot\Mumsga{\Psi_{g_{l,k}}}{\lvec{g}_{l,k}},
	\label{eq:app_mu_mmd_Psi_g_g}\\
\begin{split}
	\Cmsgb{\Psi_{g_{l,k}}}{\lvec{g}_{l,k}} &= \frac{1}{\tilde{Z}_{\Psi_{g_{l,k}}}}\cdot\msgp{u_k}{\Psi_{g_{l,k}}}(1)\cdot\vartheta(1)\cdot\Big(\Cmsga{\Psi_{g_{l,k}}}{\lvec{g}_{l,k}}+\Mumsga{\Psi_{g_{l,k}}}{\lvec{g}_{l,k}}\Mumsga{\Psi_{g_{l,k}}}{\lvec{g}_{l,k}}^H\Big)\\
	&\quad- \Mumsgb{\Psi_{g_{l,k}}}{\lvec{g}_{l,k}}\Mumsgb{\Psi_{g_{l,k}}}{\lvec{g}_{l,k}}^H.
	\label{eq:app_C_mmd_Psi_g_g}
\end{split}
\end{align}
By applying moment matching and the Gaussian quotient lemma to~\eqref{eq:app_m_Psi_g_g0}, we obtain for the final message update the following distribution,
\begin{equation}
	\msgp{\Psi_{g_{l,k}}}{\lvec{g}_{l,k}}(\lvec{g}_{l,k}) = \CN{\lvec{g}_{l,k}|\Mumsg{\Psi_{g_{l,k}}}{\lvec{g}_{l,k}},\Cmsg{\Psi_{g_{l,k}}}{\lvec{g}_{l,k}}},
	\label{eq:app_m_Psi_g_g}
\end{equation}
with
\begin{align}
	\Cmsg{\Psi_{g_{l,k}}}{\lvec{g}_{l,k}} &= \left(\Cmsgb{\Psi_{g_{l,k}}}{\lvec{g}_{l,k}}^{-1}-\Cmsg{\lvec{g}_{l,k}}{\Psi_{g_{l,k}}}^{-1}\right)^{-1},
	\label{eq:app_C_Psi_g_g}\\
	\Mumsg{\Psi_{g_{l,k}}}{\lvec{g}_{l,k}} &= \Cmsg{\Psi_{g_{l,k}}}{\lvec{g}_{l,k}}\left(\Cmsgb{\Psi_{g_{l,k}}}{\lvec{g}_{l,k}}^{-1}\Mumsgb{\Psi_{g_{l,k}}}{\lvec{g}_{l,k}}-\Cmsg{\lvec{g}_{l,k}}{\Psi_{g_{l,k}}}^{-1}\Mumsg{\lvec{g}_{l,k}}{\Psi_{g_{l,k}}}\right).
	\label{eq:app_mu_Psi_g_g}
\end{align}

\section{Derivation of Message-Passing Update Rules for \acs{JAC-EP}}\label{app:JAC_MP_updates}
In the following, we apply the \ac{EP} message-passing rules presented in Appendix~\ref{app:EP_graph} to the factor graph in Fig.~\ref{fig:app_JAC_FG} and show the detailed derivations for the message updates.
Although some of the message updates are the same or very similar to the ones derived in Appendix~\ref{app:MP_updates}, we derive all the messages for the \ac{JAC-EP} algorithm again for the sake of clarity.

\subsection{Message Updates for Leaf Nodes $\Psi_{u_k}$ and $\Psi_{h_{l,k}}$}\label{app:JAC_MP_leaf}
Since the prior distributions $p_{u_k}(u_k)$ and $p_{h_{l,k}}(\lvec{h}_{l,k})$ are in the same exponential family as the approximate distributions of $u_k$ and $\lvec{h}_{l,k}$, respectively, the corresponding message updates simplify significantly.
The factor-to-variable messages are constant and consist of the prior information on the variables $u_k$ and $\lvec{h}_{l,k}$  {which correspond to the distributions}
\begin{align}
	\msgp{\Psi_{u_k}}{u_k}(u_k) &= p_{u_k}(u_k),
	\label{eq:app_JAC_m_Psi_u_u}\\
	\msgp{\Psi_{h_{l,k}}}{\lvec{h}_{l,k}}(\lvec{h}_{l,k}) &= p_{h_{l,k}}(\lvec{h}_{l,k}).
	\label{eq:app_JAC_m_Psi_h_h}
\end{align}
This result is obtained by applying the message-passing rule in~\eqref{eq:EP_fac_to_var_message} while taking into account the fact that exponential family distributions are closed under multiplication.
This makes the projection operation in~\eqref{eq:EP_fac_to_var_message} superfluous.
Hence, {the distributions corresponding to the updated messages} are directly given by the factors $\Psi_{u_k}$ and $\Psi_{h_{l,k}}$, respectively, which correspond to the priors according to~\eqref{eq:app_JAC_p_u_k},~\eqref{eq:app_JAC_p_h_lk}.
The variable-to-factor messages for the leaf nodes are irrelevant in the unfolding of the algorithm and omitted here.

\subsection{Message Updates for $\Psi_{y_{l,t}}$}\label{app:JAC_MP_Psi_y}

\subsubsection{Incoming messages to factor node $\Psi_{y_{l,t}}$}
\begin{align}
	\msgp{\lvec{g}_{l,k}}{\Psi_{y_{l,t}}}(\lvec{g}_{l,k}) &\propto \msgp{\Psi_{g_{l,k}}}{\lvec{g}_{l,k}}(\lvec{g}_{l,k})\cdot\prod_{t'\neq t}\msgp{\Psi_{y_{l,t'}}}{\lvec{g}_{l,k}}(\lvec{g}_{l,k})\nonumber\\
	&\propto \CN{\lvec{g}_{l,k}|\Mumsg{\lvec{g}_{l,k}}{\Psi_{y_{l,t}}},\Cmsg{\lvec{g}_{l,k}}{\Psi_{y_{l,t}}}},
	\label{eq:app_JAC_m_g_Psi_y}
\end{align}
with
\begin{align}
	\Cmsg{\lvec{g}_{l,k}}{\Psi_{y_{l,t}}} &= \left(\Cmsg{\Psi_{g_{l,k}}}{\lvec{g}_{l,k}}^{-1}+\sum_{t'\neq t}\Cmsg{\Psi_{y_{l,t'}}}{\lvec{g}_{l,k}}^{-1}\right)^{-1},
	\label{eq:app_JAC_C_g_Psi_y}\\
	\Mumsg{\lvec{g}_{l,k}}{\Psi_{y_{l,t}}} &= \Cmsg{\lvec{g}_{l,k}}{\Psi_{y_{l,t}}}\Bigg(\Cmsg{\Psi_{g_{l,k}}}{\lvec{g}_{l,k}}^{-1}\Mumsg{\Psi_{g_{l,k}}}{\lvec{g}_{l,k}}+\sum_{t'\neq t}\Cmsg{\Psi_{y_{l,t'}}}{\lvec{g}_{l,k}}^{-1}\Mumsg{\Psi_{y_{l,t'}}}{\lvec{g}_{l,k}}\Bigg),
	\label{eq:app_JAC_mu_g_Psi_y}
\end{align}
which is obtained by applying the Gaussian product lemma multiple times.

\subsubsection{Outgoing messages from factor node $\Psi_{y_{l,t}}$}
\hfill\\\underline{Message Update $\msg{\Psi_{y_{l,t}}}{\lvec{z}_{l,kt}}$}\\
\begin{equation}
	\msgp{\Psi_{y_{l,t}}}{\lvec{g}_{l,k}}(\lvec{g}_{l,k}) \propto \frac{\proj{\mmd{\Psi_{y_{l,t}}}{\lvec{g}_{l,k}}(\lvec{g}_{l,k})}}{\msgp{\lvec{g}_{l,k}}{\Psi_{y_{l,t}}}(\lvec{g}_{l,k})},
	\label{eq:app_JAC_m_Psi_y_g0}
\end{equation}
with
\begin{align}
	&\mmd{\Psi_{y_{l,t}}}{\lvec{g}_{l,k}}(\lvec{g}_{l,k})\nonumber\\
	&\quad\propto \int\!\!\cdots\!\!\int \CN{\lvec{y}_{l,t}\Bigg|\sum_{k'=1}^K\lvec{g}_{l,k'}x_{k't},\sigma_n^2\lmat{I}_N}\cdot\msgp{\lvec{g}_{l,k}}{\Psi_{y_{l,t}}}(\lvec{g}_{l,k})\cdot\prod_{k'\neq k}\msgp{\lvec{g}_{l,k'}}{\Psi_{y_{l,t}}}(\lvec{g}_{l,k'})\,d\lvec{g}_{l,k'}\nonumber\\
	&\quad\overset{(a)}{=} \msgp{\lvec{g}_{l,k}}{\Psi_{y_{l,t}}}(\lvec{g}_{l,k})\cdot\int\!\!\cdots\!\!\int \CN{\lvec{g}_{l,k''}x_{k''t}\Bigg|\lvec{y}_{l,t}-\sum_{k'\neq k''}\lvec{g}_{l,k'}x_{k't},\sigma_n^2\lmat{I}_N}\nonumber\\
	&\qquad\cdot\prod_{k'\neq k}\CN{\lvec{g}_{l,k'}\big|\Mumsg{\lvec{g}_{l,k'}}{\Psi_{y_{l,t}}},\Cmsg{\lvec{g}_{l,k'}}{\Psi_{y_{l,t}}}}\,d\lvec{g}_{l,k'}\nonumber\\
	&\quad\overset{(b)}{=} \msgp{\lvec{g}_{l,k}}{\Psi_{y_{l,t}}}(\lvec{g}_{l,k})\cdot\int\!\!\cdots\!\!\int \CN{\lvec{g}_{l,k''}\Bigg|\left(\lvec{y}_{l,t}-\sum_{k'\neq k''}\lvec{g}_{l,k'}x_{k't}\right)\cdot x_{k''t}^{-1},\sigma_n^2\lmat{I}_N\cdot|x_{k''t}|^{-2}}\cdot|x_{k''t}|^{-2N}\nonumber\\
	&\qquad\cdot\prod_{k'\neq k}\CN{\lvec{g}_{l,k'}\big|\Mumsg{\lvec{g}_{l,k'}}{\Psi_{y_{l,t}}},\Cmsg{\lvec{g}_{l,k'}}{\Psi_{y_{l,t}}}}\,d\lvec{g}_{l,k'}\nonumber\\	
	&\quad\overset{(c)}{=} \msgp{\lvec{g}_{l,k}}{\Psi_{y_{l,t}}}(\lvec{g}_{l,k})\cdot\int\!\!\cdots\!\!\int \CN{\lvec{g}_{l,k''}|\gvec{\mu}_\text{tmp},\lmat{C}_\text{tmp}}\,d\lvec{g}_{l,k''}\nonumber\\
 &\qquad\cdot\CN{\lvec{0}\Bigg|\left(\lvec{y}_{l,t}-\sum_{k'\neq k''}\lvec{g}_{l,k'}x_{k't}\right)\cdot x_{k''t}^{-1}-\Mumsg{\lvec{g}_{l,k''}}{\Psi_{y_{l,t}}},\sigma_n^2\lmat{I}_N\cdot|x_{k''t}|^{-2}+\Cmsg{\lvec{g}_{l,k''}}{\Psi_{y_{l,t}}}}\nonumber\\
 &\qquad\cdot|x_{k''t}|^{-2N}\cdot\prod_{k'\neq\{k,k''\}}\CN{\lvec{g}_{l,k'}\big|\Mumsg{\lvec{g}_{l,k'}}{\Psi_{y_{l,k'}}},\Cmsg{\lvec{g}_{l,k'}}{\Psi_{y_{l,k'}}}}\,d\lvec{g}_{l,k'}\nonumber\\
	&\quad\overset{(d)}{=} \msgp{\lvec{g}_{l,k}}{\Psi_{y_{l,t}}}(\lvec{g}_{l,k})\cdot\int\!\!\cdots\!\!\int \CN{\lvec{y}_{l,t}\Bigg|\sum_{k'\neq k''}\lvec{g}_{l,k'}x_{k't}+\Mumsg{\lvec{g}_{l,k''}}{\Psi_{y_{l,t}}}x_{k''t},\sigma_n^2\lmat{I}_N+\Cmsg{\lvec{g}_{l,k''}}{\Psi_{y_{l,t}}}|x_{k''t}|^2}\nonumber\\
	&\qquad\cdot\prod_{k'\neq\{k,k''\}}\CN{\lvec{g}_{l,k'}\big|\Mumsg{\lvec{g}_{l,k'}}{\Psi_{y_{l,k'}}},\Cmsg{\lvec{g}_{l,k'}}{\Psi_{y_{l,k'}}}}\,d\lvec{g}_{l,k'}\nonumber\\
	&\quad= \dots\nonumber\\
	&\quad= \msgp{\lvec{g}_{l,k}}{\Psi_{y_{l,t}}}(\lvec{g}_{l,k})\cdot\CN{\lvec{g}_{l,k}x_{kt}\Bigg|\lvec{y}_{l,t}-\sum_{k'\neq k}\Mumsg{\lvec{g}_{l,k'}}{\Psi_{y_{l,t}}}x_{k't},\sigma_n^2\lmat{I}_N+\sum_{k'\neq k}\Cmsg{\lvec{g}_{l,k'}}{\Psi_{y_{l,t}}}|x_{k't}|^2}.
	\label{eq:app_JAC_mmd_Psi_y_g}
\end{align}
where $(a)$ is obtained by a basic transformation of the Gaussian distribution, $(b)$ is obtained by the Gaussian scaling lemma, $(c)$ is obtained by the Gaussian product rule~\eqref{eq:app_Gaussian_product} with $\gvec{\mu}_1=\left(\lvec{y}_{l,t}-\sum_{k'\neq k''}\lvec{g}_{l,k'}x_{k't}\right)\cdot x_{k''t}^{-1}$, $\gvec{\mu}_2=\Mumsg{\lvec{g}_{l,k''}}{\Psi_{y_{l,t}}}$, $\lmat{C}_1=\sigma_n^2\lmat{I}_N\cdot|x_{k''t}|^{-2}$, and $\lmat{C}_2=\Cmsg{\lvec{g}_{l,k''}}{\Psi_{y_{l,t}}}$ which yields $\lmat{C}_\text{tmp}$ and $\gvec{\mu}_\text{tmp}$ according to~\eqref{eq:app_Gaussian_product_C} and~\eqref{eq:app_Gaussian_product_mu}, respectively, and $(d)$ is obtained by integrating over $\lvec{g}_{l,k''}$ and applying the Gaussian scaling lemma as well as a basic transformation of the Gaussian distribution.
The final result in~\eqref{eq:app_JAC_mmd_Psi_y_g} is obtained by repeatedly applying the above described steps for all $k'\neq k$.
Since $\msgp{\lvec{g}_{l,k}}{\Psi_{y_{l,t}}}(\lvec{g}_{l,k})$ is Gaussian distributed, we can conclude by utilizing the Gaussian product lemma that $\mmd{\Psi_{y_{l,t}}}{\lvec{g}_{l,k}}(\lvec{g}_{l,k})$ is Gaussian distributed as well.
Hence, the projection operation in~\eqref{eq:app_JAC_m_Psi_y_g0} is superfluous since it projects $\mmd{\Psi_{y_{l,t}}}{\lvec{g}_{l,k}}(\lvec{g}_{l,k})$ onto a Gaussian distribution which is the \ac{EP} exponential family approximation choice of $\lvec{g}_{l,k}$.
Therefore, the denominator of~\eqref{eq:app_JAC_m_Psi_y_g0} cancels with the first term in~\eqref{eq:app_JAC_mmd_Psi_y_g}, and the final message update rule is given by {the distribution}
\begin{equation}
	\msgp{\Psi_{y_{l,t}}}{\lvec{g}_{l,k}}(\lvec{g}_{l,k}) = \CN{\lvec{g}_{l,k}|\Mumsg{\Psi_{y_{l,t}}}{\lvec{g}_{l,k}},\Cmsg{\Psi_{y_{l,t}}}{\lvec{g}_{l,k}}},
	\label{eq:app_JAC_m_Psi_y_g}
\end{equation}
with
\begin{align}
	\Mumsg{\Psi_{y_{l,t}}}{\lvec{g}_{l,k}} &= \left(\lvec{y}_{l,t}-\sum_{k'\neq k}\Mumsg{\lvec{g}_{l,k'}}{\Psi_{y_{l,t}}}x_{k't}\right)\cdot x_{kt}^{-1},
	\label{eq:app_JAC_mu_Psi_y_g}\\
	\Cmsg{\Psi_{y_{l,t}}}{\lvec{g}_{l,k}} &= \left(\sigma_n^2\lmat{I}_N+\sum_{k'\neq k}\Cmsg{\lvec{g}_{l,k'}}{\Psi_{y_{l,t}}}|x_{k't}|^2\right)\cdot |x_{kt}|^{-2}.
	\label{eq:app_JAC_C_Psi_y_g}
\end{align}

\subsection{Message Updates for $\Psi_{g_{l,k}}$}\label{subsubsec:MP_Psi_g}

\subsubsection{Incoming messages to factor node $\Psi_{g_{l,k}}$}
\begin{align}
	\msgp{u_k}{\Psi_{g_{l,k}}}(u_k) &\propto \msgp{\Psi_{u_k}}{u_k}(u_k)\cdot\prod_{l'\neq l}\msgp{\Psi_{g_{l',k}}}{u_{k}}(u_k),
	\label{eq:app_JAC_m_u_Psi_g}\\
	\msgp{\lvec{h}_{l,k}}{\Psi_{g_{l,k}}}(\lvec{h}_{l,k}) &= \msgp{\Psi_{h_{l,k}}}{\lvec{h}_{l,k}}(\lvec{h}_{l,k}),
	\label{eq:app_JAC_m_h_Psi_g}\\
	\msgp{\lvec{g}_{l,k}}{\Psi_{g_{l,k}}}(\lvec{g}_{l,k})&\propto \prod_{t=1}^{T_p}\msgp{\Psi_{y_{l,t}}}{\lvec{g}_{l,k}}(\lvec{g}_{l,k})\nonumber\\
	&\propto \CN{\lvec{g}_{l,k}|\Mumsg{\lvec{g}_{l,k}}{\Psi_{g_{l,k}}},\Cmsg{\lvec{g}_{l,k}}{\Psi_{g_{l,k}}}},
	\label{eq:app_JAC_g_Psi_g}
\end{align}
with
\begin{align}
	\Cmsg{\lvec{g}_{l,k}}{\Psi_{g_{l,k}}} &= \left(\sum_{t=1}^{T_p}\Cmsg{\Psi_{y_{l,t}}}{\lvec{g}_{l,k}}^{-1}\right)^{-1},
	\label{eq:app_JAC_C_g_Psi_g}\\
	\Mumsg{\lvec{g}_{l,k}}{\Psi_{g_{l,k}}} &= \Cmsg{\lvec{g}_{l,k}}{\Psi_{g_{l,k}}}\left(\sum_{t=1}^{T_p}\Cmsg{\Psi_{y_{l,t}}}{\lvec{g}_{l,k}}^{-1}\Mumsg{\Psi_{y_{l,t}}}{\lvec{g}_{l,k}}\right),
	\label{eq:app_JAC_mu_g_Psi_g}
\end{align}
which is obtained by applying the Gaussian product lemma multiple times.

\subsubsection{Outgoing messages from factor node $\Psi_{g_{l,k}}$}
\hfill\\\underline{Message Update $\msg{\Psi_{g_{l,k}}}{u_k}$}\\
\begin{equation}
	\msgp{\Psi_{g_{l,k}}}{u_k}(u_k) \propto \frac{\proj{\mmd{\Psi_{g_{l,k}}}{u_k}(u_k)}}{\msgp{u_k}{\Psi_{g_{l,k}}}(u_k)},
	\label{eq:app_JAC_m_Psi_g_u0}
\end{equation}
with
\begin{align}
	\mmd{\Psi_{g_{l,k}}}{u_k}(u_k)&\propto \int\int\delta(\lvec{g}_{l,k}-\lvec{h}_{l,k}u_k)\cdot\msgp{u_k}{\Psi_{g_{l,k}}}(u_k)\cdot\msgp{\lvec{g}_{l,k}}{\Psi_{g_{l,k}}}(\lvec{g}_{l,k})\cdot\msgp{\lvec{h}_{l,k}}{\Psi_{g_{l,k}}}(\lvec{h}_{l,k})\,d\lvec{g}_{l,k}\,d\lvec{h}_{l,k}\nonumber\\
	&\overset{(a)}{=} \msgp{u_k}{\Psi_{g_{l,k}}}(u_k)\cdot\int\msgp{\lvec{g}_{l,k}}{\Psi_{g_{l,k}}}(\lvec{h}_{l,k}u_k)\cdot\msgp{\Psi_{h_{l,k}}}{\lvec{h}_{l,k}}(\lvec{h}_{l,k})\,d\lvec{h}_{l,k}\nonumber\\
	&= \msgp{u_k}{\Psi_{g_{l,k}}}(u_k)\cdot\vartheta(u_k),
	\label{eq:app_JAC_mmd_Psi_g_u}
\end{align}
where $(a)$ is obtained by the sifting property of the Dirac delta function and using~\eqref{eq:app_JAC_m_h_Psi_g}.
The final result in~\eqref{eq:app_JAC_mmd_Psi_g_u} is obtained by considering the fact that $u_k$ is a binary random variable with $u_k\in\{0,1\}$, utilizing~\eqref{eq:app_JAC_m_Psi_h_h} and the Gaussian product rule, and, then, integrating over $\lvec{h}_{l,k}$ with
\begin{equation}
	\vartheta(u_k) = \CN{\lvec{0}|\Mumsg{\lvec{g}_{l,k}}{\Psi_{g_{l,k}}},\Cmsg{\lvec{g}_{l,k}}{\Psi_{g_{l,k}}}+\gmat{\Xi}_{l,k}u_k}.
	\label{eq:app_JAC_theta_mmd_Psi_g_u}
\end{equation}
The projection operation in~\eqref{eq:app_JAC_m_Psi_g_u0} is superfluous since $\mmd{\Psi_{g_{l,k}}}{u_k}(u_k)$~\eqref{eq:app_JAC_mmd_Psi_g_u} is already categorically distributed.
Hence, the denominator of~\eqref{eq:app_JAC_m_Psi_g_u0} cancels with the first term in~\eqref{eq:app_JAC_mmd_Psi_g_u}.
The final message update rule is given by {the distribution}
\begin{equation}
	\msgp{\Psi_{g_{l,k}}}{u_k}(u_k) \propto \vartheta(u_k).
	\label{eq:app_JAC_m_Psi_g_u}
\end{equation}

\noindent\underline{Message Update $\msg{\Psi_{g_{l,k}}}{\lvec{h}_{l,k}}$}\\
\begin{equation}
	\msgp{\Psi_{g_{l,k}}}{\lvec{h}_{l,k}}(\lvec{h}_{l,k}) \propto \frac{\proj{\mmd{\Psi_{g_{l,k}}}{\lvec{h}_{l,k}}(\lvec{h}_{l,k})}}{\msgp{\lvec{h}_{l,k}}{\Psi_{g_{l,k}}}(\lvec{h}_{l,kt})},
	\label{eq:app_JAC_m_Psi_g_h0}
\end{equation}
with
\begin{align}
	\mmd{\Psi_{g_{l,k}}}{\lvec{h}_{l,k}}(\lvec{h}_{l,k})&\propto \sum_{u_k}\int\delta(\lvec{g}_{l,k}-\lvec{h}_{l,k}u_k)\cdot\msgp{u_k}{\Psi_{g_{l,k}}}(u_k)\cdot\msgp{\lvec{g}_{l,k}}{\Psi_{g_{l,k}}}(\lvec{g}_{l,k})\cdot\msgp{\lvec{h}_{l,k}}{\Psi_{g_{l,k}}}(\lvec{h}_{l,k})\,d\lvec{g}_{l,k}\nonumber\\
	&\overset{(a)}{=} \msgp{u_k}{\Psi_{g_{l,k}}}(0)\cdot\msgp{\lvec{g}_{l,k}}{\Psi_{g_{l,k}}}(\lvec{0})\cdot\msgp{\Psi_{h_{l,k}}}{\lvec{h}_{l,k}}(\lvec{h}_{l,k})+\msgp{u_k}{\Psi_{g_{l,k}}}(1)\cdot\msgp{\lvec{g}_{l,k}}{\Psi_{g_{l,k}}}(\lvec{h}_{l,k})\nonumber\\
	&\quad\cdot\msgp{\Psi_{h_{l,k}}}{\lvec{h}_{l,k}}(\lvec{h}_{l,k})\nonumber\\
\begin{split}
	&= \msgp{u_k}{\Psi_{g_{l,k}}}(0)\cdot\vartheta(0)\cdot\msgp{\Psi_{h_{l,k}}}{\lvec{h}_{l,k}}(\lvec{h}_{l,k})+\msgp{u_k}{\Psi_{g_{l,k}}}(1)\cdot\vartheta(1)\\
	&\quad\cdot\CN{\lvec{h}_{l,k}|\Mumsga{\Psi_{g_{l,k}}}{\lvec{h}_{l,k}},\Cmsga{\Psi_{g_{l,k}}}{\lvec{h}_{l,k}}},
\end{split}
	\label{eq:app_JAC_mmd_Psi_g_h}
\end{align}
where $(a)$ is obtained by making the sum explicit for $u_k$ and  utilizing the sifting property of the Dirac delta function and~\eqref{eq:app_JAC_m_h_Psi_g}.
The final result in~\eqref{eq:app_JAC_mmd_Psi_g_h} is obtained by using~\eqref{eq:app_JAC_m_Psi_h_h}, \eqref{eq:app_JAC_theta_mmd_Psi_g_u}, and the Gaussian multiplication lemma with
\begin{align}
	\Cmsga{\Psi_{g_{l,k}}}{\lvec{h}_{l,k}} &= \left(\Cmsg{\lvec{g}_{l,k}}{\Psi_{g_{l,k}}}^{-1}+\gmat{\Xi}_{l,k}^{-1}\right)^{-1},
	\label{eq:app_C_tmp_mmd_Psi_g_h}\\
	\Mumsga{\Psi_{g_{l,k}}}{\lvec{h}_{l,k}} &= \Cmsga{\Psi_{g_{l,k}}}{\lvec{h}_{l,k}}\Cmsg{\lvec{g}_{l,k}}{\Psi_{g_{l,k}}}^{-1}\Mumsg{\lvec{g}_{l,k}}{\Psi_{g_{l,k}}}.
	\label{eq:app_mu_tmp_mmd_Psi_g_h}
\end{align}
The normalization constant for $\mmd{\Psi_{g_{l,k}}}{\lvec{h}_{l,k}}(\lvec{h}_{l,k})$ is given by
\begin{align}
	\tilde{Z}_{\Psi_{g_{l,k}}} &= \int\msgp{u_k}{\Psi_{g_{l,k}}}(0)\cdot\vartheta(0)\cdot\msgp{\Psi_{h_{l,k}}}{\lvec{h}_{l,k}}(\lvec{h}_{l,k})\nonumber\\
	&\qquad+ \msgp{u_k}{\Psi_{g_{l,k}}}(1)\cdot\vartheta(1)\cdot\CN{\lvec{h}_{l,k}|\Mumsga{\Psi_{g_{l,k}}}{\lvec{h}_{l,k}},\Cmsga{\Psi_{g_{l,k}}}{\lvec{h}_{l,k}}}\,d\lvec{h}_{l,k}\nonumber\\
	&= \msgp{u_k}{\Psi_{g_{l,k}}}(0)\cdot\vartheta(0)+\msgp{u_k}{\Psi_{g_{l,k}}}(1)\cdot\vartheta(1).
	\label{eq:app_Z_Psi_g}
\end{align}
According to~\eqref{eq:app_JAC_mmd_Psi_g_h}, $\mmd{\Psi_{g_{l,k}}}{\lvec{h}_{l,k}}(\lvec{h}_{l,k})$ is a Gaussian mixture in two components with mean vector and covariance matrix
\begin{align}
	\Mumsgb{\Psi_{g_{l,k}}}{\lvec{h}_{l,k}} &= \frac{1}{\tilde{Z}_{\Psi_{g_{l,k}}}}\cdot\msgp{u_k}{\Psi_{g_{l,k}}}(1)\cdot\vartheta(1)\cdot\Mumsga{\Psi_{g_{l,k}}}{\lvec{h}_{l,k}},
	\label{eq:app_JAC_mu_mmd_Psi_g_h}\\
\begin{split}
	\Cmsgb{\Psi_{g_{l,k}}}{\lvec{h}_{l,k}} &= \frac{1}{\tilde{Z}_{\Psi_{z_{l,kt}}}}\Big(\msgp{u_k}{\Psi_{g_{l,k}}}(0)\cdot\vartheta(0)\cdot\gmat{\Xi}_{l,k}+\msgp{u_k}{\Psi_{g_{l,k}}}(1)\cdot\vartheta(1)\\
	&\qquad\qquad\quad\cdot\big(\Cmsga{\Psi_{g_{l,k}}}{\lvec{h}_{l,k}}+\Mumsga{\Psi_{g_{l,k}}}{\lvec{h}_{l,k}}\Mumsga{\Psi_{g_{l,k}}}{\lvec{h}_{l,k}}^H\big)\Big)-\Mumsgb{\Psi_{g_{l,k}}}{\lvec{g}_{l,k}}\Mumsgb{\Psi_{g_{l,k}}}{\lvec{g}_{l,k}}^H.
	\label{eq:app_JAC_C_mmd_Psi_g_h}
\end{split}
\end{align}
Note that the estimated posterior distribution~\eqref{eq:approx_factor_beta_factorization} of $\lvec{h}_{l,k}$ is proportional to the product of $\msgp{\Psi_{g_{l,k}}}{\lvec{h}_{l,k}}(\lvec{h}_{l,k})$ and $\msgp{\Psi_{h_{l,k}}}{\lvec{h}_{l,k}}(\lvec{h}_{l,kt})$ which is proportional to $\proj{\mmd{\Psi_{g_{l,k}}}{\lvec{h}_{l,k}}(\lvec{h}_{l,k})}$ according to~\eqref{eq:app_JAC_m_h_Psi_g} and~\eqref{eq:app_JAC_m_Psi_g_h0}.
Hence, the final estimate of $\lvec{h}_{l,k}$ can be computed via~\eqref{eq:app_JAC_mu_mmd_Psi_g_h} after the last iteration of the \ac{EP} algorithm.

By applying moment matching and the Gaussian quotient lemma to~\eqref{eq:app_JAC_m_Psi_g_h0}, we obtain for the final message update the following distribution,
\begin{equation}
	\msgp{\Psi_{g_{l,k}}}{\lvec{h}_{l,k}}(\lvec{h}_{l,k}) = \CN{\lvec{h}_{l,k}|\Mumsg{\Psi_{g_{l,k}}}{\lvec{h}_{l,k}},\Cmsg{\Psi_{g_{l,k}}}{\lvec{h}_{l,k}}},
	\label{eq:app_JAC_m_Psi_g_h}
\end{equation}
with
\begin{align}
	\Cmsg{\Psi_{g_{l,k}}}{\lvec{h}_{l,k}} &= \left(\Cmsgb{\Psi_{g_{l,k}}}{\lvec{h}_{l,k}}^{-1}-\gmat{\Xi}_{l,k}^{-1}\right)^{-1},
	\label{eq:app_C_Psi_g_h}\\
	\Mumsg{\Psi_{g_{l,k}}}{\lvec{h}_{l,k}} &= \Cmsg{\Psi_{g_{l,k}}}{\lvec{h}_{l,k}}\Cmsgb{\Psi_{g_{l,k}}}{\lvec{h}_{l,k}}^{-1}\Mumsgb{\Psi_{g_{l,k}}}{\lvec{h}_{l,k}}.
	\label{eq:app_JAC_mu_Psi_g_h}
\end{align}

\noindent\underline{Message Update $\msg{\Psi_{g_{l,k}}}{\lvec{g}_{l,k}}$}\\
\begin{equation}
	\msgp{\Psi_{g_{l,k}}}{\lvec{g}_{l,k}}(\lvec{g}_{l,k}) \propto \frac{\proj{\mmd{\Psi_{g_{l,k}}}{\lvec{g}_{l,k}}(\lvec{g}_{l,k})}}{\msgp{\lvec{g}_{l,k}}{\Psi_{g_{l,k}}}(\lvec{g}_{l,kt})},
	\label{eq:app_JAC_m_Psi_g_g0}
\end{equation}
with
\begin{align}
	\mmd{\Psi_{g_{l,k}}}{\lvec{g}_{l,k}}(\lvec{g}_{l,k})&\propto \sum_{u_k}\int\delta(\lvec{g}_{l,k}-\lvec{h}_{l,k}u_k)\cdot\msgp{u_k}{\Psi_{g_{l,k}}}(u_k)\cdot\msgp{\lvec{g}_{l,k}}{\Psi_{g_{l,k}}}(\lvec{g}_{l,k})\cdot\msgp{\lvec{h}_{l,k}}{\Psi_{g_{l,k}}}(\lvec{h}_{l,k})\,d\lvec{h}_{l,k}\nonumber\\
	&\overset{(a)}{=} \msgp{u_k}{\Psi_{g_{l,k}}}(0)\cdot\msgp{\lvec{g}_{l,k}}{\Psi_{g_{l,k}}}(\lvec{g}_{l,k})\cdot\delta(\lvec{g}_{l,k})+\msgp{u_k}{\Psi_{g_{l,k}}}(1)\cdot\msgp{\lvec{g}_{l,k}}{\Psi_{g_{l,k}}}(\lvec{g}_{l,k})\nonumber\\
	&\quad\cdot\msgp{\Psi_{h_{l,k}}}{\lvec{h}_{l,k}}(\lvec{g}_{l,k})\nonumber\\
\begin{split}
	&= \msgp{u_k}{\Psi_{g_{l,k}}}(0)\cdot\msgp{\lvec{g}_{l,k}}{\Psi_{g_{l,k}}}(\lvec{g}_{l,k})\cdot\delta(\lvec{g}_{l,k})+ \msgp{u_k}{\Psi_{g_{l,k}}}(1)\cdot\vartheta(1)\\
	&\quad\cdot\CN{\lvec{g}_{l,k}|\Mumsga{\Psi_{g_{l,k}}}{\lvec{g}_{l,k}},\Cmsga{\Psi_{g_{l,k}}}{\lvec{g}_{l,k}}},
	\label{eq:app_JAC_mmd_Psi_g_g}
\end{split}
\end{align}
where $(a)$ is obtained by making the sum explicit for $u_k$ and  utilizing the sifting property of the Dirac delta function and~\eqref{eq:app_JAC_m_h_Psi_g}
The final result in~\eqref{eq:app_JAC_mmd_Psi_g_g} is obtained by using~\eqref{eq:app_JAC_m_Psi_h_h}, \eqref{eq:app_JAC_theta_mmd_Psi_g_u}, and the Gaussian multiplication lemma with
\begin{align}
	\Cmsga{\Psi_{g_{l,k}}}{\lvec{g}_{l,k}} &= \Cmsga{\Psi_{g_{l,k}}}{\lvec{h}_{l,k}},
	\label{eq:app_JAC_C_tmp_mmd_Psi_g_g}\\
	\Mumsga{\Psi_{g_{l,k}}}{\lvec{g}_{l,k}} &= \Mumsga{\Psi_{g_{l,k}}}{\lvec{h}_{l,k}}.
	\label{eq:app_JAC_mu_tmp_mmd_Psi_g_g}
\end{align}
The mean vector and covariance matrix of the Bernoulli-Gaussian distribution $\mmd{\Psi_{g_{l,k}}}{\lvec{g}_{l,k}}(\lvec{g}_{l,k})$ in~\eqref{eq:app_JAC_mmd_Psi_g_g} is given by
\begin{align}
	\Mumsgb{\Psi_{g_{l,k}}}{\lvec{g}_{l,k}} &= \frac{1}{\tilde{Z}_{\Psi_{g_{l,k}}}}\cdot\msgp{u_k}{\Psi_{g_{l,k}}}(1)\cdot\vartheta(1)\cdot\Mumsga{\Psi_{g_{l,k}}}{\lvec{g}_{l,k}},
	\label{eq:app_JAC_mu_mmd_Psi_g_g}\\
\begin{split}
	\Cmsgb{\Psi_{g_{l,k}}}{\lvec{g}_{l,k}} &= \frac{1}{\tilde{Z}_{\Psi_{g_{l,k}}}}\cdot\msgp{u_k}{\Psi_{g_{l,k}}}(1)\cdot\vartheta(1)\cdot\Big(\Cmsga{\Psi_{g_{l,k}}}{\lvec{g}_{l,k}}+\Mumsga{\Psi_{g_{l,k}}}{\lvec{g}_{l,k}}\Mumsga{\Psi_{g_{l,k}}}{\lvec{g}_{l,k}}^H\Big)\\
	&\quad- \Mumsgb{\Psi_{g_{l,k}}}{\lvec{g}_{l,k}}\Mumsgb{\Psi_{g_{l,k}}}{\lvec{g}_{l,k}}^H.
	\label{eq:app_JAC_C_mmd_Psi_g_g}
\end{split}
\end{align}
By applying moment matching and the Gaussian quotient lemma to~\eqref{eq:app_JAC_m_Psi_g_g0}, we obtain for the final message update the following distribution,
\begin{equation}
	\msgp{\Psi_{g_{l,k}}}{\lvec{g}_{l,k}}(\lvec{g}_{l,k}) = \CN{\lvec{g}_{l,k}|\Mumsg{\Psi_{g_{l,k}}}{\lvec{g}_{l,k}},\Cmsg{\Psi_{g_{l,k}}}{\lvec{g}_{l,k}}},
	\label{eq:app_JAC_m_Psi_g_g}
\end{equation}
with
\begin{align}
	\Cmsg{\Psi_{g_{l,k}}}{\lvec{g}_{l,k}} &= \left(\Cmsgb{\Psi_{g_{l,k}}}{\lvec{g}_{l,k}}^{-1}-\Cmsg{\lvec{g}_{l,k}}{\Psi_{g_{l,k}}}^{-1}\right)^{-1},
	\label{eq:app_JAC_C_Psi_g_g}\\
	\Mumsg{\Psi_{g_{l,k}}}{\lvec{g}_{l,k}} &= \Cmsg{\Psi_{g_{l,k}}}{\lvec{g}_{l,k}}\left(\Cmsgb{\Psi_{g_{l,k}}}{\lvec{g}_{l,k}}^{-1}\Mumsgb{\Psi_{g_{l,k}}}{\lvec{g}_{l,k}}-\Cmsg{\lvec{g}_{l,k}}{\Psi_{g_{l,k}}}^{-1}\Mumsg{\lvec{g}_{l,k}}{\Psi_{g_{l,k}}}\right).
	\label{eq:app_JAC_mu_Psi_g_g}
\end{align}

\bibliographystyle{IEEEtran}
\bibliography{IEEEabrv,references}

\end{document}